\documentclass[11pt,a4paper]{article}

\pdfoutput=1
\usepackage{jcappub}
\usepackage{blindtext}
\usepackage{hyperref}
\usepackage{amsmath,amssymb}
\usepackage{float}
\usepackage{microtype}

\usepackage{graphicx}
\usepackage{bm}
\usepackage{latexsym}
\usepackage{epsfig}
\usepackage{psfrag}
\usepackage{color}
\usepackage[dvipsnames]{xcolor}
\usepackage{subfigure}
\usepackage{graphicx}

\usepackage{amssymb}
\usepackage{amsmath}
\usepackage{bm}
\usepackage{latexsym}
\usepackage{epsfig}
\usepackage{psfrag}
\usepackage{amsmath}
\usepackage[normalem]{ulem}
\usepackage{textcomp}
\usepackage{color}
\usepackage{pstricks}
\usepackage[utf8]{inputenc}
\usepackage{array}
\usepackage{multirow}
\usepackage{comment}
\usepackage{mathalpha}

\usepackage{tabularray}
\makeatletter
\gdef\@fpheader{}
\makeatother

\def\beq{\begin{equation}}
\def\eeq{\end{equation}}
\def\bea{\begin{eqnarray}}
\def\eea{\end{eqnarray}}
\def\be{\begin{equation}}
\def\ee{\end{equation}}

\def\bse{\begin{subequations}}
\def\ese{\end{subequations}}

\def\cP{\mathcal{P}}

\def\ee{\eta_{\rm end}}

\def\Mpl{M_{_{\mathrm{Pl}}}}
\def\f{\frac}
\def\l{\left}
\def\r{\right}
\def\d{\mathrm{d}}
\def\e{{\mathrm{e}}}

\def\nn{\nonumber}
\def\wf{w_{\mathrm{\phi}}}
\def\tre{T_{\mathrm {RH}}}

\def\tbh{T_{\mathrm {BH}}}
\def\tev{T_{\mathrm {ev}}}

\def\min{M_{\mathrm {in}}}

\def\mbh{M_{\mathrm {BH}}}

\def\ogw{\Omega_{_{\mathrm{GW}}}}

\def\Hi{H_\mathrm{inf}}
\def\hend{H_\mathrm{end}}
\def\hre{H_{_\mathrm{RH}}}

\def\hin{H_\mathrm{in}}
\def\hev{H_\mathrm{ev}}

\def\Are{A_{_\mathrm{RH}}}

\def\ain{a_\mathrm{in}}
\def\aend{a_\mathrm{end}}
\def\are{a_{_\mathrm{RH}}}

\def\aev{a_\mathrm{ev}}

\def\kend{k_{\rm end}}
\def\kre{k_{\rm RH}}
\def\kin{k_{\rm in}}

\def\kbh{k_{\rm BH}}
\def\kev{k_{\rm ev}}
\def\kuv{k_{\rm UV}}

\def\rhor{\rho_{\rm R}}
\def\rhobh{\rho_{\rm BH}}
\def\rhophi{\rho_{\rm \phi}}
\def\rhore{\rho_{\rm RH}}
\def\betac{\beta_{\rm c}}
\def\ns{n_{\rm s}}

\def\cs{c_{\rm s}}

\allowdisplaybreaks

\begin{document}
\title{Probing the early universe with future GW observatories}

\author{Suvashis Maity$^{a}$ and}
\emailAdd{$^a$suvashis@physics.iitm.ac.in}
\author{Md Riajul Haque$^{b}$}
\emailAdd{$^b$riaj.0009@gmail.com}

\affiliation{Centre for Strings, Gravitation and Cosmology,
Department of Physics, Indian Institute of Technology Madras, 
Chennai~600036, India}


\abstract
{ One of the fundamental characteristics of slow roll inflation is its generation of tensor perturbations, which manifest as stochastic gravitational waves (GWs). 
Slow roll inflation results in a nearly scale-invariant GW spectrum that maintains its scale invariance as it transitions into the radiation-dominated era. However, introducing an intermediate reheating phase can modify the spectral tilt, depending on the equation of state governing that particular epoch. These GWs, especially on smaller scales, are anticipated to be observable by forthcoming GW detectors. In this study, we initially delineate the parameter space encompassing the inflationary energy scale, reheating temperature, and equation of state in a model-independent manner, focusing on the spectra detectable by GW detectors such as LISA, ET, DECIGO, and BBO. We also examine the implications for the $\alpha$-attractor model of inflation and explore the observational constraints on $n_s-r$ prediction in the light of GW detection. Then, we point out the probable ranges for various non-gravitational and gravitational coupling between the inflaton and Standard Model particles considering the perturbative reheating. If one assumes PBHs were formed during the early reheating era, such detection of GW signal also sheds light on the probing PBH parameters. Note that for the case of PBH domination, we also consider the contribution of the induced GWs due to the density fluctuation in PBH distribution, which helps to decode the phase of early PBH domination. Finally, to test the production of other cosmological relics through future GW missions, we consider dark matter produced via gravitational interaction in the early universe.}

\maketitle


\section{Introduction \label{sec: Introduction}}

In modern cosmology, observations on the different cosmological scales are considered to be the outcome of various processes that occurred during the early stages of cosmic evolution. Inflation has been shown to be one of the most successful paradigms, explaining both the very isotropic character of Cosmic Microwave Background (CMB) and the minuscule anisotropies exhibited. The observational data on the large scales can be well-fitted by a nearly scale-invariant scalar power spectrum resulting from slow-roll inflation driven by a canonical scalar field. At the pivot scale of $0.05~{\rm Mpc}^{-1}$, the amplitude of the scalar power spectrum, $A_s$ is restricted to be $A_s\simeq 2.09\times 10^{-9}$ \cite{Planck:2018vyg}. Although the primary tensor spectrum cannot yet be directly detected on these scales, observational constraints provide an upper bound on the tensor spectrum strength at the pivot scale, measured by the tensor-to-scalar ratio, which is the ratio of the tensor power spectrum to the scalar power spectrum. 

 Inflation offers an exclusive mechanism for generating GWs of quantum mechanical origin in addition to the many astrophysical mechanisms that can also generate GWs (for discussions on the formation of primary and secondary GWs, see, for example, Refs~\cite{Giovannini:1998bp,Giovannini:1999bh,Riazuelo:2000fc,
Seto:2003kc,Boyle:2007zx,Stewart:2007fu,Li:2021htg,Artymowski:2017pua,Caprini:2018mtu,Bettoni:2018pbl,Figueroa:2019paj,Bernal:2020ywq,
Vagnozzi:2020gtf,Mishra:2021wkm,Haque:2021dha,Benetti:2021uea,Giovannini:2022eue,Easther:2006gt,Ananda:2006af,Baumann:2007zm,Espinosa:2018eve,Domenech:2019quo,Ragavendra:2020sop,Inomata:2023zup,Franciolini:2023pbf,Firouzjahi:2023lzg,Maity:2024odg,Domenech:2020ssp,Papanikolaou:2020qtd,Domenech:2021wkk,Papanikolaou:2022chm,Bhaumik:2022pil,Bhattacharya:2020lhc,Braglia:2024kpo,Braglia:2024siw,Soman:2024zor,Barman:2023ktz} and references therein). The tensor perturbations generated from vacuum fluctuations are amplified during inflation and subsequently evolve through the different phases of the universe until they are detected by GW detectors today. As a result, the genesis and dynamics of primordial GWs are convoluted into their spectrum today. They bear the marks of the process that gives rise to them, namely the processes responsible for driving inflation and the initial conditions that give rise to the perturbations. On the other hand, they also carry the imprints of the dynamics of the subsequent cosmological phases as the GWs propagate through those phases. A natural consequence after inflation is the decay of the inflaton into the Standard Model (SM) particles, which finally ushers in the period of radiation dominance. In this article, we will study the evolution of primordial GWs, focusing on the consequences caused by the reheating period. Our specific goal is to understand the reheating mechanism from the GW spectrum today with the help of forthcoming GW detectors.

 Reheating is characterized by the dominance of inflaton energy, with energy transfer facilitated by coupling between the inflaton and SM particles, eventually damping the coherent oscillations through particle production. While many studies often assume instantaneous reheating, a more realistic picture involves finite-duration reheating, lasting several e-folds and potentially extending until the time of Big Bang Nucleosynthesis (BBN). Various mechanisms facilitate energy transfer from the inflaton to SM particles, including non-gravitational couplings \cite{Drewes:2017fmn,Garcia:2020eof,Haque:2020zco,Garcia:2020wiy}, gravitational interaction \cite{Haque:2022kez,Clery:2021bwz,Clery:2022wib,Co:2022bgh,Ahmed:2022tfm,Haque:2023zhb}. In addition, ultralight primordial black holes (PBHs) can, in principle, form during the early reheating phase. Evaporation of those PBHs leads to the production of SM particles, and eventually, reheating occurs\footnote{Note that for such a reheating scenario where PBHs are responsible for the production of SM particles, we consider a very weak coupling between the inflation and SM field to drive such production subdominant and always consider the inflation equation of state greater or equal to that of the relativistic particles to make the inflaton redshift process equal or faster than radiation.}~\cite{RiajulHaque:2023cqe,Haque:2024cdh,Barman:2024slw}. PBHs can be demonstrated to be formed by gravitational collapse if the amplitude of the local density fluctuations is sufficiently strong above a threshold value $\delta_c$ ($\frac{\delta \rho}{\rho} \gtrsim \delta_c \sim 1$). Several methods have been examined in the literature to produce such a large local density fluctuation, taking into account various physically motivated scenarios: quantum fluctuations generated during inflation through single field \cite{PhysRevD.48.543,PhysRevD.50.7173,Yokoyama:1998pt,Saito:2008em,Garcia-Bellido:2017mdw}, multi-fields \cite{Yokoyama:1995ex,Randall:1995dj,Garcia-Bellido:1996mdl,Pi:2017gih}, collapse of cosmic string loops \cite{MacGibbon:1990zk,Jenkins:2020ctp,Helfer:2018qgv,Matsuda:2005ez,Lake:2009nq}, collapse of domain walls \cite{Rubin:2000dq,Rubin:2001yw}, bubble collision during phase transition \cite{KodamaPTP1979} and many more~\cite{Lu:2024xnb,Lu:2024zwa,Flores:2024lng,Flores:2023zpf,Flores:2024eyy,Ballesteros:2024hhq}. Instead of digging into the formation mechanism, we study this PBH evolution in a model independent manner and investigate decoding its parameters in the light of GW detectors. To do so, we further consider another source of GWs due to the density fluctuations in PBH distribution, which is only effective when PBH decays after it dominates the universe~\cite{Domenech:2020ssp,Papanikolaou:2020qtd,Domenech:2021wkk,Papanikolaou:2022chm,Bhaumik:2022pil}. 
 
  The inception of GW astronomy commenced with the landmark detections of GWs originating from binary mergers by LIGO/VIRGO \cite{LIGOScientific:2016aoc,LIGOScientific:2016dsl,LIGOScientific:2016wyt,LIGOScientific:2017bnn,LIGOScientific:2017ycc,LIGOScientific:2017vox}. Subsequent studies have endeavored to discern whether black holes could have a primordial origin. Recently, Pulsar Timing Array (PTA) collaborations, including NANOGrav~\cite{NANOGrav:2023gor, NANOGrav:2023hde}, EPTA (incorporating data from InPTA) \cite{Antoniadis:2023lym,EPTA:2023sfo}, PPTA \cite{Zic:2023gta, Reardon:2023gzh}, and CPTA~\cite{Xu:2023wog}, jointly confirmed the presence of a stochastic GW background around nano Hz frequency. We adhere to the scenario where perturbations arise from vacuum fluctuations during single-field slow roll inflation, yielding a nearly scale-invariant tensor spectrum. However, such spectra are unlikely to be observed scale invariant for modes re-entering during the radiation-dominated (RD) era. We shall also clarify that any tilt to the primary GW (PGW) spectra introduced by the reheating is also challenging for PTA to detect due to the limitations of the extensions of the reheating era set by BBN, and the PTA frequency band stands just above the BBN scale. So our focus lies in exploring the prospects of detecting PGWs  through the forthcoming detectors that lie in higher frequencies than PTA, such as 
LISA (Laser Interferometer Space Antenna) \cite{Amaro-Seoane:2012aqc, Bartolo:2016ami}, 
DECIGO (Deci-hertz Interferometer Gravitational-Wave Observatory) \cite{Kawamura_2011, Kawamura:2019jqt}, 
BBO (Big Bang Observer) \cite{Crowder:2005nr, Corbin:2005ny, Baker:2019pnp},
and ET (Einstein telescope) \cite{Sathyaprakash:2012jk}. Out of the four detectors, LISA, DECIGO, and BBO are space-based detectors that consist of laser interferometers in a triangular configuration, whereas ET is a ground-based detector. The frequency bands in which these detectors operate are different, and together, the observational range spans from $10^{-5}$-$10^2$ Hz. PGWs offer a robust probe for determining the energy scale of inflation, and the parameters associated with the reheating phase, and their spectrum exhibits weaker dependence on inflationary dynamics in most of the slow roll models of inflation, manifesting in a power law behavior in the spectrum depending on the post-inflationary dynamics namely the phase of reheating.

 Let us briefly discuss what insights can be gleaned from the detection of these PGWs regarding the early universe. The spectral tilt of the PGW spectrum can provide information about the Equation of State (EoS) during reheating. This, in turn, helps to constrain the shape of the potential near its minimum and sheds light on the dynamics of reheating. Furthermore, for a fixed scale of inflation, the detection allows us to determine the timescale over which the inflaton field dominates, which offers for determining the reheating temperature, or the EoS governs the background before the onset of the RD era. The GW spectra also offer insights into the inflationary energy scale. Moreover, such detection through future GW missions aids in determining the coupling of the inflaton to SM and enlightens in probing the PBH-dominated phase if one assumes PBHs are present in the very early universe. Other than decaying into SM particles, inflaton can also transfer its energy into other beyond standard model particles, such as dark matter (DM). Hence, the detection of GWs will also shed light on the nature of the DM (in this context, see Ref. \cite{Ghoshal:2024gai}, where the authors discussed non-thermal DM and baryogenesis scenarios through GW detection).
 
The paper is organized as follows. 
In Section \ref{sec: pgw}, we delve into the generation of tensor perturbations stemming from vacuum fluctuations and explore how the reheating era influences the spectrum as the modes evolve through different phases.
Section \ref{sec: detector} is dedicated to examining various detectors and outlining the methodology employed in our analysis.
Moving on to Section \ref{sec: inflationary model}, we focus on a slow roll inflationary model, specifically the $\alpha$-attractor E model, and elucidate the relationship between the model parameters and the parameters describing the reheating phase.
Section \ref{sec: reheating} begins with an introductory discussion on the fundamentals of reheating dynamics. Subsequently, we explore distinct reheating scenarios, encompassing the production of SM particles through both non-gravitational and gravitational coupling. Additionally, we delve into a reheating scenario wherein the radiation bath originates from the evaporation of PBHs. Here, we also discuss the induced GW spectrum by fluctuations in the PBH number density, which is only important in the case of PBH domination.
In Section \ref{sec: DM production}, we analyze the mechanism behind the production of DM through gravitational interaction.
Section \ref{sec: results} is devoted to discussing the outcomes of our study, specifically the potential parameter space that can be probed through the detection of GWs using the Forecast of LISA, ET, DECIGO, and BBO.
Finally, in Section \ref{sec: conclusions}, we provide a brief summary and draw conclusions based on our findings.

Let us make a few clarifying remarks on the conventions and notations that we shall adopt in this work. 
We shall work with natural units such that $\hbar=c=1$, and set the reduced Planck mass to be $\Mpl=\l(8\pi G\r)^{-1/2} \simeq 2.4 \times10^{18}\, \mathrm{GeV}$.
The cosmic time, conformal time, the number of e-fold, and the scale factor are denoted as $t$, $\eta$, $N$, and $a$, respectively.
We shall adopt the signature of the metric to be~$(-,+,+,+)$.


\section{Evolution of primary GWs through reheating \label{sec: pgw}}
{\black
In this section, we shall briefly recall the discussion on the generation of the tensor perturbation during inflation and its propagation through different epochs until today. The primary GWs are considered to be generated as tensor fluctuations during inflation in the absence of any source term. Taking into account the tensor perturbations, the line element describing the spatially flat FLRW universe can be expressed as \cite{Maggiore:1999vm}
\bea 
\d s^2=a^2(\eta)\l[-\d\eta^2+(\delta_{ij}+h_{ij})\,\d x^i\d x^j \r],
\eea 
where $h_{ij}$ is the tensor perturbations considered to be transverse and traceless, i.e., $\partial_i h_{ij}=h^i_i=0$.
We write the equation of motion of the tensor perturbation in the Fourier space, given by
\bea 
h_k''+2\frac{a'}{a}h_k'+k^2h_k=0.
\eea
One can further define the Mukhanov Sasaki variable given by $u_k=a\,h_k$, and the equation of motion  of the Mukhanov Sasaki variable is as follows
\cite{MUKHANOV1992203,Martin:2003bt,Martin:2004um,Bassett:2005xm,Sriramkumar:2009kg,Baumann:2008bn,Baumann:2009ds}
\bea 
u_k''+\l(k^2-\f{a''}{a} \r)u_k =0.
\eea 
This equation for the mode can be solved in terms of Bessel's functions, and the solution depends on the background evolution, which is controlled by the Hubble parameter.
The power spectrum is defined through the two-point correlation in the Fourier space as
\bea 
\cP_{T}(k)=4\f{k^3}{2\pi^2}|h_k|^2,
\label{eq: pt}
\eea 
where the factor $4$ arises due to the polarization. Note that the spectrum is calculated on the super-Hubble scales during inflation. We assume the most simple scenario of the slow roll inflation considering the de Sitter approximation. In such cases, the scale factor during inflation and during reheating behaves as  
\bea 
a_{\rm I}(\eta) &=& \f{\aend}{1-\Hi\aend(\eta-\ee)},\\ 
\are(\eta) &=& \aend\l[1+\f{\Hi\,(1+3\wf)}{2}\aend(\eta-\ee) \r]^{\frac{2}{1+3\wf}},
\eea 
where $\Hi$ is the Hubble parameter during inflation taken to be constant as we assume de Sitter inflation. The scale factor is chosen in such a way that $a_{\rm I}(\ee)=\are(\ee)=\aend$, i.e., continuous across the end of inflation.
If we redefine the conformal time as $\tau=\eta-\ee-(1/\aend \Hi)$ during inflation and $\tau=\eta-\ee-\frac{2}{\aend \Hi\,(1+3\wf)}$ during reheating, the scale factors reduce to the following form
\bea 
a_{\rm I}(\tau) &=& -\f{1}{\tau\Hi}  ,~~~ 
\are(\tau) = \aend \l( \f{\tau}{\tau_{\rm end}}\r)^{\frac{2}{1+3\wf}},
\eea
with $\tau_{\rm end}=-1/\aend \Hi$. Using these forms of the scale factor, we can find the solution to the mode function during inflation. At the end of inflation, the mode is given as 
\bea 
h_k(\tau_{\rm end})&=& \f{\sqrt{2}}{\Mpl} \f{i\Hi}{\sqrt{2k^3}} \l(1-i\f{k}{\kend}\r)\e^{ik/\kend},
\label{eq: mode inflation}
\eea 
where $\kend$ is the mode that leaves the Hubble radius at the end of inflation and $\kend=-1/\ee$. Note that this solution is obtained by using the Bunch-Davies initial condition. 
The corresponding power spectrum at the end of inflation can be obtained using the solution of Eq.~\eqref{eq: mode inflation}
in Eq.~\eqref{eq: pt}
\bea 
\cP_T(k) &=& \f{2\Hi^2}{\pi^2\Mpl^2} \l(1+\f{k^2}{\kend^2}\r).
\eea 
It is clear from the above equation that in the limit $k\ll \kend$ de Sitter inflation generates a scale-invariant tensor power spectrum. We shall mention that slow roll inflation will introduce a very small red tilt to the power spectrum, which can be neglected. In this work, we shall stick to the scale-invariant tensor power spectrum during inflation at all scales. It is also to be noted that our analysis assumes a sharp transition from inflation to reheating and reheating to radiation domination. To follow the  evolution of the tensor perturbation in the post-inflationary era, it is convenient to write the tensor perturbation as
\bea 
h_k(\eta)= h_k(\tau_{\rm end})~\chi_k(\eta)\,,
\eea 
where $h_k(\tau_{\rm end})$ is the mode function  given in Eq.~\eqref{eq: mode inflation} and $\chi_k(\eta)$ is the transfer function which satisfies the following equation of motion \cite{Haque:2021dha,Mishra:2021wkm}
\bea 
\chi_k''+2\f{a'}{a}\chi_k'+k^2\chi_k=0.
\eea 
 We define a quantity $A=a/\aend$, with $\aend$ being the scale factor at the end of inflation. Using the relation $\d/\d \tau=(\aend A^2H)~\d/\d A$ we can write
\bea 
\f{\d^2\chi_k}{\d A^2}+ \l[ \f{4}{A}+ \f{1}{H}\f{\d H}{\d A}\r]\f{\d \chi_k}{\d A} +\f{(k/\kend)^2}{(H/\Hi)^2A^4}\chi_k =0.
\label{eq: eom transfer function1}
\eea 
 During reheating the Hubble parameter behaves as $H=\Hi A^{-(3/2)(1+\wf)}$ and the equation of motion for $\chi_k$ reduces to 
\bea 
\f{\d^2\chi_k}{\d A^2}+  \f{5-3\wf}{2A}\f{\d \chi_k}{\d A} +\f{(k/\kend)^2}{A^{1-3\wf}}\chi_k =0 .
\label{eq: eom transfer function reh}
\eea 
We obtain the solution of $\chi_k$ from the above equation in terms of the Bessel functions as 
\bea 
\chi_k^{\rm RH}(A)= A^{-\nu} \l[  C_{1k}~ J_{-\f{\nu}{\gamma}}\l(\f{k}{\gamma \kend}A^{\gamma}\r)  +C_{2k}~ J_{\f{\nu}{\gamma}}\l(\f{k}{\gamma \kend }A^{\gamma}\r)  \r],
\label{eq: sol transfer function reh}
\eea 
where $\nu$ and $\gamma$ are functions of the EoS given as $\nu=(3/4)(1-\wf)$ and $\gamma=(1/2)(1+3\wf)$. The coefficients $C_{1k}$ and $C_{2k}$ can be found by matching the solution and its derivative at the end of inflation, $\chi_k(A=1)=1$ and $\chi_k'(A=1)=-(k/\kend)^2/[1-i(k/\kend)]$ 
\bea 
C_{1k} &=& \f{\pi\,\pi}{2\gamma \,\kend} \l[ \f{k}{ik-\kend} J_{\f{\nu}{\gamma}}\l(\f{k}{\gamma \kend}\r) - J_{\f{\gamma+\nu}{\gamma}}\l(\f{k}{\gamma \kend}\r)\r] {\rm cosec}\l(\f{\pi\nu}{\gamma}\r),\nn\\
C_{2k} &=& \f{\pi\,\pi}{2\gamma \,\kend} \l[ \f{k}{ik-\kend} J_{-\f{\nu}{\gamma}}\l(\f{k}{\gamma \kend}\r) - J_{-\f{\gamma+\nu}{\gamma}}\l(\f{k}{\gamma \kend}\r)\r] {\rm cosec}\l(\f{\pi\nu}{\gamma}\r).\nn
\eea 
During the RD era, the Hubble parameter behaves as $H=\hre(\Are/A)^2$ with $\Are $ and $\hre$ are scale factor and Hubble parameter at the end of reheating, respectively. It is also straightforward to see that $\hre=\hend\Are^{-(3/2)(1+\wf)}$. One can find the EOM of the transfer function during RD takes the following form 
\bea 
\f{\d^2\chi_k}{\d A^2}+  \f{2}{A}\f{\d \chi_k}{\d A} +\f{(k/\kre)^2}{\Are^{2}}\chi_k =0 ,
\label{eq: eom transfer function rd}
\eea 
where $\kre = \are \hre$ is the mode that re-enters the Hubble radius at the end of the reheating. The solution of the above equation is found to be  
\bea 
\chi_k^{\rm RD}(A)= \f{1}{A} \l[D_{1k}~\exp\l(-\f{ikA}{\kre\Are}\r) + D_{2k}~\exp\l(\f{ikA}{\kre\Are}\r)\r],
\eea 
where as usual, the coefficients $D_{1k}$ and $D_{2k}$ are calculated by matching the solution of the mode and its derivative across the reheating and have the following form
\bea 
D_{1k} &=& \f{\Are}{2}\e^{ik/\kre} \l[\l(1+\f{i\kre}{k}\r) \chi_k^{\rm RH}(\Are) + i\f{\kre}{k}\Are\f{\chi_k^{\rm RH}(\Are)}{\d A}  \r] = i\f{\Are\kre}{2k} {\cal E}_{1k}, \nn\\
D_{2k} &=& \f{\Are}{2}\e^{-ik/\kre} \l[\l(1-\f{i\kre}{k}\r) \chi_k^{\rm RH}(\Are) - i\f{\kre}{k}\Are\f{\chi_k^{\rm RH}(\Are)}{\d A}  \r] = -i\f{\Are\kre}{2k} {\cal E}_{2k}.\nn
\eea 
It is straightforward to see from the above relations that $D_{1k}=D_{2k}^{\ast}$. 

With the solutions in hand for the transfer functions during reheating and during the RD era, we shall now calculate the energy density of the GWs.
At any given time $\eta $, the quantity is given by 
\bea 
\rho_{_{\rm GW}}(\eta) = \f{\Mpl^4}{4a^2} \l(\f{1}{2} \langle {\hat h_{ij}'^2}\rangle  +\f{1}{2}\langle {(\partial\hat h_{ij})^2}\rangle\r),
\eea 
and the energy density of the GWs per logarithmic interval is given by the relation $\rho_{_{\rm GW}}(k,\eta) = \d\rho_{_{\rm GW}}(\eta)/\d\ln k$. We find in Fourier space, it follows
\bea 
\rho_{_{\rm GW}}(k,\eta) &=& \f{\Mpl^2}{2a^2} \f{k^3}{2\pi^2} \l(k^2|h_k|^2+|h_k'|^2\r) =\f{\Mpl^4}{2\pi^2a^2}k^3 |h_k(\tau_{\rm end})|^2 \l( \f{k^2}{2}|\chi_k^{\rm RD}| + \f{1}{2} |{\chi_k^{\rm RD}}'|   \r)\nn\\
&=& \f{\Mpl^2k^2}{8\aend^2A^4} \cP_T(k) \Big[(|D_{1k}|^2+|D_{2k}|^2)\l\{ 2+\l(\f{\Are\kre}{Ak}\r)\r\}\nn\\ 
&& + \l(\f{\Are\kre}{Ak}\r)\Big\{ D_{1k}D_{2k}^\ast \l(1+\f{2ikA}{\kre\Are}\r) \exp\l(-\f{2ikA}{\kre\Are}\r) 
\nn\\&&\quad\quad\quad\quad 
+D_{1k}^\ast D_{2k} \l(1-\f{2ikA}{\kre\Are}\r) \exp\l(\f{2ikA}{\kre\Are}\r)
\Big\}\Big].
\label{eq: gw energy density1}
\eea 
We are interested in the spectrum of the GW energy density at the present time. During the late RD era, at the limit, $A/\Are\gg 1$, the first term in Eq.~\eqref{eq: gw energy density1} dominates over the other terms, and the energy density reduces to
\bea 
\rho_{_{\rm GW}}(k,\eta) &=& \f{\Mpl^2k^2}{4\aend^2A^4} \cP_T(k) \l(|D_{1k}|^2+|D_{2k}|^2\r).
\label{eq: gw energy density2}
\eea 
The dimensionless density parameter of GWs is described by the ratio between the energy density of GWs to the critical energy density of the universe, $\rho_c(\eta)=3\Mpl^2H^2(\eta)$. Using the expression of the energy density from Eq.~\eqref{eq: gw energy density2} and after some simplifications, it can be shown that
\bea 
\ogw(k,\eta)= \f{\rho_{_{\rm GW}}(k,\eta)}{\rho_c(\eta)}
 = \f{\cP_T(k)}{48} (|{\cal E}_{1k}|^2+|{\cal E}_{2k}|^2).
\eea 
After entering the Hubble radius, the energy density of GWs behaves as the energy density of radiation, and it falls as $a^{-4}$. At the present time, the density parameter of GWs can be obtained in terms of the current radiation energy density as 
\bea 
\ogw(k)h^2=c_g \,\Omega_{_{\rm rad,0}} h^2 ~\ogw(k,\eta),
\eea
where $c_g= (g_{\ast,{\rm eq}}/g_{\ast,{0}}) (g_{\ast S,0}/g_{\ast S,{\rm eq}})^{4/3}$, where $g_{\ast,{\rm eq}}$, $g_{\ast,{0}}$ and $g_{\ast S,{\rm eq}}$, $g_{\ast S,0}$ represent the number of relativistic degrees of freedom and entropy degrees of freedom at equality and today respectively. The quantity $\Omega_{_{\rm rad,0}} h^2 =4.16\times 10^{-5}$ (i.e., including photons and all three species of neutrinos). One can show at the limit $k\ll\kre$, $|{\cal E}_{1k}|^2=|{\cal E}_{2k}|^2 \simeq 1$ and at the limit $k\gg\kre$,
$
|{\cal E}_{1k}|^2=|{\cal E}_{2k}|^2=\f{4\gamma^2}{\pi} \Gamma^2\l(1+\f{\nu}{\gamma}\r) \l(\f{k}{2\gamma\kre}\r)^{n_{_{\rm GW}}}
$
where the spectral index of the GW spectrum $n_{_{\rm GW}}=1-(2\nu/\gamma)=-2(1-3\wf)/(1+3\wf)$.
We find that for $k\geq 2\gamma\kre$, there is a relative redshift based on the background EoS, which introduces a spectral tilt to the spectrum. Thus, for  $\wf>1/3$, we expect a blue titled spectrum for those wave numbers re-entering during reheating; on the other hand, for $\wf<1/3$, the GW spectrum will be red-tilted. 
Finally, the GW spectrum today in the limit $k\ll \kre$ can be written as $\ogw^{\rm rad}(k)= \Omega_{_{\rm rad,0}} h^2  \Hi^2/(12\pi^2\Mpl^2)$ and at the limit $k\gg\kre$ 
\bea 
\ogw(k) &=& \ogw^{\rm rad}(k)\f{4\gamma^2}{\pi} \Gamma^2\l(1+\f{\nu}{\gamma}\r) \l(\f{k}{2\gamma\kre}\r)^{n_{_{\rm GW}}}.
\label{eq: gw energy density final}
\eea

\begin{figure}[t!]
    \centering
    \includegraphics[scale=.39]{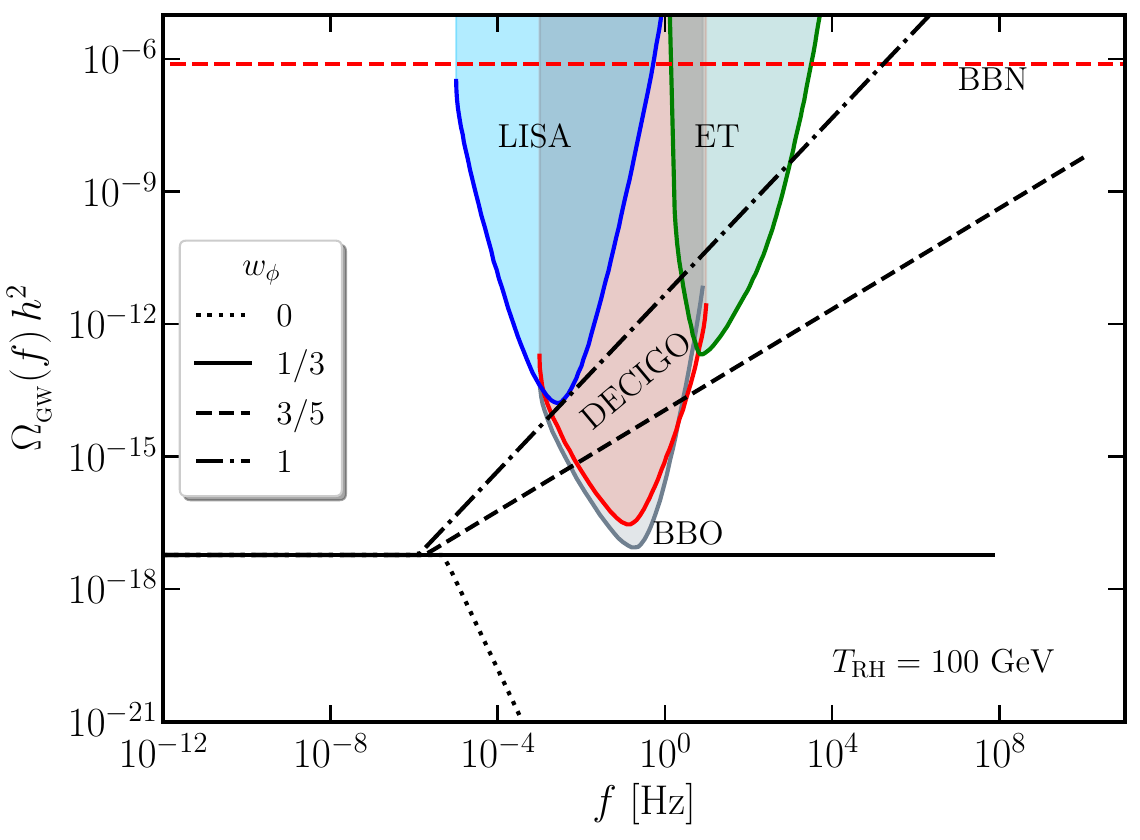}
    \includegraphics[scale=.39]{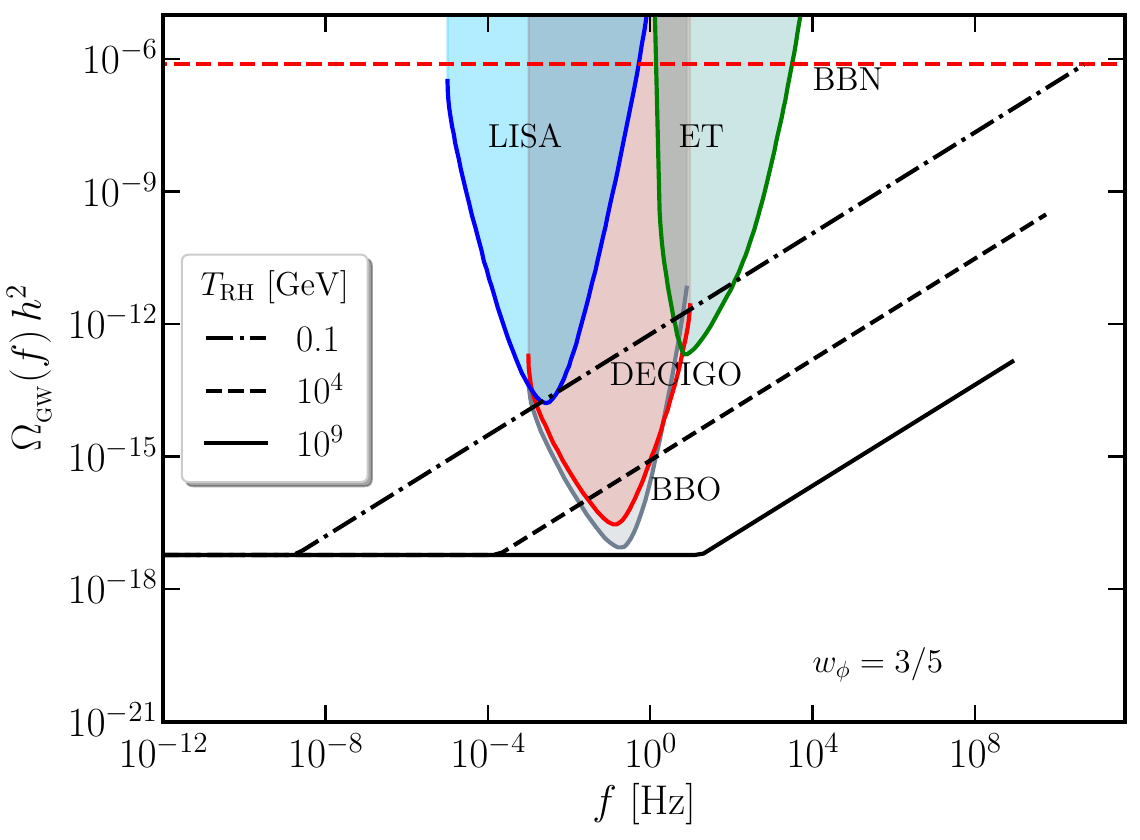}
    \caption{ {The spectrum of GWs is plotted in terms of frequency. On the left panel, we plot $\ogw$ for $\tre=100$ GeV for three different values of $\wf$, $1/3$, $3/5$, and $1$. On the right panel, we fix $\wf=3/5$ and choose three different values of $\tre=(0.1,\,10^4,\,10^9)$ GeV. }}
    \label{fig: pgw spectrum}
\end{figure}
In Fig.~\ref{fig: pgw spectrum}, we illustrate the behavior of $\ogw$ in terms of the frequency for various $\wf$ and $\tre$. We see that for $\wf=1/3$, the spectrum remains scale invariant as the GWs also behave as radiation. We see that a steeper EoS will produce a steeper spectrum, and with the variation of the inflation equation state $\wf=(0,\,1)$, the spectral tilt of the GWs varies from $(-2\to 1)$. Moreover, from the right panel of Fig.~\ref{fig: pgw spectrum}, it can be understood that the $\tre$ controls the duration of the tilted spectrum. As we decrease $\tre$, the duration for the tilted spectrum increases. However, for a stiff equation, the smaller value of $\tre$ can be restricted if one considers the additional relativistic degrees of freedom coming from the GWs.  Now, we shall discuss how small $\tre$ we can consider in order to be consistent with the $\Delta N_{\rm eff}$ constraints once we fix the EoS and the energy scale of the inflation.  $\Delta N_{\rm eff}$ is the additional relativistic degrees of freedom at the time of BBN or CMB decoupling. In the case of the GWs the expression of $\Delta N_{\rm{eff}}$ can be written as \cite{Jinno:2012xb}
\begin{equation}
    \Delta N_{\rm{eff}}=\frac{\rho_{_{\rm GW}}}{\,\rho_{\nu}}
   =\frac{8}{7}\left(\frac{11}{4}\right)^{\frac{4}{3}}\frac{\rho_{_{\rm GW}}}{\rho_{\rm \gamma}}\,,
   \label{Eq: neff}
\end{equation}
where $\rho_\nu$ and $\rho_{\rm \gamma}$ are the energy density of single SM neutrino species and the photon energy density, respectively. 
We have used the relation between the neutrino and photon temperature, $T_{\rm \nu} = \left({4}/{11}\right)^{{1}/{3}}\, T_{\rm \gamma}$ in Eq.~\eqref{Eq: neff}. The above equation provides the restriction on the current GW energy density as
\begin{align}
  \int_{k_{_{\rm RH}}}^{k_{\rm f}}\frac{dk}{k}\Omega_{_{\rm GW}}h^2(k)\leq \frac{7}{8}\left(\frac{4}{11}\right)^{4/3}\Omega_{\rm \gamma}h^2\,\Delta{\rm N_{\rm eff}},
  \label{eq: deltaneff}
\end{align}
where $\Omega_{\rm \gamma}h^2\simeq 2.47\times10^{-5}$ is the photon relic density at the present day.
For a blue-tilted spectrum, the maximum energy contribution comes from the mode $\kend$ that exists the Hubble radius at the end of inflation and reenters immediately after that. We shall mention the fact that the relevance of the $\Delta{\rm N_{\rm eff}}$ constraints on the GW energy density comes for a blue-tilted power spectrum, i.e., for $\wf>1/3$. In such a case, upon substitution of Eq.~\eqref{eq: gw energy density final}  in Eq.~\eqref{eq: deltaneff} and performing the integration, one can find
\begin{align}
  \int_{k_{_\text{BBN}}}^{\kend}\frac{\d k}{k}\Omega_{_{\rm GW}}h^2(k)\simeq \Omega_{_{\rm GW}}h^2 \zeta(\wf )\,\left(\frac{\kend}{\kre}\right)^\frac{6\wf -2}{1+3\wf },
  \label{eq: BBNapprox}
\end{align}
where $\zeta(\wf )=(1+3\wf )^{{4}/({1+3\wf })}\Gamma^2\left(\frac{5+ 3\wf }{2+6\wf } \right)\frac{(1+3\wf )}{2\pi(3\wf-1)}$. The ratio between $\kend$ and $\kre$ for a $\wf$ dominated reheating phase can be expressed as 
\bea
\frac{\kend}{\kre}=\left(\frac{30\rho_{\rm end}}{\pi^2 g_{\ast S,{\rm RH}}}\right)^{\frac{1+3\,\wf }{6\,(1+\wf )}}\,\tre^{-\frac{2}{3}\frac{(1+3\,\wf )}{(1+\wf )}}.
\label{eq: kend kre}
\eea
where $\rho_{\rm end}=3\,\Mpl ^2\,H_{\rm end}^2$ represents the inflaton energy density at the end of the inflation and for a de Sitter inflation $\hend=\Hi$ (constant).
Combining Eq.~\eqref{eq: deltaneff}, Eq.~\eqref{eq: BBNapprox} and Eq.~\eqref{eq: kend kre}, we obtain the restriction on the lower bound of the reheating temperature as
\bea 
\tre \geq \left(\frac{\Omega^{\rm rad}_{_{\rm GW}}h^2\, \zeta(\wf )}{5.61\times 10^{-6}\,\Delta N_{\rm eff}}\right)^{\frac{3\,(1+\wf )}{4\,(3\,\wf -1)}} \left(\frac{30\rho_{\rm end}}{\pi^2 g_{\ast S} (\tre)}\right)^{\frac{1}{4}}=T_{\rm RH}^{\rm GW},
\label{eq:BBNrestriction}
\eea
where $g_{\ast S} (\tre)$ is the relativistic degrees of freedom associated with the entropy calculated at the end of the reheating. By setting the temperature at the BBN energy scale, \( T_{\rm RH}^{\rm GW} \sim T_{\rm BBN} \) (4 MeV), we observe that the BBN constraint on primordial gravitational waves (PGWs) becomes relevant only when \( w_\phi \geq 0.60 \), considering $\rho_{\rm end}\sim 10^{64}\,{\rm GeV^4}$. We refer to this new lower limit on the reheating temperature due to PGWs as \( T_{\rm RH}^{\rm GW} \).
For example, when \( w_\phi = 0.6 \) (\( n = 8 \)), the expression for \( T_{\rm RH}^{\rm GW} \) can be written as:
\bea
T_{\rm RH}^{\rm GW}\sim 4 \,{\rm GeV}\left(\frac{\rho_{\rm end}}{ 10^{64}\,\rm GeV^4}\right)^{\frac{7}{4}}\,,
\eea
\begin{table}[t!]
  \begin{center}
    \begin{tblr}{|c|c|}
          \hline
 $n(w_\phi)$ & $T_{\rm RH}^{\rm GW}$ (GeV) \\ [0.5ex] 
 \hline\hline
 8 (0.60) & $4.0$ \\ \hline
  10 (0.67) & $6.9\times 10^2$ \\ \hline
   12 (0.71) & $9.8\times 10^3$ \\ \hline
 14 (0.75) & $4.8\times 10^4$ \\  \hline
  16 (0.78) & $1.4\times 10^5$ \\ \hline
   18 (0.80) & $3.0\times 10^5$ \\ \hline
     20 (0.82) & $5.2\times 10^5$ \\
 \hline
    \end{tblr}
    \caption{\label{GWtre}{ Numerical values of the $T_{\rm RH}^{\rm GW}$ for a fixed value of inflationary energy scale defined at the end of the inflation,  $\rho_\phi^{\rm end}=10^{64}$ $\rm GeV^4$}.}
  \end{center}
\end{table}
where we choose $\Delta N_{\rm eff}\simeq 0.28$, using the current Planck data~\cite{Planck:2018jri}.
For \( \rho_{\rm end} = 10^{64} \, \text{GeV}^4 \), the reheating temperature \( T_{\rm RH}^{\rm GW} \) is found to be 4 GeV. Fixing the same $\rho_{\rm end}$, the numerical values of $T_{\rm RH}^{\rm GW}$ for different $w_\phi$ shown in Tab.~\ref{GWtre}. Note that, conversely, by using Eq. \ref{eq:BBNrestriction}, a constraint can be placed on both \( \Hi \) and \( r \) once the reheating parameters, such as \( T_{\rm re} \) and \( w_\phi \), are fixed.

In Fig.~\ref{fig: pgw spectrum}, along with the GW spectrum, we also plot the projected sensitivity curves of various future GW detectors.
We see that, in some cases, the spectrum can be observed by the detectors. Fig.~\ref{fig: pgw spectrum} also shows that some of the cases where the spectrum can be detected are ruled out as the amplitude of the spectrum hits the BBN bound for $k<\kend$. 
In the next section, we shall discuss in detail various future GW observatories and their associated noise and sensitivity curves and the conditions of a primordial GW spectrum to be detected by them.

\section{Noise and sensitivity of detectors \label{sec: detector}}
Let us now discuss the method we require to examine the detectability of the signal. As mentioned, we consider three space-based detectors, LISA, BBO, and DECIGO, and one ground-based detector, ET.
All the detectors will be operating in different frequency bands,  i.e., LISA will detect signals between $10^{-5}-1$ Hz, whereas BBO and DECIGO will detect in the range $10^{-3}-10^2$ Hz. ET will detect in the slightly higher frequency range of $1-10^4$ Hz. Each space-based detector is proposed to be operational for four years, whereas the observational time for the ground-based detector is five years.
\begin{figure}[t!]
    \centering
    \includegraphics[scale=.37]{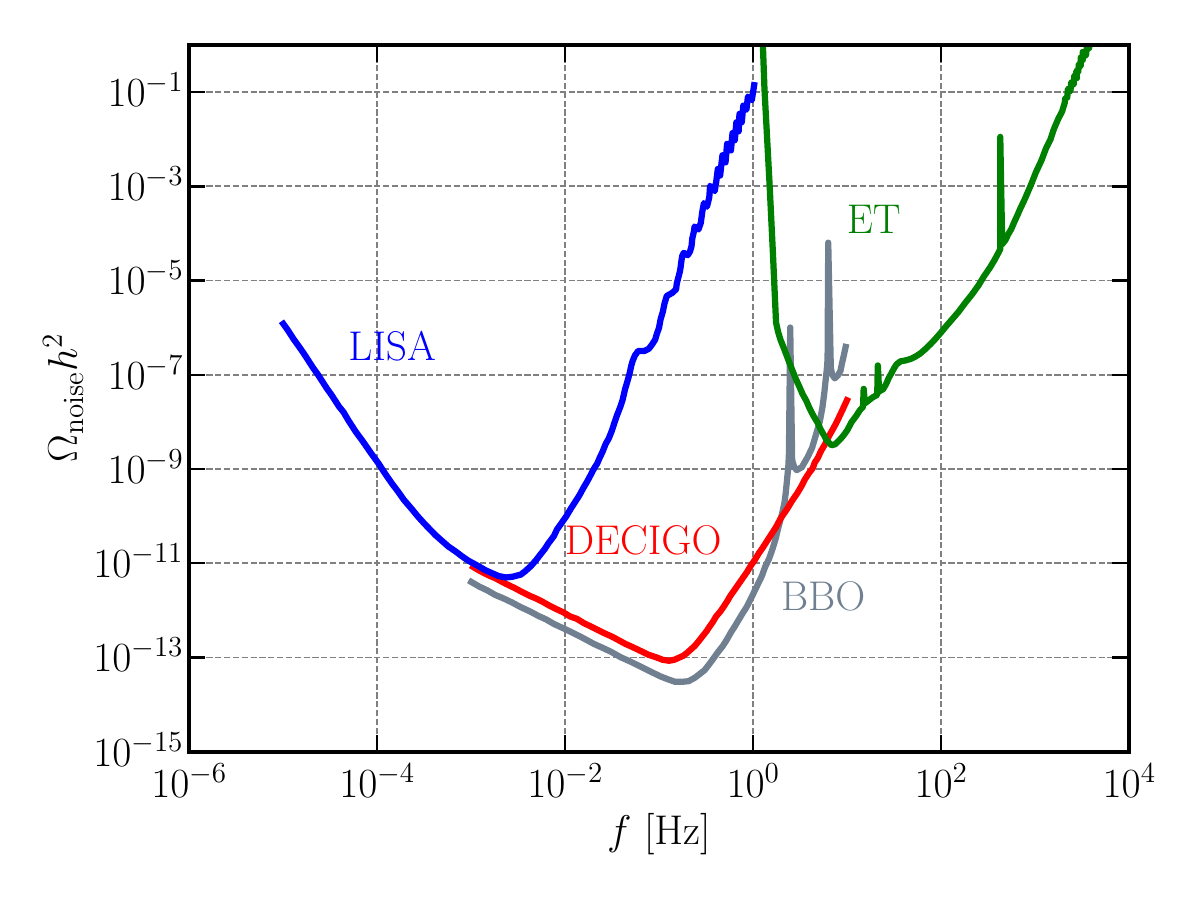}
    \includegraphics[scale=.37]{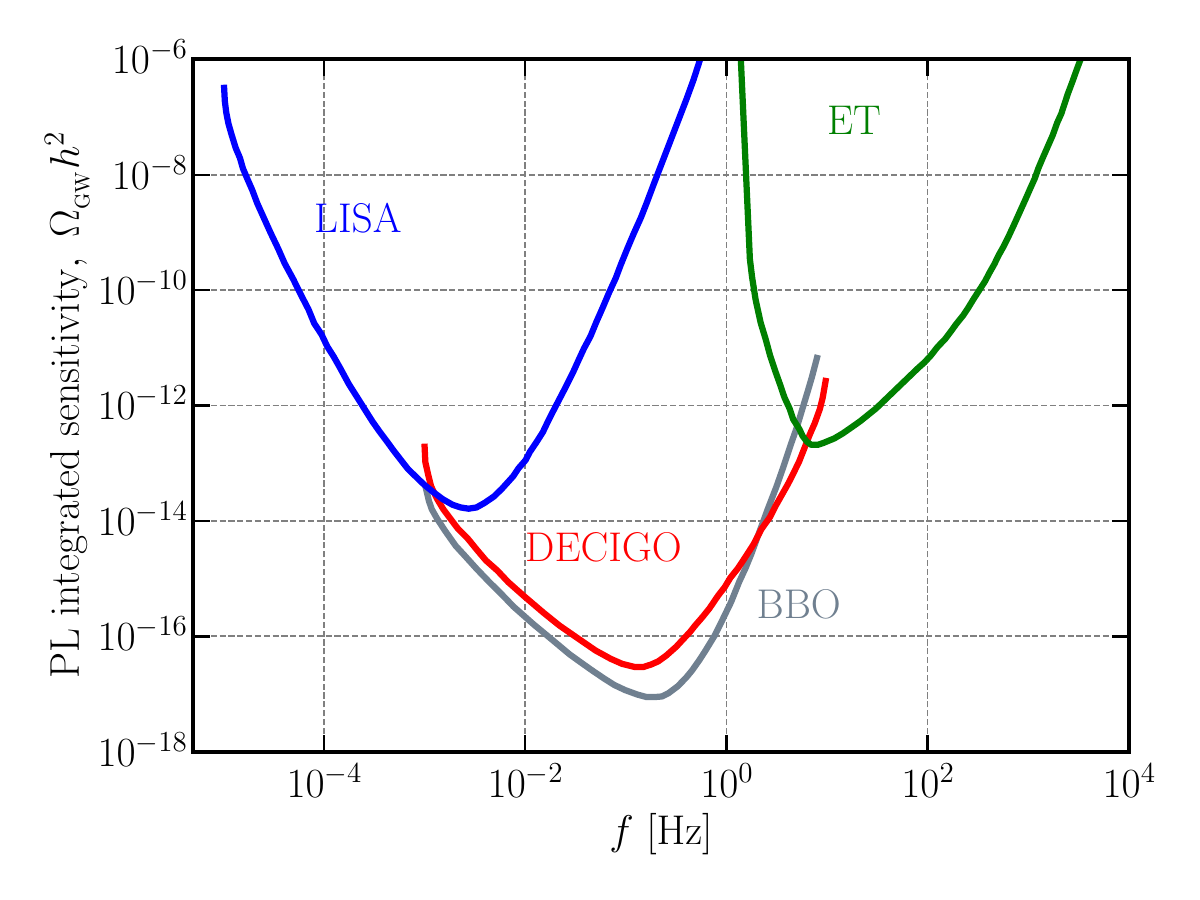}
    \caption{ The noise (left panel) and the sensitivity curves (right panel) are plotted here in terms of frequency for various GW detectors such as LISA, DECIGO, BBO, and ET. The sensitivity curves are obtained by integrating over the noise curve for a power-law GW spectrum \cite{Schmitz:2020rag}.}  
    \label{fig: noice_sensitivity}
\end{figure}

Each experiment has associated noise data that can be used to determine whether a particular GW spectrum has adequate amplitude to be detected by the detector or not. The left plot in Fig.~\ref{fig: noice_sensitivity} contains such noise curves for the detectors of our interest plotted in terms of frequency. Before going into the details, let us write the quantity associated with the detection called signal-to-noise ratio (SNR), defined as \cite{Schmitz:2020syl,Schmitz:2020rag} 
\bea 
\rho=\l[ 2\,t_{\rm obs} \int_{f_{\rm min}}^{f_{\rm max}} \d f
\l(\f{\ogw(f)h^2}{\Omega_{_\mathrm{noise}}(f)h^2}\r)^2\r]^{1/2}.
\label{eq: snr}
\eea
In this expression, $\ogw\,h^2$ is the GW strength generated from a particular scenario, and  $\Omega_{_\mathrm{noise}}h^2$ is the noise data of the corresponding detector of our interest. $[f_{\rm min},f_{\rm max}]$ is the operational frequency domain of the detector. $t_{\rm obs}$ is the time scale of the observations. The factor $2$ arises due to the fact that we are measuring cross-correlation.

One can adopt one of the following two approaches for the analysis. The first one we shall refer to as the SNR approach, where for each given spectrum of $\ogw$, we need to evaluate the quantity in Eq.~\eqref{eq: snr} using the noise data provided for a particular observation, $\Omega_{_{\rm eff}}$. If the value of $\rho$ is above some threshold value $\rho_{\rm th}$ for a given $\ogw$, we conclude that the detector will detect the signal. Tab.~\ref{tb: decector list} contains such values of $\rho_{\rm th}$ along with observation time and frequency range of operation for the detectors we consider here. 
\begin{table}[t!]
  \begin{center}
    \begin{tblr}{|c|c|c|c|}
      \hline
         ~~\bf Detectors ~~& \bf frequency (Hz)& ~\bf $\bm t_{\bf  obs}$ (years)~& ~~$\bm \rho_{\bf th}$~~\\ \hline\hline
         LISA & $10^{-5}-1$& $4$ & $10$\\ \hline
         ET &$1-10^4$& 5 & 5\\ \hline
         DECIGO &$10^{-3}-10^2$& 4& 10\\ \hline
         BBO &$10^{-3}-10^2$& 4 & 10\\ \hline
    \end{tblr}
    \caption{\label{tb: decector list}{ The frequency range of operation, observation time, and threshold value of $\rho$ are listed here for LISA, ET, DECIGO, and BBO \cite{Schmitz:2020syl}.}}
  \end{center}
\end{table}
There is yet another method known as the power law integrated sensitivity (PLIS), which is used if the predicted spectrum from the theory is a power law in nature. In this method, one calculates the sensitivity curve in the beginning from the noise data by using a power law spectrum, $\ogw(f)=\ogw(f_\ast)\l({f}/{f_\ast} \r)^b$, where $b$ defines the spectral index of the spectrum, $f_\ast$ is a reference frequency and $\ogw(f_\ast)$ is the amplitude of the spectrum at that reference point. 
The right panel of Fig.~\ref{fig: noice_sensitivity} illustrates such sensitivity curves derived from corresponding noise data.
The following job is to check whether the spectra obtained from the theory pass through the sensitivity region. This can be seen in Fig.~\ref{fig: pgw spectrum}, where one or multiple detectors can observe the spectral energy density given in dashed and dash-dotted lines.
We shall also mention that a similar method called peak integrated sensitivity (PIS) can be used in the case of peaked spectrum to obtain the sensitivity curve in a similar way to PLIS. However, we shall restrict ourselves to the PLIS method if the GW spectra have simple power law behavior, which is the case for PGWs. 
Both the SNR and PLIS methods have their advantages and disadvantages. 
One of the disadvantages of the SNR method is that it is impossible to conclude the detectability just by looking at the noise curve (left plot in Fig.~\ref{fig: noice_sensitivity}). One needs to perform the integration in Eq.~\eqref{eq: snr} for each spectrum to draw the conclusion, which makes it computationally intense and time-consuming. 
Meanwhile, PLIS is computationally faster if the predicted spectrum from the theory is a power law in nature. This method also provides a visual understanding of the spectrum and its detection, which is also evident from Fig.~\ref{fig: pgw spectrum}. However, if the spectra have different power law behaviors, it is incorrect to justify all the spectra with one sensitivity curve. In our analysis, we have both power law and peaked spectrum. Thus, we shall follow the method of calculating SNR.


\section{Inflationary model of interest: $\alpha$-attractor E-model \label{sec: inflationary model}}

Let us start our discussion by considering a specific model of inflation that permits slow-roll inflation and generates a scale-invariant tensor power spectrum, discussed in Sec.~\ref{sec: pgw}, and the behavior at the minima describes the reheating phase. We consider one such model, the so-called $\alpha$-attractor E-model where the potential $V(\phi)$ has the following form: \cite{Starobinsky:1980te,Starobinsky:1983zz,Kallosh:2013hoa,Kallosh:2013yoa}
    \begin{equation} \label{a}
V(\phi) = \Lambda^4 \left[  1 - \exp\l({ -\sqrt{\f{2}{3\alpha}}\frac{\phi}{\Mpl} }\r) \right]^{n}\,,
\end{equation}
where 
$\Lambda$ has the dimension of mass, which is constrained by the COBE normalized value from CMB.
If we expand the above potential around the minima, as the dominant term, it can be expressed in a power law form 
\bea\label{potentialminima}
V(\phi) = \Lambda^4\left(\frac{2}{3\alpha\Mpl^2}\right)^{n/2}\,\phi^{n}=\lambda\frac{\phi^n}{\Mpl^{n-4}} \, ,\label{eq: oscillate v phi}
\eea
where $\lambda$ is a dimensionless constant related to $\Lambda$ as $\lambda=[2/(3\alpha)]^{n/2}\left({\Lambda}/{\Mpl}\right)^4$. 
For slow-roll inflation, this quantity can be expressed in terms of the amplitude, $A_s$, and spectral index, $n_s$ of the scalar power spectrum and the tensor-to-scalar ratio, $r$ as (see,
for instance, Ref.\cite{Drewes:2017fmn})
\bea
 \lambda = & &\l(\f{2}{3\alpha}\r)^{n/2} {\left(\frac{3\pi^2 r A_s}{2}\right) }  
  \left[\frac{n^2+n+\sqrt{n^2+3\alpha(2+n)(1-n_s)}}{n(2+n)}\right]^{n}.
 \eea
From the condition at the end of the inflation, the first slow roll parameter $\epsilon_1$ needs to be
$
\epsilon_1(\phi_{\rm end})={(2\, \Mpl^2)^{-1}}\l({V_\phi}/{V}\r)^2_{\phi=\phi_{\rm end}}=1 ,  $
where $\phi_{\rm end}$ is the value of the inflation field at the end of the inflation. Inverting the above relation, we find
\bea\label{infpotentialfield}
\phi_{\rm end}=\Mpl  \sqrt{\frac{3\alpha}{2}}~ \ln\left(\frac{n}{\sqrt{3\alpha}}+1\right)\,.
\eea
Upon substitution, the value of $\phi_{\rm end}$ form Eq.~\eqref{infpotentialfield} into Eq.~\eqref{potentialminima}, the expression of the potential at the end of inflation takes the following form

\bea \label{Eq:vend}
V(\phi_{\rm end})= \lambda\,\Mpl^4 \l(\frac{3\alpha}{2}\r)^{n/2} \left(\frac{n}{n+\sqrt{3\alpha}}\right)^{n} .
\eea
The inflaton energy density at the end of inflation, which provides the initial condition for the subsequent reheating dynamics, turns out as 
\bea
\rho_{\rm end}=3\Mpl^2 H_{\rm inf}^2\sim \frac{3}{2}V(\phi_{\rm end}).
\label{eq: rhoend alpha}
\eea
One can use the slow roll approximation to find out the number of e-folds $N_k$ between the point where the mode corresponding to the pivot scale leaves the Hubble radius to the end of inflation. Where $\phi_k$ is the field value associated with the pivot scale and can be expressed as
\bea 
N_k &=& \f{3\alpha}{4n} \l[\exp\l(\sqrt{\f{2}{3\alpha}} \f{\phi_k}{\Mpl}\r)  - \exp\l(\sqrt{\f{2}{3\alpha}}\f{\phi_{\rm end}}{\Mpl}\r) -\sqrt{\f{2}{3\alpha}} \f{\phi_k-\phi_{\rm end}}{\Mpl}\r]\,,
\label{eq: Nk}
\eea 
and the spectral index takes the form 
\bea 
n_s &=& 1-\f{2\,n\,}{3\alpha}
\l[2\exp\l(\sqrt{\f{2}{3\alpha}} \f{\phi_k}{\Mpl}\r)+n\r]
 {\l[\exp\l(\sqrt{\f{2}{3\alpha}} \f{\phi_k}{\Mpl}\r)-1\r]^{-2}}.
\label{eq: ns}
\eea 
From Eq.~\eqref{eq: ns} we express $\phi_k$ in terms of $n_s$ as
\bea 
\phi_k&=& \Mpl \sqrt{\f{3\alpha}{2}} \ln\l[1+\f{2n+\sqrt{4n^2 +6\alpha n(2+n)(1-n_s)}}{3\alpha(1-n_s)}\r].
\eea 
Finally, the tensor-to-scalar ratio is given by
\bea 
r &=& \f{48\alpha n^2 (1-n_s)^2}{\l[2n+\sqrt{4n^2 +6\alpha n(2+n)(1-n_s)}\r]^2 }.
\label{eq: ttos}
\eea 
In slow roll approximation, the tensor to scalar ratio can be connected to the inflationary energy scale as $\Hi=\pi\Mpl\sqrt{rA_s/2}$. To summarize, we derived important quantities that are relevant to our analysis considering the $\alpha$-attractor E-model. In the next section, we will briefly discuss different reheating scenarios.  


\section{Brief overview of the dynamics of reheating and different possibilities\label{sec: reheating}}

 After the end of inflation, the inflaton field oscillates around the minima of the potential. In most of the inflation models, inflationary potential behaves as $V(\phi)\sim \phi^n$ around the minima (which we can see from Eq.~\eqref{eq: oscillate v phi}), and the exponential expansion period is followed by a phase of inflation oscillation. As a natural consequence, this oscillating field of inflaton decays mainly into the SM particles, which leads to a standard RD Universe. This intermediate phase, where the transformation of energy from inflaton to radiation takes place, is known as reheating. The oscillation of the inflaton is time-dependent and can be decomposed as 
\bea \label{Eq:phi}
\phi(t)=\phi_0(t) \mathcal{P}(t)\,,
\eea
with the amplitude of the oscillation, $\phi_0(t)$ and the periodicity represented by $\mathcal{P}(t)$. In Fourier space $\mathcal{P}(t)$ can be expressed in terms of the Fourier modes $\mathcal{P}(t)=\sum_\mu \mathcal{P}_\mu \e^{i \mu \omega t}$, where $w$ represents the frequency of the oscillation calculated to be
 \bea \label{Eq:w}
\omega= m_{\phi}(t)\sqrt{\frac{\pi n}{2(n-1)}}\frac{\Gamma\left(\frac{1}{2}+\frac{1}{n}\right)}{\Gamma\left(\frac{1}{n}\right)}\,,
\eea
where $m_\phi$ is the effective mass of the inflaton, $\phi_0(t), $ encodes the effect of both redshift and decay, assuming that the time scale of the oscillation is negligible compared to the redshift and decay time scale. The average of the single oscillation leads to $\langle\dot{\phi}^2\rangle\simeq \langle\phi \, V_\phi(\phi)\rangle$, that in turn approximate inflaton energy density $\rho_\phi=\langle({\dot \phi}^2/2) + V(\phi)\rangle \sim V(\phi_0)$. Under this assumption, we have the average inflaton EoS~\cite{Garcia:2020eof,Bernal:2019mhf}
 \bea
 w_\phi=\frac{P_\phi}{\rho_\phi}=\frac{\langle({\dot \phi}^2/2) - V(\phi)\rangle}{\langle({\dot \phi}^2/2) + V(\phi)\rangle }\simeq \frac{n-2}{n+2}\,,
 \label{eq: were n}
 \eea
where we assume the potential of the form given in Eq.~\eqref{eq: oscillate v phi} at the minima. Then the effective mass of the inflaton $m_\phi(t)$ can be expressed as 
\bea \label{Eq: inflaton mass}
m^2_\phi(t)=V_{\phi\phi}(\phi_0(t))=n\,(n-1)\lambda\, \Mpl ^2\left(\frac{\phi_0(t)}{\Mpl }\right)^{n-2}\,.
\eea
Once we incorporate the decay term in the equation of motion of the inflaton, one can find the evolution as
\bea
\ddot{{\phi}}+\left(3\,H\,+\Gamma_\phi(t)\right)\,\dot{\phi}+V_\phi(\phi)=0,
\eea
where $\Gamma_\phi(t)$ is the time-dependent decay term. The above equation of motion can be re-written in terms of energy density as
\begin{eqnarray}
&& \dot{\rho_{\rm \phi}}+3\,H\,(1+w_\phi)\,\rho_{\rm\phi}= -\Gamma_\phi(t)\,\rho_{\rm \phi}\,(1+w_{\rm \phi}) \,.  \label{Boltzman1} 
\end{eqnarray}
On the other hand, the evolution of the radiation density sourced by the inflaton decay or scattering  can be written as
\begin{eqnarray}\label{Eq:Boltzmann-rad}
&&\dot{\rho}_{ \rm R}+4\,H\,\rho_{ \rm R}=\Gamma_{ \rm \phi}(t)\,\rho_{ \rm \phi}\,(1+w_{\phi})\,,
\label{Boltzman2} 
\end{eqnarray}
where the Hubble parameter is followed by the Friedmann equation, where the total energy budget must be the sum of the components of both inflaton and radiation
\begin{eqnarray}
&&H^2=\frac{\rho_{ \rm \phi}+\rho_{\rm  R}}{3\,\Mpl^2}\label{Boltzman3}\,.
\end{eqnarray}
The end of reheating is defined at the point $\are$ when $\rho_\phi (\are)=\rho_R(\are)$. Hence, one can apply the condition of entropy conservation  $g_{\ast S}\,a^3T^3$= constant from the end of reheating to the present day. Using this relation we can relate the reheating temperature $\tre$ to the present temperature $T_0$ and the number of e-folds during inflation $N_{\rm inf}$ and reheating $N_{_{\rm RH}}$ as
\bea
\tre &=& \l(\f{43}{11\,g_{\ast S,{\rm RH}}}\r)^{{1}/{3}}T_0 \f{\Hi}{k_\ast}\e^{-(N_{\rm inf}+N_{_{\rm RH}})},
\eea 
where $k_\ast$ is the wavenumber of the mode that leaves the Hubble radius at the pivot scale and $g_{\ast S,{\rm RH}}$ is the relativistic degrees of freedom associated with the entropy evaluated at the end of reheating. $N_{\rm RH}$ is the e-folding number associated with the duration of reheating and present CMB temperature $T_0=2.725$ K. One can further use the relation between the radiation energy density and its temperature, $\rho_{\rm R}=(\pi^2/30)g_\ast T^4$, and write the reheating temperature as
\bea 
\tre &=& \l[\f{3\Mpl^2\Hi^2}{(\pi^2/30)g_{\ast,{\rm RH}}}\r]^{1/4} \e^{-\f{3}{4}N_{_{\rm RH}}(1+\wf)}\,,
\eea 
where $g_{\ast,{\rm RH}}$ represents the relativistic degrees of freedom associated with the thermal bath measured at the end of reheating. To obtain the above relation, we assume that during reheating $\rho_\phi \propto a^{-3(1+\wf)}$, ignoring the decay term, which is only important at the very end of reheating.  
The evolution of the radiation component is sensitive to the nature of the coupling between inflaton and radiation field, it may be gravitational or non-gravitational. Further, if one assumes PBHs are formed during the early stage of reheating, those PBHs are ultralight, and their evaporation process may also change the reheating history. In our subsequent discussion, we will focus on those different possibilities, starting with non-gravitational interactions between inflaton and daughter fields. After that, we will move our discussion to such a scenario where gravity is responsible for reheating the universe; here, we took two different possibilities: 1) reheating is controlled by the non-minimal coupling to the curvature, and 2) decay of PBHs with and without dominating the total energy budget of the Universe.


\subsection{Reheating through non-gravitational interactions}

As stated earlier, reheating happens due to the decay or annihilation of inflaton into daughter particles, specifically into the SM degrees of freedom. In the case of non-gravitational interaction, we assume three  phenomenological decay or annihilation processes with the following interaction Lagrangian~\cite{Garcia:2020wiy},
\bea
\mathcal{L}_{\rm int} \supset  \begin{cases} 
y_\phi\phi \bar{f}f & \phi \to \bar{f}f \\
 g_\phi \phi bb & \phi \to b b\\
  \sigma_\phi \phi^2 b^2 & \phi \phi\to b b\\
\end{cases},
\label{RH:procs}
\eea
where $f(b)$ is used to denote fermionic (bosonic) particles. The couplings $y_\phi$ and $g_\phi$ are the dimensionless Yukawa coupling and the dimensionful bosonic coupling, and $\sigma_\phi$ is the four-point dimensionless coupling.
\begin{figure}[t]
\centering
\includegraphics[scale=1.3]{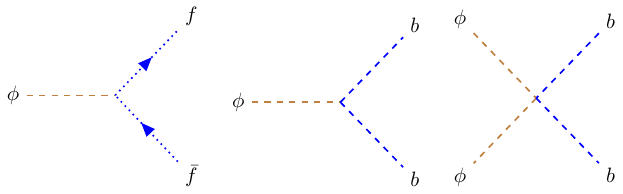}
\caption{Feynman diagrams for the production of radiation particle through three different non-gravitational interactions : (i) $\phi\rightarrow\bar ff$, (ii) $\phi\rightarrow bb$, and (iii) $\phi\phi\rightarrow bb$.}
\label{Fig:Feynmann}
\end{figure}
The Feynman diagrams associated with the interactions mentioned above are shown in Fig.~\ref{Fig:Feynmann}. Note that in this analysis, although we have restricted ourselves to these three scenarios, that can be extended to include more inflaton-radiation couplings. For these three processes we consider, which include inflaton decays into a pair of fermions, scalars, and the four-point process, the time-dependent decay rate takes the following form
\bea
\label{Eq:gammaphibis}
\Gamma_{\phi} (t)= 
\begin{cases}
\frac{y_{\rm eff}^2\,m_\phi(t)}{8\pi}\,, & \phi\rightarrow \bar{f}f\,,\\
\frac{g_{\rm eff}^2}{8\pi\,m_\phi(t)}\,,& \phi\rightarrow bb\,,\\
\frac{\sigma_{\rm eff}^2}{8\pi}\,\frac{\rho_\phi(t)}{m_\phi^3(t)}\,,& \phi\phi\rightarrow bb\,.
\end{cases}
\eea
Using the expression of $m_\phi (t)$ from Eq.~\eqref{Eq: inflaton mass}, where the time-dependence only includes the envelope $\phi_0(t)$, one can find the expression for decay rate to be
\bea \label{Eq:Gammafinal}
\Gamma_\phi(t)=\gamma_\phi(t)\left(\frac{\rho_\phi}{\Mpl ^4}\right)^p\,,
\eea
with
\bea
\label{Eq:gamma}
\gamma_{\phi} (t),\,p= 
\begin{cases}
\frac{y_{\rm eff}^2\,\sqrt{n\,(n-1)}\,\lambda^{1/n}\,\Mpl }{8\pi},~~\,\frac{1}{2}-\frac{1}{n}\, & \text{for }\phi\rightarrow \bar{f}f\,,\\
\frac{g_{\rm eff}^2}{8\pi\,\sqrt{n\,(n-1)}\,\lambda^{1/n}\,\Mpl },\,~~~\frac{1}{n}-\frac{1}{2}\, & \text{for }\phi\rightarrow bb\,,\\
\frac{\sigma_{\rm eff}^2\,\Mpl }{8\pi\,[n\,(n-1)]^{3/2}\,\lambda^{3/n}}\,,~~~~~\frac{3}{n}-\frac{1}{2}\,& \text{for } \phi\phi\rightarrow bb\,,
\end{cases}
\eea
where recall that $n$ is the power of $\phi$ with which the potential behaves at its minima.

Now using the expressions of $\Gamma_\phi(t)$ and  $\gamma_\phi(t)$ form Eq.~\eqref{Eq:Gammafinal} and Eq.~\eqref{Eq:gamma}  to Eq.~\eqref{Boltzman2} and after integration, one can find the radiation energy density behaves as 
\bea\label{Eq: radiation energy}
\rho_R=\frac{2\,n}{n+8-6np}\,\frac{\gamma_\phi\,\rho_{\rm end}^{p+1}}{\hend\,\Mpl^{4p}}\left(\frac{\aend}{a}\right)^4\left[\left(\frac{a}{\aend}\right)^{\frac{n+8-6np}{n+2}}-1\right]\,,
\eea
where $\rho_{\rm end}$ is the inflaton energy density at the end of reheating. The reheating point $\are$, which is calculated when $\rho_\phi(\are)=\rho_R(\are)$, takes the following form:
\bea \label{Eq:arh1}
\frac{\are}{\aend}=\left(\frac{n+8-6np}{2n}\,\frac{\Mpl ^{4p-1}\,\rho_{\rm end}^{\frac{1}{2}-p}}{\sqrt{3}\,\gamma_\phi}\right)^{\frac{n+2}{3n-6np}}\,.
\eea
The above equation is true for $8+n-6np>0$. However, for $8+n-6np<0$, we have
\bea \label{Eq:arh2}
\frac{a_{\rm RH}}{a_{\rm end}}=\left(\frac{6np-n-8}{2n}\,\frac{\Mpl ^{4p-1}\,\rho_{\rm end}^{\frac{1}{2}-p}}{\sqrt{3}\,\gamma_\phi}\right)^{\frac{n+2}{2n-8}}\,.
\eea
Finally substitute the expression for $a_{\rm RH}/\aend$ in Eq.~\eqref{Eq: radiation energy}, one can find reheating temperature 
 for $8+n-6np>0$ as
\bea 
T_{\rm RH} &=& \l(\f{30}{\pi^2\,g_{\ast,{\rm RH}}}\r)^{1/4}  
\l(\f{n+8-6np}{2n}\f{\Mpl^{4p-1}}{\sqrt{3}\gamma_\phi}\r)^{\f{1}{2(2p-1)}}\,,
\label{eq:tre-ng1}
\eea 
where $g_{\ast, \rm RH}$ represents the relativistic degrees of freedom associated with the thermal bath. For SM particles $g_{\ast,\rm RH}\simeq 106.75$. Whereas, for $8+n-6np<0$
\bea 
T_{\rm RH} = \l(\f{30}{\pi^2\,g_{\ast,{\rm RH}}}\r)^{1/4}
\rho_{\rm end}^{\f{6np-n-8}{8(n-4)}}
\l(\f{6np-n-8}{2n}\f{\Mpl^{4p-1}}{\sqrt{3}\gamma_\phi}\r)^{-\f{3n}{4(n-4)}}\,.
\label{eq:tre-ng2}
\eea 
\begin{table}
\centering
 \begin{tabular}{||c | c |c |c |c|c |c ||} 
 \hline
 $n\,(w_\phi)$ & $\sum \mu^3\lvert \mathcal{P}_\mu\rvert^2$ & $\sum \mu\lvert\mathcal P_\mu\rvert^2$ & $\sum \mu\lvert(\mathcal P^2)_\mu\rvert^2$  & $\frac{y_{\rm eff}}{y_\phi}$& $\frac{g_{\rm eff}}{g_\phi}$& $\frac{\sigma_{\rm eff}}{\sigma_\phi}$\\ [0.5ex] 
 \hline\hline
 2 (0.0) & $\frac{1}{4}$ & $\frac{1}{4}$ & $\frac{1}{8}$ & 1 & 1 & 1\\ 
 \hline
 4 (1/3) & 0.241  & 0.229& 0.125 & 0.71 & 1.42& 3.64\\
 \hline
 10 (2/3)& 0.257 & 0.205 & 0.120& 0.50 & 2.14& 15.6 \\
 \hline
 14 (3/4) & 0.270 & 0.198 & 0.117&0.44 & 2.49& 25.8\\
 \hline
 20 (9/11) & 0.287  & 0.191& 0.114&  0.38 & 2.92& 44.0 \\ 
 \hline
 \end{tabular}
 \caption{Numerical values of the Fourier sums present in the effective couplings:}\label{fouriersum}
\end{table}
One important point is to note that such a scenario ($8+n-6np<0$) will appear in the case of $\phi\to f\bar{f}$ process for $n>7$ and $n<2.5$ for $\phi\phi\to bb$. However, for bosonic reheating with $\phi\to bb$, this scenario will not appear.\\
The ratio between the oscillation-induced effective coupling parameters $y_{\rm eff}$, $g_{\rm eff}$, and $\sigma_{\rm eff}$ over their respective Lagrangian values $y_\phi$, $g_\phi$ and $\sigma_\phi$ is estimated to be \cite{Garcia:2020wiy}
\begin{eqnarray}{\label{t2}}
&&\left(\frac{y_{\rm eff}}{y_\phi}\right)^2=(n+2)(n-1)\,\left(\frac{w}{m_\phi}\right)^3\sum^{\infty}_{\mu=1}\mu^3\,\lvert\mathcal P_\mu\rvert^2\,,\\
&&\left(\frac{g_{\rm eff}}{g_\phi}\right)^2=(n+2)(n-1)\,\frac{w}{m_\phi}\sum^{\infty}_{\mu=1}\mu\,\lvert\mathcal P_\mu\rvert^2\,,\\
&&\left(\frac{\sigma_{\rm eff}}{\sigma_\phi}\right)^2=n\,(n+2)\,(n-1)^2\,\frac{w}{m_\phi}\sum^{\infty}_{\mu=1}\mu\,\lvert\mathcal (\mathcal P^2)_\mu\rvert^2\,.
\end{eqnarray}

In Tab.~\ref{fouriersum}, we have tabulated the Fourier sum associated with various coupling constants for different values of the background EoS. Note that for matters like reheating ($n=2$), the effective coupling is exactly the same with their lagrangian values. However, as we increase the background EoS, the oscillation effect effectively enhances the bosonic production rate and reduces the fermionic production rate. As an example, increasing the $n$ value from $2\to 10$ ($w_\phi$ changes from $0\to 2/3$), the effective bosonic coupling with respect to its tree label values gets doubled $g_{\rm eff} \simeq 2\, g_\phi$ for $\phi\to bb$ and 15 times greater $\sigma_{\rm eff} \simeq 15\, \sigma_\phi$ for $\phi\phi\to bb$, whereas that of the fermionic coupling is demolished by half $y_{\rm eff} \simeq 0.5\, y_\phi$.

Till now, we have discussed reheating through non-gravitational interaction between inflaton and daughter fields. However, gravity can play a significant role in reheating the universe (see, for instance, Refs. \cite{Haque:2022kez,Clery:2021bwz,Clery:2022wib,Co:2022bgh,Ahmed:2022tfm,RiajulHaque:2023cqe,Haque:2024cdh}). In the next section, we discuss such possibilities of reheating related to non-minimal coupling to gravity and evaporation of the ultra-light PBHs.

\subsection{Gravitational reheating}
\begin{figure}[t]
\hskip 2.5 cm
\includegraphics[scale=1.5]{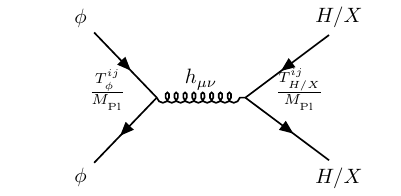}
\caption{Feynman diagram for the production of Higgs and DM through the exchange of graviton from inflaton condensate.}
\label{Fig:Feynmann2}
\end{figure}
In this section, we are interested in a scenario where purely gravitational interaction is responsible for reheating. Thus, we assume the direct coupling between inflation and SM particles is either very weak or not present. The FLRW metric can be expanded across the flat Minkowski spacetime, $g_{ij}=\eta_{\ij}+\frac{2\,h_{ij}}{\Mpl}$, we find the interaction between all matter fields and the gravitational field. The Lagrangian associated with the purely gravitational interaction is given by \cite{Choi:1994ax,Holstein:2006bh} 
\bea
\sqrt{-g}\,\mathcal{L}_{\rm int}=-\frac{1}{\Mpl}h_{ ij}\left(T_{\rm \phi}^{ij}+T_{\rm SM}^{ ij}+T_{\rm DM}^{ij}\right)\,,
\eea
where the subscript SM is used for Standard Model fields, DM represents the DM, and $h_{ij}$ is the canonically normalized tensor
perturbation. The expression for the stress-energy tensor relies on the spin of the field. For scalar spin-zero particles ($S$) such as inflaton, and Higgs boson, the form of the stress-energy tensor is given by  
\bea
T_{0}^{ij}=\partial^i S\partial^i S-g^{ij}\left[\frac{1}{2}\partial^i S\partial_i S-V(S)\right]\,,
\eea
where $V(S)$ is the potential associated with the scalar particle.
For now, let us focus only on the dynamics of reheating and leave the discussion for the DM production for the next section. The production rate associated with the SM particles can be written as
\bea
\Gamma_{\rm \phi}\,\rho_{\rm \phi}\,(1+w_{\rm \phi})=R_{\rm n}^{\rm 0}=\frac{\mathcal{N}_{\rm h}\,\rho_{\rm \phi}^2}{16\pi\,\Mpl ^4}\sum^{\infty}_{\mu=1}2\,\mu\,\omega\,\lvert\mathcal P_{2\mu}^n \rvert^2.
\eea
\begin{table}
\centering
 \begin{tabular}{||c|c|c||} 
 \hline 
 n & $\alpha_n^\xi(\alpha=1)$& $\alpha_n^\xi(\alpha=10)$ \\
 \hline\hline
 $6$ & $0.00038 + 0.015 \,\xi^2$ & $0.00017+0.007\,\xi^2$
\\  \hline
 $8$ & $0.00105+0.041 \,\xi^2$ & $0.00043+0.017\,\xi^2$
\\  \hline
 $10$ & $0.00192+0.073\,\xi^2$ & $0.00075+ 0.029\,\xi^2$
\\  \hline
 $12$ & $0.00288+ 0.107\,\xi^2$ & $0.00108+ 0.040\,\xi^2$
\\  \hline
 $14$ & $0.00383+0.140\,\xi^2$ & $0.00141+0.051\,\xi^2$
\\  \hline
 $16$ & $0.00476+0.171\,\xi^2$ & $0.00172+0.061\,\xi^2$
\\  \hline
 $18$ & $0.00562+0.198\,\xi^2$ & $0.00199+0.070\,\xi^2$
\\  \hline
 $20$ & $0.00636+0.223\,\xi^2$ & $0.00224+0.078\,\xi^2$
\\  \hline
 \end{tabular}
\caption{Numerical values of $\alpha_n^\xi$ relevant for production rate in case of gravitational reheating.}\label{fouriersum-grav}
\end{table}
\noindent Note that in our analysis, we restricted ourselves to the Standard Model; we took $\mathcal{N}_{\rm h}=4$ for the real scalar present in the SM, and we neglected the Higgs mass. Such a production mechanism may give rise to reheating temperatures above the BBN energy scale ($T_{\rm RH}\geq4$ MeV \cite{Kawasaki:1999na,Kawasaki:2000en,Hasegawa:2019jsa}), but inconsistent with the $\Delta N_{\rm eff}$ constraints due to the overproduction of the inflationary GWs \cite{Haque:2022kez,Barman:2022qgt}. This leads us to move beyond the minimal scenario by introducing non-minimal coupling. For the production of radiation, we assume non-minimal coupling of Higgs to curvature of the form
\bea
\mathcal{L}_{\rm \xi}=-\frac{\xi}{2}\lvert H\rvert^2\,\mathcal{R}\,,
\label{eq:lagrangian_gravitational}
\eea
where $\mathcal{R}$ is the Ricci scalar. This produces an effective coupling between the Higgs boson and the inflaton, which can be written as
\bea
\mathcal{L}_{\rm \xi}^{\rm \phi H}=\frac{\xi}{\Mpl ^2}\left[2V(\phi)-\frac{1}{2}g^{ij}\partial_{ i}\phi\,\partial_{ j}\phi\right]\lvert H\rvert^2\,.
\eea
For coupling strength $\xi\geq \mathcal{O}(1)$, the non-minimal coupling dominates over the above-described minimal production mechanism and successfully reheats the universe as well as consistent with the $\Delta N_{\rm eff}$ constraints at BBN. This non-minimal coupling of Higgs boson to curvature leads to the production rate
\bea\label{Eq: R-non-minimal}
R_{\rm n}^{\rm \xi}=\frac{\mathcal{N}_{\rm h}\,\xi^2}{4\pi\,\Mpl ^4}\sum^{\infty}_{\mu=1}2\,\mu\,\omega\,\Big| 2\,\mathcal{P}_{2\mu}^n\,\rho_{\rm \phi}+\frac{(\mu\,\omega)^2}{2}\phi_{\rm 0}^2\,\lvert\mathcal{P}_\mu\rvert^2\Big|^2\,.
\eea
In the total production rate, we need to add the standard gravitational contribution associated with the exchange of gravitons to the contribution from the non-minimal coupling to gravity, thus $R_{\rm n}^{\rm t}=R_{\rm n}^{\rm 0}+R_{\rm n}^{\rm \xi}$. Substituting the $\phi_{\rm 0}$ dependence and the expression for $\omega$ (see, Eq.~\eqref{Eq:w}) in Eq.~\eqref{Eq: R-non-minimal}, one can find the total production rate
\bea \label{Eq: Rtotal}
R_{\rm n}^{\rm t}=\alpha_n^\xi\,\Mpl ^5\left(\frac{\rho_{\rm \phi}}{\Mpl ^4}\right)^{\frac{5n-2}{2n}}\,.
\eea
The numerical values of $\alpha_{n}^\xi$ for different values of $n$ are tabulated in Tab.~\ref{fouriersum-grav}. Finally, solving Eq.~\eqref{Eq:Boltzmann-rad} with the production rate mentioned above, we have radiation energy density
\bea \label{Eq: rad-nonminimal}
\rho_{\rm R} (a)\simeq \alpha_n^\xi\,
\frac{n+2}{8n-14}\sqrt{3}\Mpl ^4\left(\frac{\rho_{\rm end}}{\Mpl ^4}\right)^{\frac{2n-1}{n}}\left(\frac{\aend}{a}\right)^4\,,
\eea
and reheating temperature 
\bea
T_{\rm RH}=\l(\f{30}{\pi^2\,g_{\ast,\rm RH}}\r)^{1/4}\,\Mpl \left(\frac{\rho_{\rm end}}{\Mpl ^4}\right)^{\frac{4n-7}{4\,(n-4)}}\left(\frac{\alpha_n^\xi\sqrt{3}\,(n+2)}{8n-14}\right)^{\frac{3n}{4(n-4)}}\,.
\label{eq:tre-grav}
\eea
In the next section, we move our discussion toward another possibility where the evaporation of ultra-light black holes is controlling the dynamics of reheating.
\subsection{Reheating due to the evaporation of PBHs \label{subsec: Pbh-reheating}}

In this section, we shall discuss another possibility where the evaporation of (PBHs) leads to the dominant contribution to the radiation bath over the contribution from the inflaton that can successfully reheat the universe. Note that here, we are only interested in the scenario where PBHs are formed during reheating. 
Our objective is to study the evolution of the PBHs without considering the details of the PBH formation. The formation of PBHs is generally characterized by two quantities: the formation mass and the density. Once we assume the ultralight PBHs are formed after inflation during the phase of reheating, Boltzmann equations for different energy components can be written as
\bea 
\f{\d \rhor}{\d a}+ 4\f{\rhor}{a}&=&-\f{\rhobh}{\mbh}\f{\d\mbh}{\d a} +\f{\Gamma_\phi\rhophi(1+\wf)}{aH}\label{eq: rhor bz}\,,\\
\f{\d \rhobh}{\d a} +3\f{\rhobh}{a} &=& \f{\rhobh}{\mbh} \f{\d\mbh}{\d a} \label{eq: rhobh bz}\,,\\
\f{\d \mbh}{\d a} &=& -\epsilon \f{\Mpl^4}{\mbh^2}\f{1}{aH},
\label{eq: mbh bz}
\eea 
where $\epsilon=({27}/{4}){(\pi/480)~g_\ast(\tbh)}$
and the factor $27/4$ arises due to the greybody factor. The assumption we made is that the greybody factor is calculated in the geometrical-optics limit \cite{Arbey:2019mbc,Cheek:2021odj, Baldes:2020nuv}. $\rhor$ and $\rhobh$ are the radiation and PBH energy density, respectively. The change in PBH energy density comes from the dissipation of the mass due to Hawking evaporation and the dilution due to the expansion of the universe. 
The mass of a PBH can be found at any time by solving Eq.~\eqref{eq: mbh bz} as,
\bea 
\mbh=\min\l( 1-\Gamma_{\rm BH}(t-t_{\rm in})\r)^{1/3},
\label{eq:mbh}
\eea 
where $\Gamma_{\rm BH}=3\,\epsilon\,\Mpl^4/\min^3$ and $t_{\rm in}$ is the time associated with the formation point. One can find the lifetime of the BHs, $t_{\rm ev}=1/\Gamma_{\rm BH}$.
The quantity $\min$ is the formation mass of PBH related to the horizon size at the point of formation 
\bea 
\min=\gamma \f{4}{3}\f{\pi}{\hin^3}\rhophi({\ain}) = \f{4\pi\gamma\Mpl^2}{\hin},
\label{eq: mbhin}
\eea 
where $\gamma=\wf^{3/2}$ is the efficiency of the collapse. 
One can further obtain the time at formation to be $t_{\rm in}=\min/[6(1+wf)\pi\gamma\Mpl^2]$.
There are two different scenarios that can occur in the context of PBH reheating based on the value of $\beta$. $\beta$ is the fraction of total energy density that forms PBH, given by $ \rhobh(\ain)/\rho_{\rm tot}(\ain)$, where we assume $\rho_{\rm tot}\simeq \rhophi$. There exists a critical value of $\beta$, say $\betac$, above which the PBH energy density dominates over the inflaton energy density, and the reheating happens through the evaporation of PBHs. We obtain
\bea 
\betac &=& \l[ \f{\epsilon}{2\pi(1+\wf)\gamma}\f{\Mpl^2}{\min^2}\r]^{\f{2\wf}{1+\wf}}.
\label{eq:betac}
\eea 
On the other hand, even for $\beta<\betac$, reheating can happen through the evaporation of PBHs, where PBH energy density is always sub-dominated compared to the background inflaton energy density. However, for such a scenario, there are certain assumptions. Let us briefly discuss two possible scenarios (for detailed analysis, see Refs.\cite{RiajulHaque:2023cqe,Haque:2024cdh,Barman:2024slw}):
\begin{itemize}
    \item  For $\beta>\beta_{\rm c}$, the energy density of PBHs will dominate over the background energy density before its decay. Here, the evaporation end is the point of reheating $\aev=\are$ and the radiation temperature associated with the evaporation point
\bea 
\tev=\tre= \l(\f{360\,\epsilon^2}{\pi^2\,g_{\ast,\rm RH}}\r)^{1/4}\l(\f{\Mpl}{\min}\r)^{\frac{3}{2}}\Mpl\,.
\eea 
\item For $\beta<\beta_{\rm c}$, unlike the previous case, PBHs are formed and evaporate during inflaton domination, and they never dominate the total energy component of the universe. Note that, in this scenario, reheating through the evaporation of the PBHs is only possible if the inflaton EoS follows a stiff fluid, $w_\phi>1/3$, and the production rate associated with the inflaton decay or scattering is subdominant, $\Gamma_\phi \rho_\phi(1+w_\phi) < -\frac{\rho_{\rm BH}}{M_{\rm BH}} \frac{dM_{\rm BH}}{dt}$. In this context, one can find the reheating temperature \cite{Haque:2024cdh}
\beq \label{Eq:Pbhreheat}
T_{\rm RH}\sim \left(\frac{1440}{g_{\ast,\rm RH}}\right)^{\frac{1}{4}}\left(\frac{\epsilon}{2\,(1+w_\phi)\,\pi\,\gamma^{3w_\phi}}\right)^\frac{1}{2\,(1-3\,w_\phi)}\,\beta^{\frac{3\,(1+w_\phi)}{4\,(3\,w_\phi-1)}} \left(\frac{\Mpl}{M_{\rm in}}\right)^{\frac{3\,(1-w_\phi)}{2\,(1-3\,w_\phi)}}\,\Mpl\,.
\eeq
\end{itemize}

\subsubsection{Induced gravitational wave from PBH density fluctuations during PBH domination}

The ultralight PBHs can contribute in many ways to the generation of GWs. For example, inducing large curvature perturbations that result in PBH formation \cite{Baumann:2007zm,Espinosa:2018eve,Domenech:2019quo,Ragavendra:2020sop,Inomata:2023zup,Franciolini:2023pbf,Firouzjahi:2023lzg,Maity:2024odg} (also see~\cite{Flores:2024eyy,Ballesteros:2024hhq}), by evaporating gravitons \cite{Fujita:2014hha}, the inhomogeneous distribution of PBHs leads to the density fluctuations and generate GWs ~\cite{Domenech:2020ssp,Papanikolaou:2020qtd,Domenech:2021wkk,Papanikolaou:2022chm,Bhaumik:2022pil,Bhaumik:2022zdd,Papanikolaou:2022hkg,Papanikolaou:2024kjb,Domenech:2024wao,Bhaumik:2024qzd}. Note that since the ultralight PBHs form very near to the end of inflation, the scalar-induced secondary GWs (SGWs) corresponding to the formation of the PBHs lie at a very high frequency, close to the wave number associated with the end of inflation. Moreover, the stochastic GW background, due to the evaporation of PBHs into gravitons, also constitutes very high-frequency GWs. Thus, in this section, we ignore those possibilities and focus only on the induced GWs from PBH density fluctuation \footnote{Note that for induced GWs from PBH density fluctuation in a PBH domination scenario, we assume the formation of the PBHs in a radiation like background, $\wf=1/3$ and leave the analysis for  PBHs, which formed in a general $w$ dominated background, for our future analysis.}. Our final objective is to link the GW amplitude and frequency to PBH parameters like formation mass $M_{\rm in}$ and energy fraction $\beta$, such that every possible GW detection may be seen as a probe for the early PBH domination. 

We consider a gas of PBHs with the same mass randomly distributed with a Poissonian spatial distribution in space~\cite{Papanikolaou:2020qtd}. The PBH density fluctuation, which is considered to have a Poissonian distribution, will act as the isocurvature fluctuations. These isocurvature perturbations become adiabatic perturbations when PBHs dominate the Universe's energy density, and these can produce GWs in the second order. Hence, in this case of PBH domination, along with the PGWs, this induced GW spectrum will be essential since density fluctuations of the PBHs are substantial on scales nearly equal to the mean separation of PBHs at the formation time. During PBH domination, we assume a generic power spectrum in wavenumber $k$ of fluctuations in gravitational potential $\Phi$
\bea
\mathcal{P}_\Phi (k)=\mathcal{A}_\Phi \left(\frac{k}{k_{\rm UV}}\right)^m\Theta(k_{\rm UV}-k)\,,
\eea
where the amplitude of the power spectrum is $\mathcal{A}_\Phi$ and the index of the spectrum donted by $m$. $k_{\rm UV}$ indicates the ultraviolet cutoff scale of the power spectrum.  The induced GW energy density for a given primordial spectrum of fluctuations can, therefore, be computed by integrating across the internal momenta and multiplying with the appropriate kernel, as shown in  Refs.~\cite{Papanikolaou:2020qtd,Domenech:2020ssp,Domenech:2024wao,Bhaumik:2024qzd}. The contribution to the induced GWs at the peak in the approximation of instantaneous evaporation can be calculated as
\bea
\Omega_{_{\rm GW}} (k_{\rm UV})\sim \int d\,\left(\frac{p}{k_{\rm UV}}\right)\,\mathcal{P}_{\rm \Phi}^2\,(p)\,\overline{I^2\left(\frac{p}{k_{\rm UV}},\,\frac{p}{k_{\rm ev}}\right)}\sim \mathcal{P}_{\rm \Phi}^2 (k_{\rm UV})\left(\frac{k_{\rm UV}}{k_{\rm ev}}\right)^7\,,
\eea
where $p$ represents the internal scalar loop momentum and $I$ is the induced GWs kernel. Finally, following the Ref.~\cite{Domenech:2020ssp}, one can find the amplitude of the peak GW emission frequency
\bea\label{eq:omegapeak}
\Omega_{_{\rm GW,\,ev}}^{\rm peak}\simeq \frac{\pi\,c_{\rm s}^{7/3}\l(1-c_{\rm s}^2\r)^2}{384\times 2^{2/3}}\left(\frac{k_{\rm UV}}{k_{\rm ev}}\right)^{7}\mathcal{A}_\Phi^2\,,
\eea
where the amplitude of the power spectrum at evaporation can be expressed as
\bea \label{eq:Aphi}
\mathcal{A}_\Phi=\f{3}{8\pi}\l(\f{3}{2}\r)^{\frac{1}{3}}\l(\f{\kbh}{k_{\rm UV}}\r)^4\l(\f{\kev}{k_{\rm UV}}\r)^{\frac{2}{3}}\,,
\eea
where $k_{\rm BH}$ and $k_{\rm ev}$ are the wave numbers defined at the transition point to the early matter domination and the evaporation end, respectively. The mean separation distance between PBHs determines the comoving wavenumber of the UV cut-off scale as
\bea \label{eq:kUV}
k_{\rm UV} = \frac{a_{\rm in}}{d_{\rm in}} = \ain\,\left(\frac{4\,\pi\,\rho_\text{BH}\left(a_\text{in}\right)}{3\,M_{\rm in}}\right)^{1/3} = \sqrt{3}\,k_{\rm in}\,\beta^{1/3}\,,
\eea
where $k_{\rm in}=a_{\rm in}\, H(a_{\rm in})$ is the wave number defined at the point of formation. One can find the mentioned wavenumber in terms of the formation mass of PBHs, $\min$, and the fraction of total energy that forms PBHs, $\beta$ as 
\bea \label{eq:wavenumber}
\kin &=& 1.6\times 10^{7}\,{\rm Hz} ~
\beta^{-1/3}\l(\f{\min}{1~{\rm g}}\r)^{-5/6},\nn\\
\kbh &=& 1.6\times 10^{7}\,{\rm Hz} ~
\beta^{\f{2}{3}}\l(\f{\min}{1~{\rm g}}\r)^{-5/6},\\
\kev &=& 5.8\times 10^{3} \,{\rm Hz}~\l(\f{\min}{1~{\rm g}}\r)^{-3/2}.\nn
\eea 
\begin{figure}
    \centering
    \includegraphics[scale=.37]{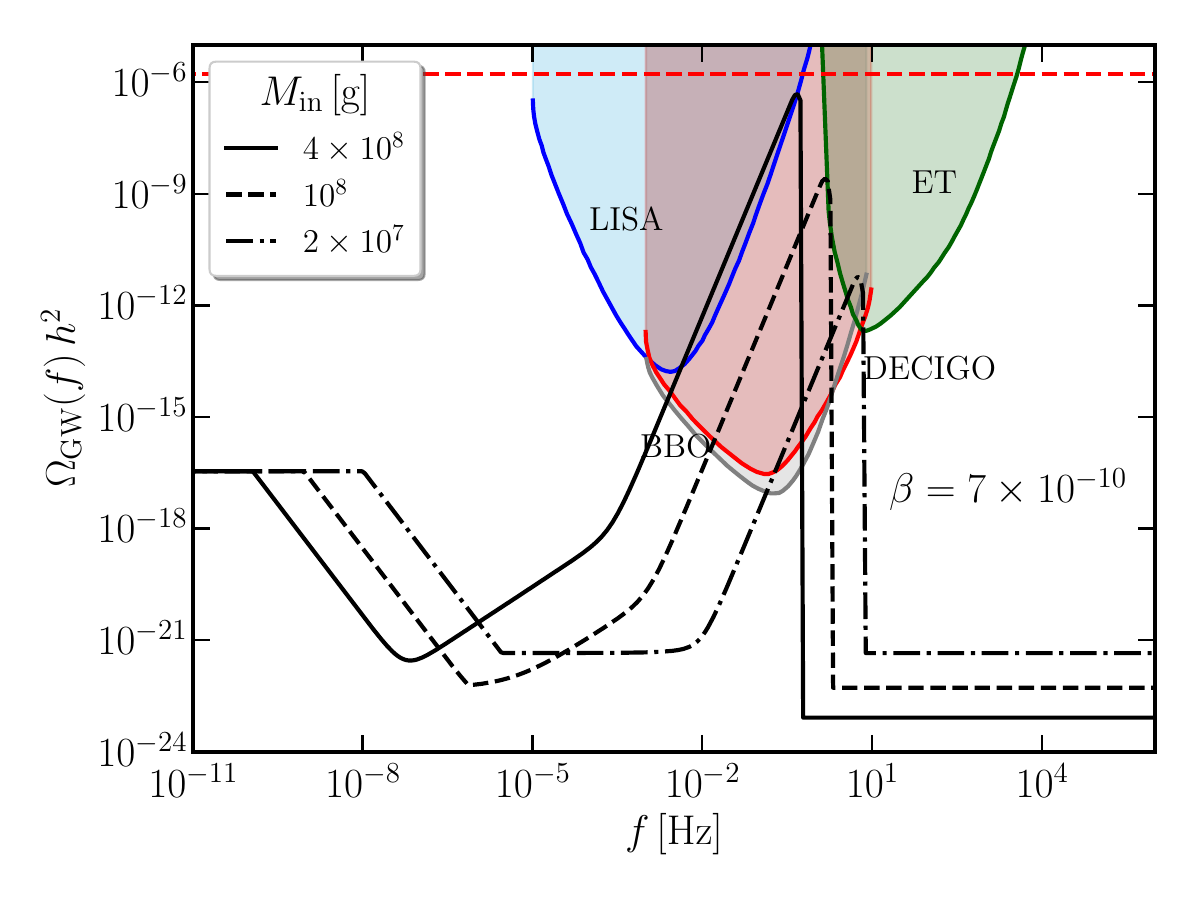}
    \includegraphics[scale=.37]{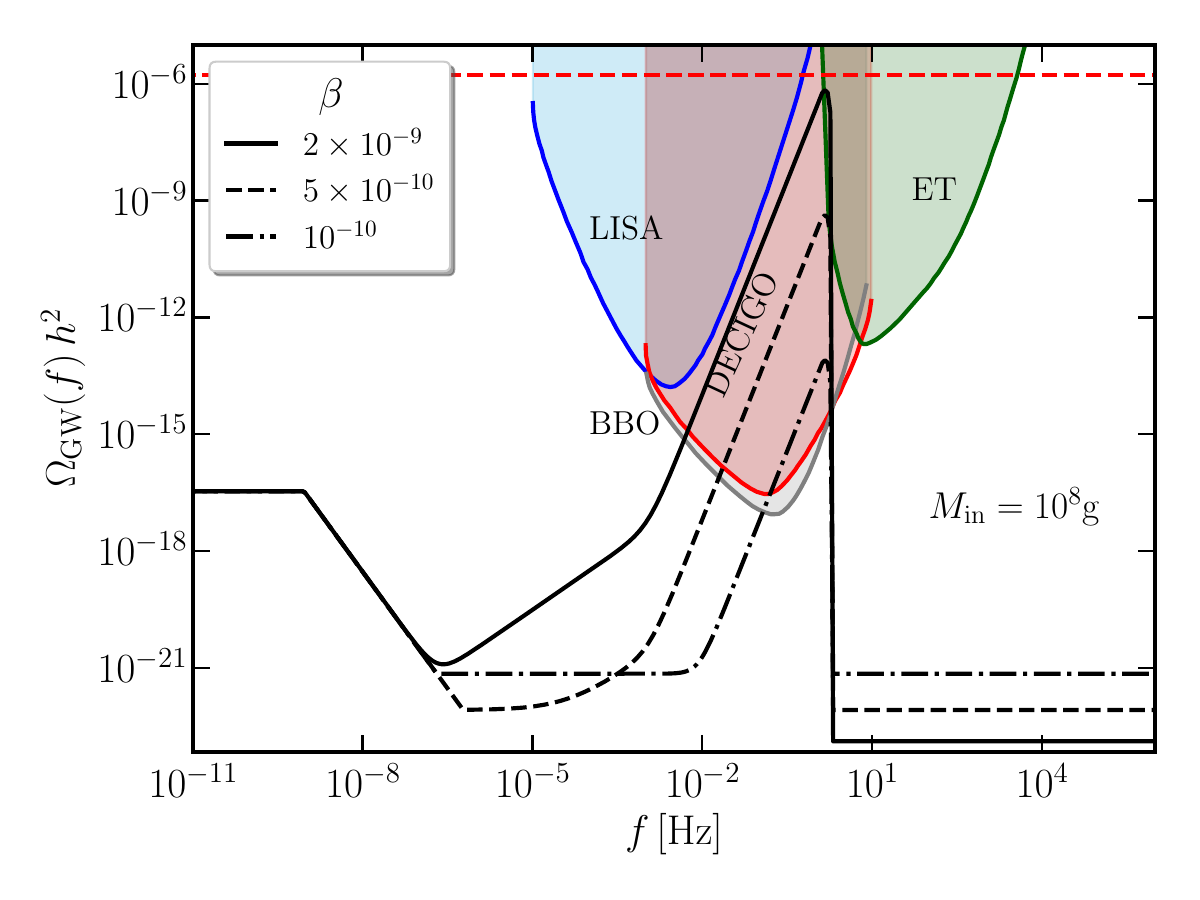}
    \caption{The spectrum of induced GWs from PBH density fluctuation, together with the spectrum of primary GWs, are plotted here.
    On the left panel, we have plotted for $\beta=7\times10^{-10}$ and for three different values of $M_{\rm in}=(4\times 10^8, \, 10^{8},\,2\times 10^{7})$ g. On the right panel, we have plotted for $M_{\rm in}=10^{8}$ g and for three different values for $\beta=2\times10^{-9},\,5\times10^{-10}$, and $10^{-10}$. Here we set the inflationary energy scale ($\Hi$) to its maximum allowed value, $4.8\times10^{13}$ GeV.}
    \label{fig:igwpbh}
\end{figure}
The spectral energy density of the induced GWs from the isocurvature due to the PBH density fluctuation as a function of frequency in terms of its peak value can be obtained as 
\bea \label{eq:finalgw}
{\ogw}_{,\rm ev} (f)
\simeq&&  \Omega_{_{\rm GW,\,ev}}^{\rm peak}(f)
 \l(\f{f}{f_{\rm UV}}\r)^{11/3} \int_{-s_0(f)}^{s_0(f)}\d s
\f{(1-s^2)^2}{(1-\cs^2s^2)^{5/3}},
\eea 
where $\cs$ is the speed of sound in the RD era, $\cs=1/\sqrt{3}$ and $f_{\rm UV}=\f{k_{\rm UV}}{2\pi}$. The limit of the above integration is taken to be 
\bea 
s_0(f)=
\begin{cases}
    1 &\f{f_{\rm UV}}{f}\geq \f{1+\cs^{-1}}{2}\\
    2\f{f_{\rm UV}}{f}-\cs^{-1} &
    \f{1+\cs^{-1}}{2}\geq \f{f_{\rm UV}}{f}
    \geq \f{\cs^{-1}}{2}\\
    0 & \f{\cs^{-1}}{2} \geq \f{f_{\rm UV}}{f}
\end{cases}.\label{eq:s0}
\eea 
However, away from the peak, at limit $k\ll \kuv$ the spectral energy density can be approximated as~\cite{Domenech:2024wao} 
\bea \label{eq:omegagw-IR}
\Omega_{_{\rm GW,\,IR}} (f)= C  \f{\cs^4 (2/3)^{1/3}}{120\pi^2} \l(\f{\kbh}{\kuv}\r)^8 \l(\f{\kuv}{\kev}\r)^{14/3} \l(\f{f}{f_{\rm UV}}\r),
\eea 
where $C \simeq 1.11$.
Also, as we have already discussed, the present-day spectral energy density of GWs can be obtained by multiplying  $c_g ~\Omega_{_{\text{rad}}}$ with the spectrum. Note that by comparing Eqs. \ref{eq:omegapeak}, \ref{eq:Aphi}, and \ref{eq:omegagw-IR}, it is clear that the amplitude of the IR tail is greatly suppressed relative to the resonant peak, specifically by a factor on the order of \( \kev / \kuv \).

In Fig.~\ref{fig:igwpbh}, we have plotted the spectrum of induced GWs from PBH density fluctuation together with the spectrum of primary GWs with the projected sensitivity of several proposed future GW experiments such as  LISA, ET, BBO, and DECIGO. We find $\Omega_{_{{\rm GW}}}\,h^2\leq 2\times 10^{-6}$, with $\Delta N_{\rm eff}\simeq 0.28$, using the current Planck data~\cite{Planck:2018jri}, which is shown by the red dashed line in each panel of Fig.~\ref{fig:igwpbh}. Here, we assume a PBH domination, which sandwiches between reheating and standard radiation domination. The spectrum of PGW has a simple power law nature, as mentioned earlier, where the peaks correspond to the induced GWs. We observe that the frequency dependency of the induced GW spectrum is $\propto f^{{11}/{3}}$ [cf. Eq.~\eqref{eq:finalgw}] and the frequency $f_{\rm UV}$ is connected to the peak frequency. Additionally, the spectrum declines rapidly above the frequency scale $2\cs\,f_{\rm UV}$ and reaches zero at the cutoff scale $2f_{\rm UV}$. Given that $f_{\rm UV}\propto k_{\rm in}$ [cf. Eq.~\eqref{eq:kUV}] and smaller mass BHs form earlier, the peak frequency shifts more toward the end of the inflation or to higher frequencies as the PBH mass drops. Once we fixed the mass of the PBHs, the amplitude of the induced GWs spectrum is highly sensitive to the PBH parameter $\beta$, i.e. $\Omega_{_{{\rm GW}}}\,h^2\propto \beta^{16/3}$ [cf. Eqs.~\eqref{eq:omegapeak} and \eqref{eq:wavenumber} ]. Thus, the amplitude of induced GW increases as $\beta$ increase, which is also visible in Fig.~\ref{fig:igwpbh}. On the left panel of Fig.~\ref{fig:igwpbh}, we have illustrated how the amplitude and the location of the induced GWs spectrum is sensitive to the $M_{\rm in}$ for a fixed value of $\beta$. Whereas in the right panel, we fixed the formation mass $M_{\rm in}=10^{8}$ g and varied the $\beta$ parameter. Note that in both plots, we assume that PBH formed in a radiation-like background, $\wf=1/3$. We shall discuss later in detail what we can infer about the detection possibilities of the PBH parameters through various future GW observatories with this GW spectrum.

\subsubsection{Effects of memory-burdened PBHs on the induced gravitational wave spectrum}

So far, in our discussion of PBH evaporation, we have assumed that PBHs lose mass solely through the Hawking evaporation process. The standard Hawking evaporation process relies on the semi-classical approximation and completely overlooks the backreaction of the emitted radiation on the quantum state of the black hole (BH) itself. In Refs. \cite{Dvali:2020wft,Dvali:2018xpy} Dvali et al., showed that this backreaction effect cannot be ignored when the energy of the emitted quanta becomes comparable to that of the BH. The key insight from \cite{Dvali:2020wft} is their application of the general phenomenon of the memory burden effect to the BH decay process, a behavior commonly observed in systems with high memory storage capacity. The central idea is that such a system is stabilized by the large amount of quantum information it holds. This means the system must transfer its memory pattern from one set of modes to another for decay to occur. Once the system has decayed by about half, the quantum information stored in memory begins to backreact and stabilize the system. As a result, the decay rate of the BH decreases as it loses mass. With this additional effect, the evolution of the PBH mass is now modified to \cite{Alexandre:2024nuo}
\bea 
\f{\d \mbh}{\d a} &=& - \f{\epsilon}{[S(\mbh)]^K}\f{\Mpl^4}{\mbh^2}\f{1}{aH},
\eea 
where $S=(1/2)(\mbh/\Mpl)^2$ is the entropy of the PBHs, and $K$ is the memory burden parameter, generally taken to be a positive number.
In general, the memory burden effect is expected to become dominant once primordial black holes (PBHs) have lost a certain fraction of their initial mass, i.e., \( \mbh = q \, M_{\rm in} \), where \( 0 < q < 1 \). It has been shown that this effect is most significant for \( q = 1/2 \)~\cite{Alexandre:2024nuo, Thoss:2024hsr}. With the inclusion of the memory burden effect, the evaporation time of PBHs is given by:
\beq
t_{\rm ev}^K \simeq \frac{q^{3+2K}}{2^K(3+2K)}\left(\frac{M_{\rm in}}{4.3\times10^{-6}~\rm{g}}\right)^{3+2K}~5.7 \times 10^{-44}~\rm{s\,,}
\eeq
Note that, to estimate the above expression, we exclude \( K \)-values very close to zero, ensuring that the duration of the burden phase always dominates over the duration of standard Hawking evaporation.  
We see that $t_{\rm ev}^K$ keeps increasing with the memory burden parameter $K$. 
Recall that we are particularly interested in the masses of PBHs that would allow them to evaporate before BBN. Since the memory burden effect prolongs the PBH lifetime, the maximum mass of PBHs that can fully evaporate before BBN decreases as the strength of this effect,
$K$, increases.
The maximum allowed mass of PBHs to evaporate before BBN is
$\min^{\rm max}\simeq6.1\times 10^3$ g for $K=1$ and $\min^{\rm max}\simeq21$ g for $K=2$.

The critical value of $\beta$ above which the PBH domination occurs is now given by 
\cite{Haque:2024eyh}
\bea 
\betac^K &=& \l[ \f{2^K (3+2K) \epsilon}{8\pi \gamma q^{3+2K}}\r]^{1/3}
\l(\f{\Mpl}{\min}\r)^{{(1+K)}}.
\label{eq:betack}
\eea 
In the expression, $\betac^K$ decreases as $k$ increases. 
This is because the memory burden effect extends the lifetime of the PBHs. As a result, even PBHs with a lower initial abundance can dominate the entire background of the universe before they evaporate. An additional constraint on the value of \( \beta \) arises in this scenario, depending on whether the memory burden effects begin before or after PBH domination. This constraint is given by: \cite{Balaji:2024hpu}
\bea
\beta_{\ast} = \sqrt{\f{3\epsilon}{16\pi\gamma (1-q^3)S(\min)}}.
\eea
Taking into account the memory burden effect, the relevant wavenumbers corresponding to the GWs induced by the PBH distribution are modified to \cite{Barman:2024iht,Balaji:2024hpu}
\bea
    \kuv &=& \kin\gamma^{-1/3}\beta^{1/3},\\
    \kev &=& \l[\f{2^k(3+2K)\epsilon}{6\pi\gamma}\r]^{1/3}
    \l(\f{\Mpl}{q\min}\r)^{\f{2}{3}(1+K)} \kin \beta^{1/3},
\eea
and
\bea 
\kbh=\begin{cases}
    \sqrt{2}\beta\kin, & {\text{for}}~\beta>\beta_{\ast},\\
    \sqrt{2}q\beta\kin, & {\text{for}}~\beta<\beta_{\ast}.
\end{cases}
\eea 
By combining these expressions, we obtain the following relation:   
\begin{align}\label{eq:kfrac}
    \frac{\kuv}{\kev}\approx 4133\,\e^{8K}\,\left(\frac{3}{3+2K}\right)^{\frac{1}{3}}\left(\frac{q\,\min}{1\,{\rm g}}\right)^{\frac{2}{3}(1+K)}\,.
\end{align}
The frequency \( f_{\rm ev} = \frac{\aev H_{\rm ev}}{2\pi} \) can be approximated as  
\begin{align}\label{eq:fev}
    f_{\rm ev} &= \frac{\sqrt{g_{\ast}(T_{\rm ev})}}{6\sqrt{10}\Mpl}\left(\frac{g_{\ast S}(T_0)}{g_{\ast S}(T_{\rm ev})}\right)^{1/3} T_{\rm ev}\,T_0 \simeq 963\,{\rm Hz}\,e^{-12K}\,\sqrt{\frac{3+2K}{3}}\,\left(\frac{q\,\min}{1\,{\rm g}}\right)^{-\left(\frac{2}{3}+K\right)}\,,
\end{align}
which leads to  
\begin{align}\label{eq:fuv}
    f_{\rm UV} \simeq 4.8\times 10^{6}\,{\rm Hz}\,e^{-4K}\,\left(\frac{3+2K}{3}\right)^{\frac{1}{6}}\,\left(\frac{q\,\min}{1\,{\rm g}}\right)^{-\left(\frac{5}{6} + \frac{K}{3}\right)}\,.
\end{align}
This implies that the modified Hawking evaporation generally results in a lower-frequency gravitational wave signature compared to the standard Hawking evaporation case. To derive the above equation, we use the temperature associated with the point of evaporation, which can be written as  
\begin{align} \label{eq:Tev1}
    T_{\rm ev} &= \Mpl\,\left(\frac{40}{\pi^2\,g_*(T_{\rm ev})}\right)^{1/4}\,\left[2^K(3+2K)\,\epsilon\left(\frac{\Mpl}{q\,M_{\rm in}}\right)^{3+2K}\right]^{1/2}\,.
\end{align}
The spectral energy density at the peak in the memory burden scenario will be modified as follows: 
\bea 
\Omega_{_{\rm GW,\,ev}}^{\rm peak}(k) =\f{q^4}{24567\sqrt{3}\pi} \l(\f{3+2K}{3\sqrt{2(1+K)}}\r)^{-\f{4}{3+2K}}
\l(\f{\kbh}{\kuv}\r)^8 
\l(\f{\kuv}{\kev}\r)^{7-\f{4}{3+2K}},
\eea 
and the spectral energy density of the induced GWs is given by 
\bea \label{eq:finalgwk}
{\ogw}_{,\rm ev} (f)
\simeq&&  \Omega_{_{\rm GW,\,ev}}^{\rm peak}(f)
 \l(\f{f}{f_{\rm UV}}\r)^{\f{11+10K}{3+2K}} \int_{-s_0(f)}^{s_0(f)}\d s
{(1-s^2)^2}\,{(1-\cs^2s^2)^{-\f{5+2K}{3+2K}}},
\eea 
where $s_0$ is defined in a similar way as given in Eq.~\eqref{eq:s0}. Upon substituting the expressions for the wave numbers, we obtain the following for the peak of the gravitational wave spectrum:  
\begin{align}\label{eq:omega-peak-ev}
& \Omega_{\rm GW,ev}^{\rm peak}\simeq \frac{4133^{-\frac{4}{3+2K}}\,q^4}{2.3\times10^{-20}}\left(\frac{3+2K}{3}\right)^{-\frac{7}{3}+\frac{4}{9+6K}}\,\beta^{\frac{16}{3}}\,e^{8\,K\,(7-\frac{4}{3+2K})}
\nonumber\\&\times 
\left(\frac{q\,M_{\rm in}}{1\,{\rm g}}\right)^{\frac{2}{3}(1+K)\left(7-\frac{4}{3+2K}\right)}
\begin{cases}
1 & \text{for}\quad \beta>\beta_\star
\\[10pt]
q^8 & \text{for}\quad \beta<\beta_\star\,.
\end{cases}
\end{align}  
The peak amplitude is exponentially enhanced with \( k \), while also being sensitive to both \( M_{\rm in} \) and \( \beta \).
The IR part of the spectral energy density is given by \cite{Domenech:2024wao}
\bea
{\ogw}_{,\rm IR} (f)
\simeq&& \f{C\,\cs^4q^4}{24\pi^2} \f{3+2K}{15+18K}
\l(\f{3}{2}\r)^{\f{2}{3+2K}} \l(\f{3\sqrt{(1+K)}}{3+2K}\r)^{\f{4}{3+2K}} \nn\\&&
\l(\f{\kuv}{\kev}\r)^{\f{14+12K}{3+2K}}
\l(\f{\kbh}{\kuv}\r)^8  \l(\f{k}{\kuv}\r).
\eea 
We would like to emphasize again that the IR part of the spectral energy density is subdominant compared to the resonant peak. As a result, it does not have a significant impact on the \( \beta-\min \) parameter space, which we derived through the SNR analysis to ensure compatibility of our induced gravitational wave spectrum with future GW missions.

\begin{figure}
    \centering
    \includegraphics[scale=.47]{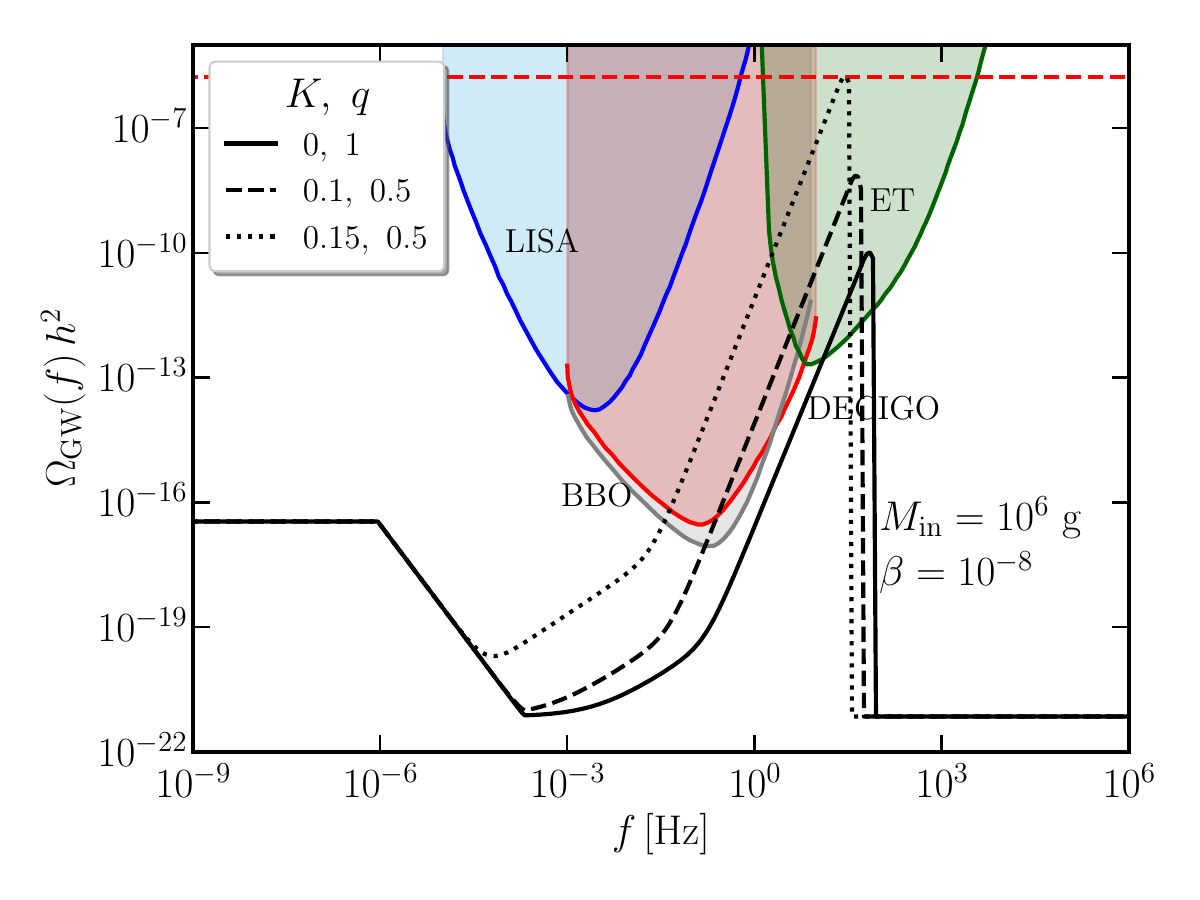}
    \caption{The effect of the memory burden parameter on the spectral energy density of induced GWs from PBH density fluctuation is illustrated.
    We have chosen $\min=10^6$ g and $\beta=10^{-8}$.
    Further, we assumed that the memory burden effect starts when $q=1/2$. The Hawking evaporation $(K=0, \, q=1)$ and the memory burden effects are shown in solid and dashed lines, respectively.}
    \label{fig:igwpbhmb}
\end{figure}

In Fig.~\ref{fig:igwpbhmb}, we show the impact of memory burden on the spectral energy density of the induced gravitational waves and compare it to the standard case with \( k=0 \) and \( q=1 \), where the memory effect is not considered. We observe that the peak of the spectrum not only shifts to lower frequencies, but its amplitude also increases significantly. This is because, for \( k=0 \), there is no enhancement in the peak amplitude, as shown by Eq.~\eqref{eq:omega-peak-ev}. Additionally, in the presence of the memory effect, the UV cutoff frequency, \( f_{\rm UV} \), experiences exponential suppression [cf. Eq.~\eqref{eq:fuv}], causing it to shift to a lower value compared to the case with \( k=0 \). Thus, we conclude that accounting for backreaction effects can actually improve the detection prospects for induced gravitational waves from PBH density fluctuations.

\section{Gravitational production of dark matter
\label{sec: DM production}}


We start our discussion with the well-studied gravitational production of DM through the exchange of a graviton\footnote{It is important to note that the evaporation of PBHs also leads to the production of DM. A comprehensive analysis would involve considering all possible production channels and comparing the contributions from each process. However, in this work, we have focused solely on the production of DM through gravitational interactions, leaving the exploration of other potential mechanisms for future investigation.} \cite{Mambrini:2021zpp,Bernal:2021kaj,Barman:2021ugy,Haque:2021mab,Clery:2021bwz,Haque:2022kez,Clery:2022wib,Ahmed:2022tfm}. Depending on the origin of the production and the spin of the final state, DM energy density varies. In principle, DM can be produced gravitationally from both inflaton and thermal bath. However, the production from inflation always dominates over the production from thermal baths for both scalar and fermionic DM, except at very high temperatures in the case of fermionic DM. With this point in mind, we shall ignore the production of DM from the thermal bath throughout our analysis. The Boltzmann equation for DM number density is given by
\beq
\frac{dn_s^\phi}{da} + 3 \frac{n_s^\phi}{a}=
\frac{R^\phi_s(a)}{H a}\,,
\label{Eq:boltzmann4}
\eeq
where the subscript $s$ refers to the type of DM that is produced, i.e., $s=0$ for scalar DM and $s=1/2$ for fermionic DM. Here $R^\phi_s$ is the production rate for scalar DM species \cite{Clery:2021bwz}, which is given as 
\begin{equation}
R^{\phi}_{0}=\frac{\rho_\phi^2}{8\pi \Mpl ^4}\,\Sigma_{0}^n\,,\,  \Sigma_{0}^n = \sum_{\mu=1}^{\infty} |{\cal P}_{\mu}^n|^2\,
   \left( 1+\frac{2\,m_X^2}{E_{\mu}^2}\right)^2\,
\left(1-\frac{4\,m_{X}^2}{E_{\mu}^2}\right)^{1/2} \,.
\label{eq:ratescalar}
\end{equation}
Similarly, for the production of fermionic DM, one can find production rate \cite{Clery:2021bwz} to be
\begin{equation}
R^{\phi}_{1/2}=\frac{\rho_\phi^2}{2\pi \Mpl ^4}\frac{m_X^2}{m_\phi^2}\,\Sigma_{1/2}^n\,, \,  \Sigma_{1/2}^n = \sum_{\mu=1}^{\infty} |{\cal P}_{\mu}^n|^2\,
    \frac{m_\phi^2}{E_{\mu}^2}\,
\left(1-\frac{4\,m_{X}^2}{E_{\mu}^2}\right)^{3/2} \,.
\label{eq:rateferm}
\end{equation}
In both Eq.~\eqref{eq:ratescalar} and Eq.~\eqref{eq:rateferm} the summation accounts for the sum over the Fourier modes associated with the inflaton potential
\bea
V(\phi)=V(\phi_0)\sum_{\mu=-\infty}^{+\infty}\mathcal{P}_\mu^n\,\e^{-i\mu\omega t}=\rho_{\rm \phi}\sum_{\mu=-\infty}^{+\infty}\mathcal{P}_\mu^n\,\e^{-i\mu\omega t}\,,
\eea
 and $E_\mu = \mu \omega$ represents the energy of the $n^{\rm th}$ oscillation mode of inflaton. Eq.~\eqref{Eq:boltzmann4} can be re-written in terms of comoving DM number density 
 \bea \label{Eq: YieldDM}
 \frac{ d(n_s a^3)}{da}=\frac{\sqrt{3}\,\Mpl}{\sqrt{\rho_{\rm end}}}a^2\left(\frac{a}{\aend}\right)^{\frac{3\,n}{n+2}}R^{\phi}_{s}(a)\,.
 \eea
 To find above equation we assume $H=\sqrt{\rho_{\rm \phi}/(3\Mpl^2)}$, which is valid for $a<\are$. 
\begin{table}
\centering
 \begin{tabular}{||c|c|c|c|c|c|c|c|c|c|c||} 
 \hline 
 $n$ & $2$& $4$& $6$ &$8$&$10$&$12$&$14$&$16$&$18$&$20 $ \\
 \hline\hline
 $\sum_0^n$& $0.063$ &  $0.063$ &  $0.056$ & $0.049$ & $0.043$ & $0.039$ & $0.035$ & $0.032$ & $0.029$ & $0.027$
 \\ \hline
 $\sum_{1/2}^n$& $0.016$&  $0.061$&  $0.101$& $0.133$& $0.157$ & $0.177$ & $0.192$ & $0.205$ & $0.216$  & $0.225$
 \\ \hline
 \end{tabular}
 \caption{\label{tb: dm grav1}{Numerical values of $\Sigma_0^n$ and $\Sigma_{1/2}^n$ are tabulated here for different values of $n$.}}
 \end{table}
 Substituting Eq.~\eqref{eq:ratescalar} into the above equation, one can find DM number density for scalar DM at $\are$
 \bea \label{Eq: DM-scalar}
 \frac{n_0^{\phi}(\are)}{\tre^3}\simeq\frac{\sqrt{3}\,(n+2)\alpha_T^{\frac{n+2}{2n}}}{48\pi(n-1)}\left(\frac{\Mpl}{\tre}\right)^{\frac{n-4}{n}}\left(\frac{\rho_{\rm end}}{\Mpl^4}\right)^{\frac{n-1}{n}}\Sigma_{0}^n\,,
 \eea
where $\alpha_T=\frac{\pi^2}{30}\,g_{\ast,\rm RH}$. For the fermionic DM we have
\begin{equation}\label{Eq: DM-fermion}
\frac{n_{1/2}^{\phi}(\are)}{\tre^3}\simeq\frac{\alpha_T^{\frac{n+2}{2n}}\sqrt{3}(n+2)}{12\pi n (n-1)\lambda^{\frac{2}{n}}}\left(\frac{m_{X}}{\Mpl}\right)^2\left(\frac{\tre}{\Mpl}\right)^{\frac{4-n}{n}}
\left(\frac{\rho_{\rm end}}{\Mpl^4}\right)^{\frac{1}{n}}\Sigma_{1/2}^n\,,
\end{equation}
 where $\Sigma_{0}^n$ and $\Sigma_{1/2}^n$ are given in Eq.~\eqref{eq:ratescalar} and Eq.~\eqref{eq:rateferm}. In Tab.~\ref{tb: dm grav1} the numerical values of $\Sigma_{0}^n$ and $\Sigma_{1/2}^n$ are tabulated for different values of $n$. We see that $\Sigma_{0}^n$ decreases with the increasing $n$ whereas $\Sigma_{1/2}^n$ increases with $n$. Finally, the DM abundance at present-day can be written as \cite{book}
\bea
\frac{\Omega_X h^2}{0.12}= 1.3\times 10^9\,\frac{g_0}{g_{\ast,\rm RH}}\,\frac{ n_s^\phi(\are)}{\tre^3}\,\frac{m_X}{\text{GeV}}\,,
\label{Eq:omegah2m}
\eea
where $g_0=3.91$ indicates the number of light species for entropy calculated at the present day. From Eqs.~\eqref{Eq:omegah2m}, \eqref{Eq: DM-fermion}, and \eqref{Eq: DM-scalar}, it is clear that the relic density of DM today depends on the inflationary energy density, reheating EoS, and reheating temperature, those factors are also important in case of GW spectrum. Therefore, the detection of PGWs by the already mentioned GW detectors opens up a probing possibility of DM parameters in an indirect manner, such as here, the mass of the DM $m_X$. In the next section, we shall focus on our results related to probing different reheating scenarios and gravitational production of DM in the light of the GW spectrum with four future GW experiments, such as LISA, ET, BBO, and DECIGO.


\section{Results \label{sec: results}}
\begin{figure}[t!]
    \centering
    \includegraphics[scale=.37]{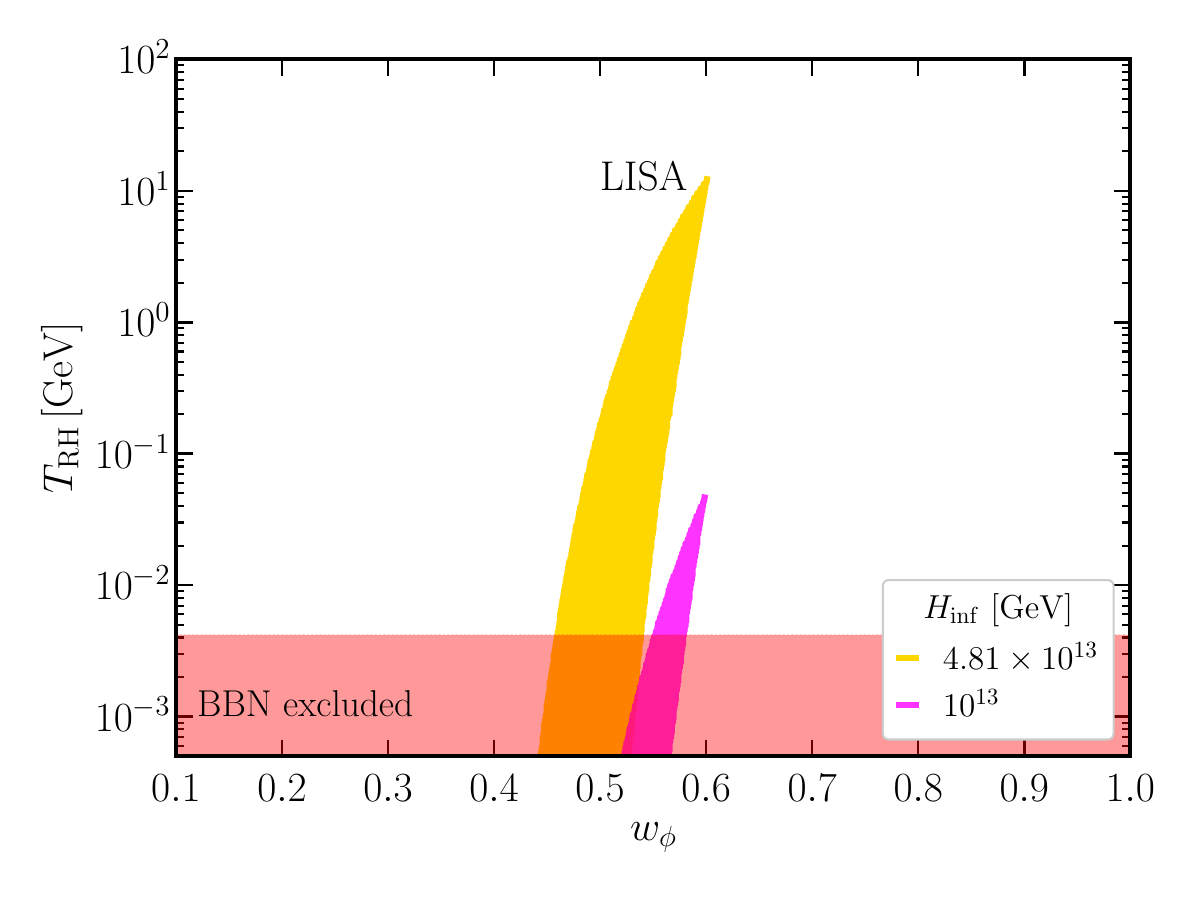}
    \includegraphics[scale=.37]{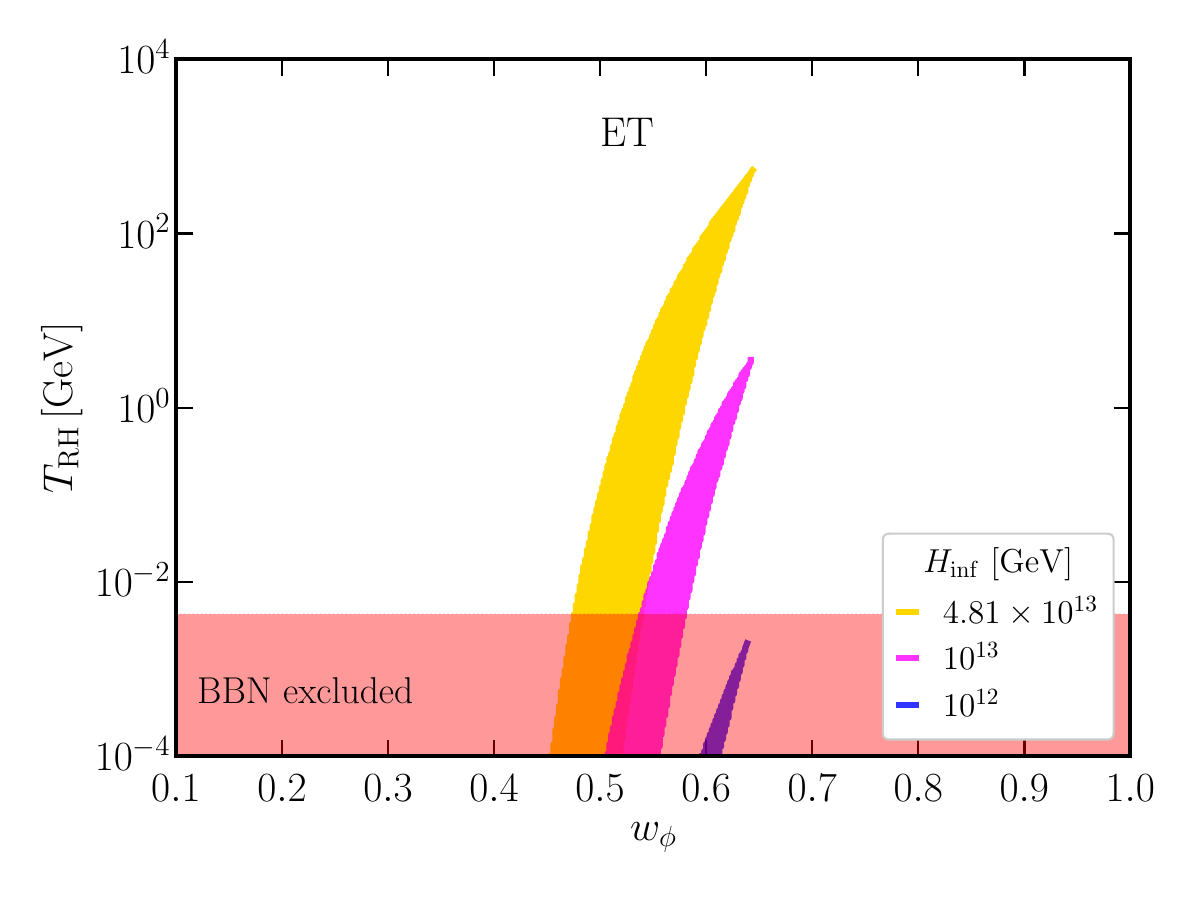}
    \includegraphics[scale=.37]{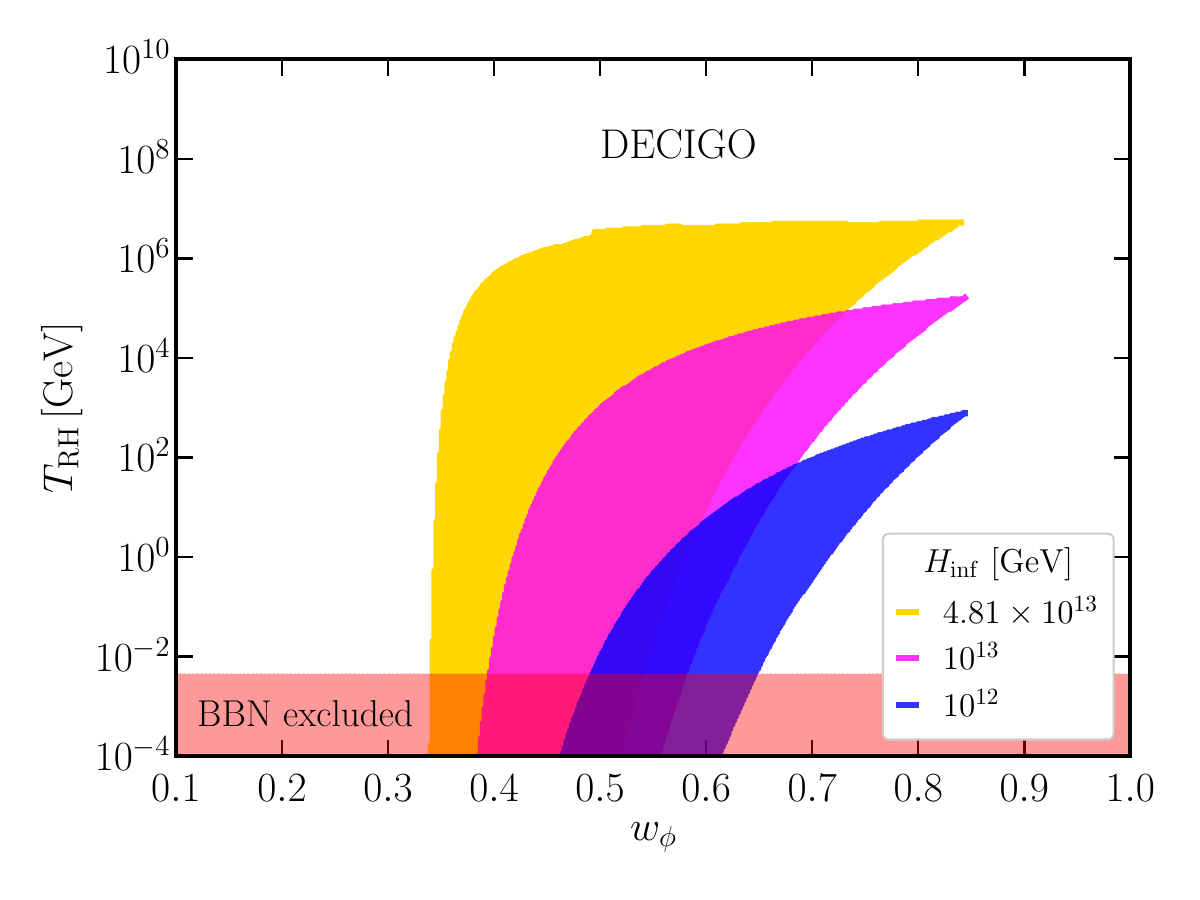}
    \includegraphics[scale=.37]{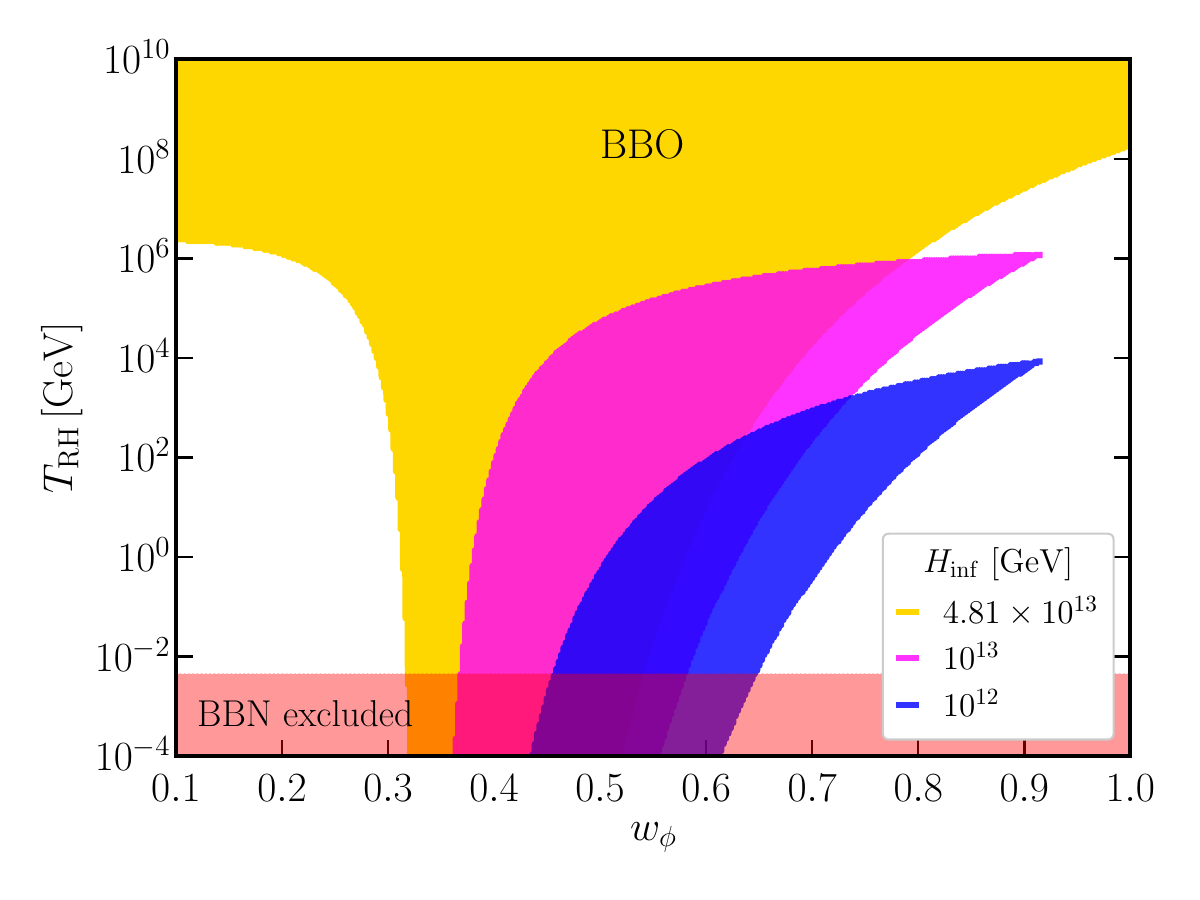}
    \caption{ The consistent region in $T_{\rm RH}-w_\phi$ parameter space from the forecast of future GW detectors (LISA, ET, DECIGO, and BBO) and $\Delta {\rm N_{\rm eff}}$ constraints for PGWs. The shaded region indicates $\rm SNR>\rho_{\rm th}$. Where for $\rho_{\rm th}$, we choose 10 for all space-based interferometers, and for ET, it is 5. We have chosen three sample values of $H_{\rm inf}= (4.8\times 10^{13},\,10^{13},\,10^{12})$ GeV. For all space-based detectors LISA, BBO, and DECIGO, we assume $t_{\rm obs}=4$ yrs, and for ground-based interferometer ET, we took $t_{\rm obs}=5$ yrs. }
    \label{fig:freeplot}
\end{figure}
We shall now discuss the parameter space for all the scenarios we have discussed that can be probed through the detection of PGWs by LISA, ET, BBO, and DECIGO.  We shall mention again that we use the SNR approach (discussed in Sec.~\ref{sec: detector}) for the analysis of the detection of a GW spectrum by future GW missions. 
There are certain constraints that we need to consider throughout our analysis. Firstly, the lowest value of $\tre$ is always set to the BBN temperature  $T_{_{\mathrm{BBN}}}\sim 4$ MeV. 
For a scale-invariant tensor spectrum during inflation, the spectral tilt of the PGWs for the wave numbers re-entering during reheating is determined by the background EoS $\wf$, which is fixed by the post-inflationary dynamics of the inflaton, and the amplitude is fixed by the energy scale of the inflation. Moreover, the frequency band for which the tilted spectrum is present depends on $\tre$, $\wf$, and the inflationary energy scale. Once we fixed the inflationary energy scale and $\wf$, lowering the reheating temperature increases the frequency range for the tilted PGWs spectrum. Note that the amplitude of the spectrum does not exceed the maximum value set by the constraints from $\Delta N_{\mathrm{eff}}$. 
The parameter space for the quantities which shall be shown in this section involves the spectrum that a detector will detect, but also, the spectrum does not violate the $\Delta N_{\mathrm{eff}}$ bound.

In this section, we shall discuss our results in the following manner.
Firstly, we take a model-independent approach to find out the reheating parameters, such as $\tre$, $\wf$, and inflationary parameters, such as the energy scale of the inflation $H_{\rm inf}$, which are consistent with the mentioned GW detectors considering PGWs. Then, we shall consider the $\alpha$-attractor model of inflation and show the $n_s-r$ predictions of this model consistent with the forecast set by different detectors. Later, we take different reheating scenarios, such as non-gravitational, gravitational, and PBH reheating, and estimate the parameter space associated with those reheating scenarios, which can be validated through the forecast of different GW detectors. We shall also demonstrate the parameter space for the DM mass that is produced from gravitational interaction. Finally, we will show how the parameter space alters with the variation of the threshold value of the detector's SNR.

\subsection{Model independent results \label{sec: MI approach}}
Let us discuss the parameters that play a role in the amplitude and tilt of the PGW spectrum and could be probed by future GW detectors. 
As mentioned earlier, the EoS during reheating, $\wf$, plays the role of modifying the spectral tilt of the PGWs spectrum, reheating temperature $\tre$, $\wf$, and inflationary energy scale $H_{\rm inf}$ fixes the frequency band for the tilt, and the amplitude of the spectrum is set by $H_{\rm inf}$. Hence, our goal is to probe those parameters using various GW detectors.

In Fig.~\ref{fig:freeplot}, we illustrate the parameter space of $\wf$ and $\tre$ for different inflationary energy scales consistent with the forecast of the four different observations, LISA, ET, DECIGO, and BBO, shown in the top left, top right, bottom left, and bottom right, respectively. Note that the shaded region indicates SNR $> \rho_{\rm th}$, which is needed for detection.
For all space-based detectors, LISA, BBO,
and DECIGO, we assume $t_{\rm obs}$ = 4 years, and for ground-based interferometer ET, we took $t_{\rm obs} = 5$ years. We choose the range of $\wf$ from matter like reheating to kination, $\wf=(0,\,1)$.  The red-shaded area corresponds to the region excluded from the BBN bound on the reheating temperature. 
One can also notice that the parametric region is bounded from both left and right sides. 
On the left, the excluded region indicates whether the signal enters the detector sensitivity. 
 The excluded region on the right side of the plots comes from the $\Delta N_{\mathrm{eff}}$ constraint.

We see that among four GW detectors, for LISA, we get a very narrow region of consistent parameter space in the $\tre-\wf$ plane. Moreover, decreasing the inflationary energy scale shrinks the parameter space further.
\begin{table}[h!]
  \begin{center}
    \begin{tabular}{||c|c|c|c|c||}
      \hline
         Detectors & LISA & ET & DECIGO & BBO \\
         \hline\hline
         $H_{\rm inf}^{\rm min}$ (GeV) &$5\times10^{12}$ & $2\times10^{12}$&$5\times10^{9}$&$10^9$\\
         \hline
         $\tre^{\rm max}$ (GeV) &15&500&$6\times 10^6$& $10^{15}$\\
         \hline
         $\wf$  &$0.45-0.6$&$0.47-0.63$&$0.33-0.84$& $0-1.0$\\
         \hline
    \end{tabular}
    \caption{The minimum value of the inflationary energy scale, the maximum value of the reheating temperature, and the ranges of $\wf$  that can be probed by LISA, ET, DECIGO, and BBO are listed here. }\label{tabl:minenergy}
  \end{center}
\end{table}
The minimum inflationary energy scale, the maximum value of the reheating temperature, and the ranges of $\wf$ that can be probed are listed in Tab.~\ref{tabl:minenergy}.
Another important point is to note that except for BBO, all other detectors we discussed need a blue tilt in the spectrum to be detected. However, even for such a scenario where the future forecast from BBO can detect a red-tilted GW spectrum, which is for the case when $\wf<1/3$, a high inflationary energy scale $H_{\rm inf}$ is required. Hence, post-inflationary dynamics with $\wf \leq 1/3$ are largely unobserved in the mentioned GW detectors with PGWs for a slow roll inflation model (i.e., tensor perturbation is roughly scale-invariant). A higher value of the spectral index is required to detect the spectrum for a lower energy scale, which refers to a steeper EoS. This can also be seen in Fig.~\ref{fig:freeplot}. To summarize, we see that a larger part of the low $\wf$ region is not favored with the detection due to the low spectral index of the GW spectrum, whether the steeper region of $\wf$ and low $\tre$ is restricted from the $\Delta N_{\rm eff}$ constraints for the PGWs. The detectors favor the region around $\wf\sim 0.5$. In the next section, we shall consider a slow roll inflationary model, i.e., the $\alpha-$ attractor model, and discuss the results. As we know, the inflaton EoS $\wf$, which defines the background dynamics of the post-inflationary era, is directly connected with the exponent of the potential, $\wf=(n-2)/(n+2)$, we show the region in $\tre-n$ plane for different values of potential parameter $\alpha$ can be probed with the forecast of different GW detectors. Then, we examine how the $n_s-r$ prediction of this model is consistent with the estimates of different GW detectors. 
\subsection{Model-dependent results: $\alpha$-attractor model of inflation \label{sec: M approach}}
\begin{figure}[t!]
    \centering
    \includegraphics[scale=.37]{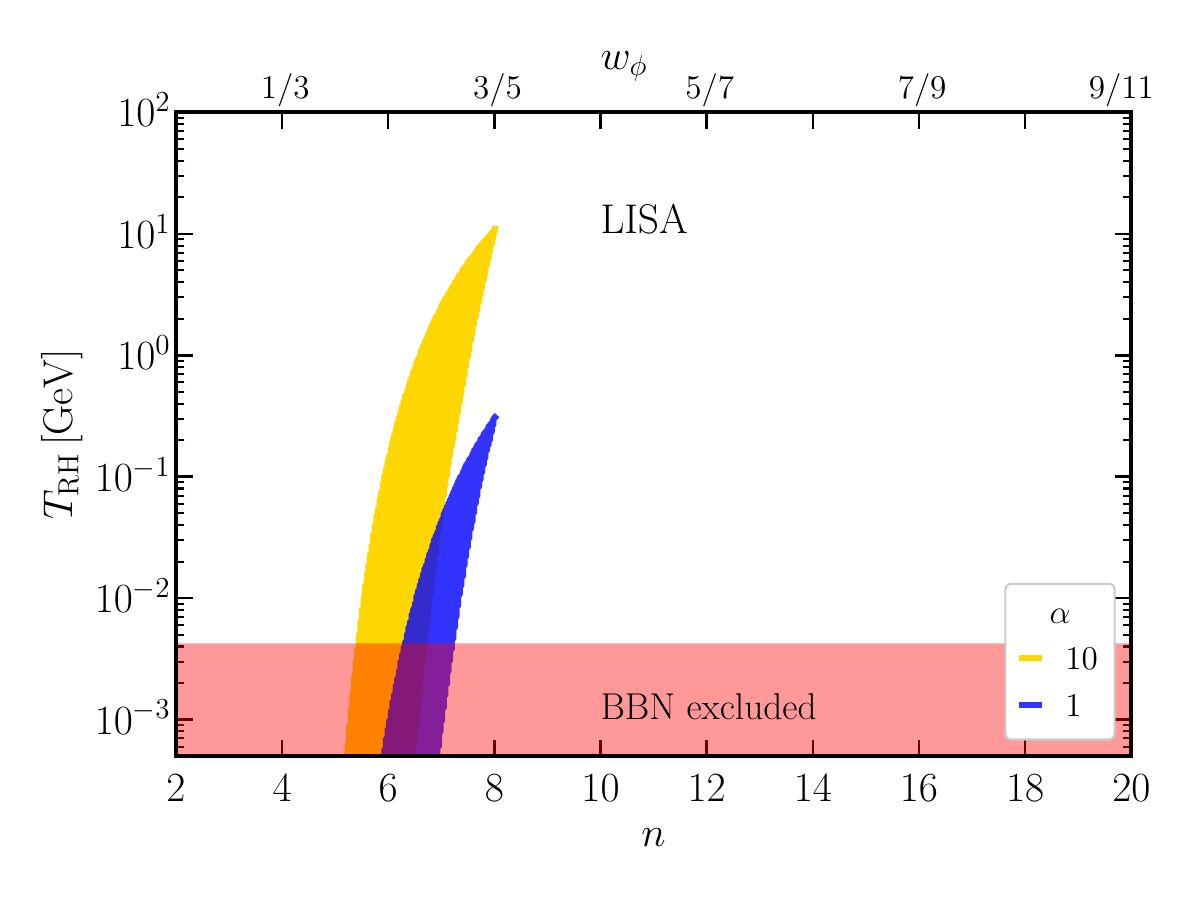}
    \includegraphics[scale=.37]{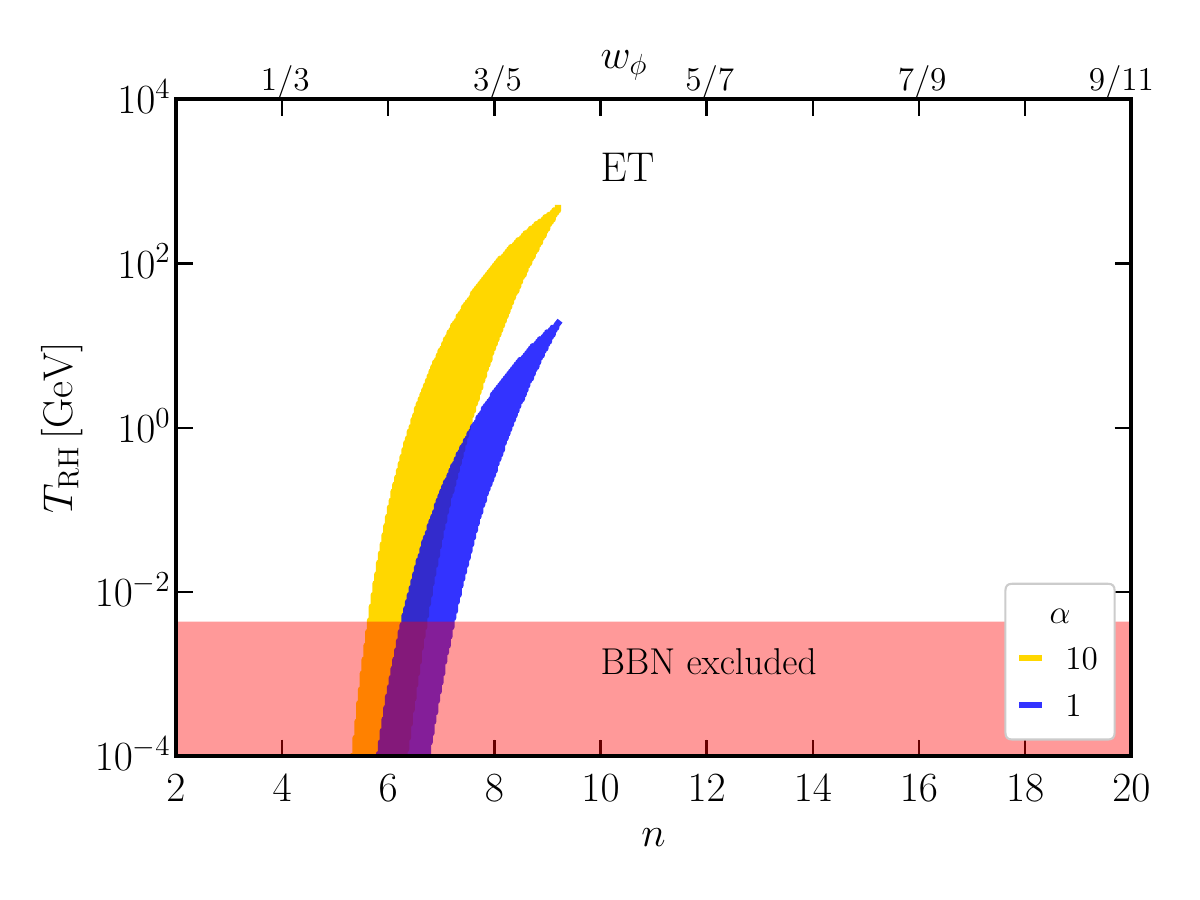}
    \includegraphics[scale=.37]{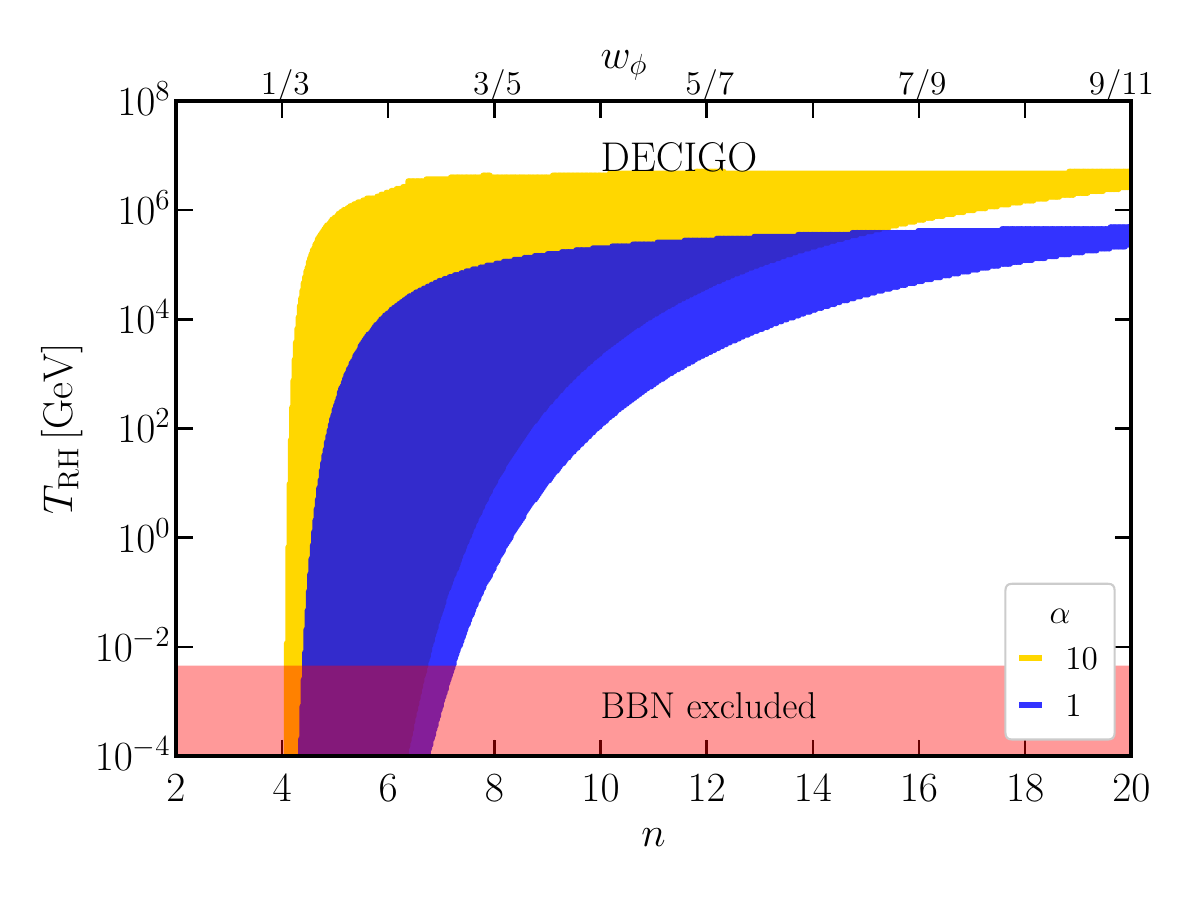}
    \includegraphics[scale=.37]{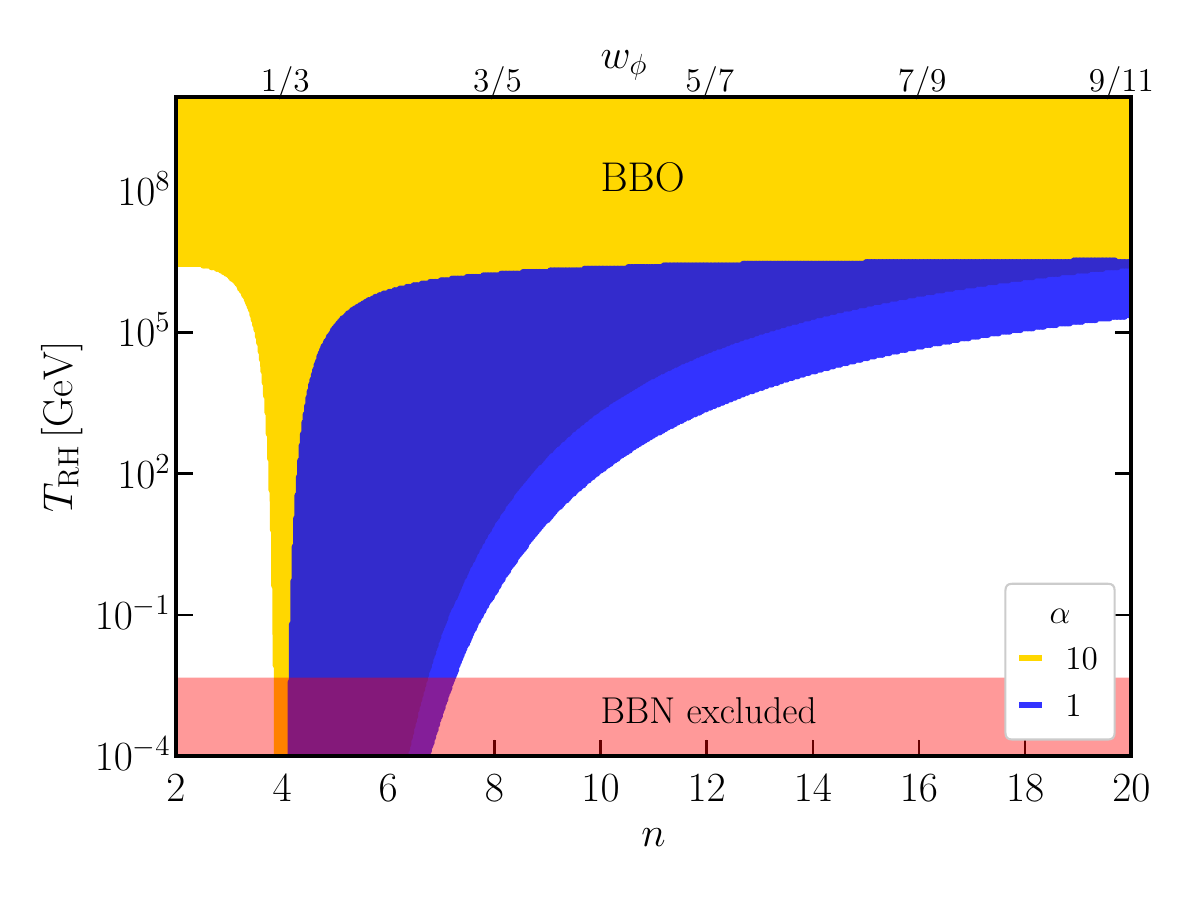}
    \caption{ The detectable region in $\tre-n$ parameter space from future GW detectors (LISA, ET, DECIGO, and BBO) and $\Delta {\rm N_{\rm eff}}$ constraints for PGWs are illustrated here for $\alpha$-attractor model of inflation. The shaded region indicates $\rm SNR>\rho_{\rm th}$. We have chosen two different values of $\alpha$, where $\alpha=1$ and $10$ are plotted in blue and yellow color. The red region is restricted from BBN bound as $\tre<4$ MeV.}
    \label{fig:modelplot}
\end{figure}

\begin{figure}[t!]
    \centering
    \includegraphics[scale=.39]{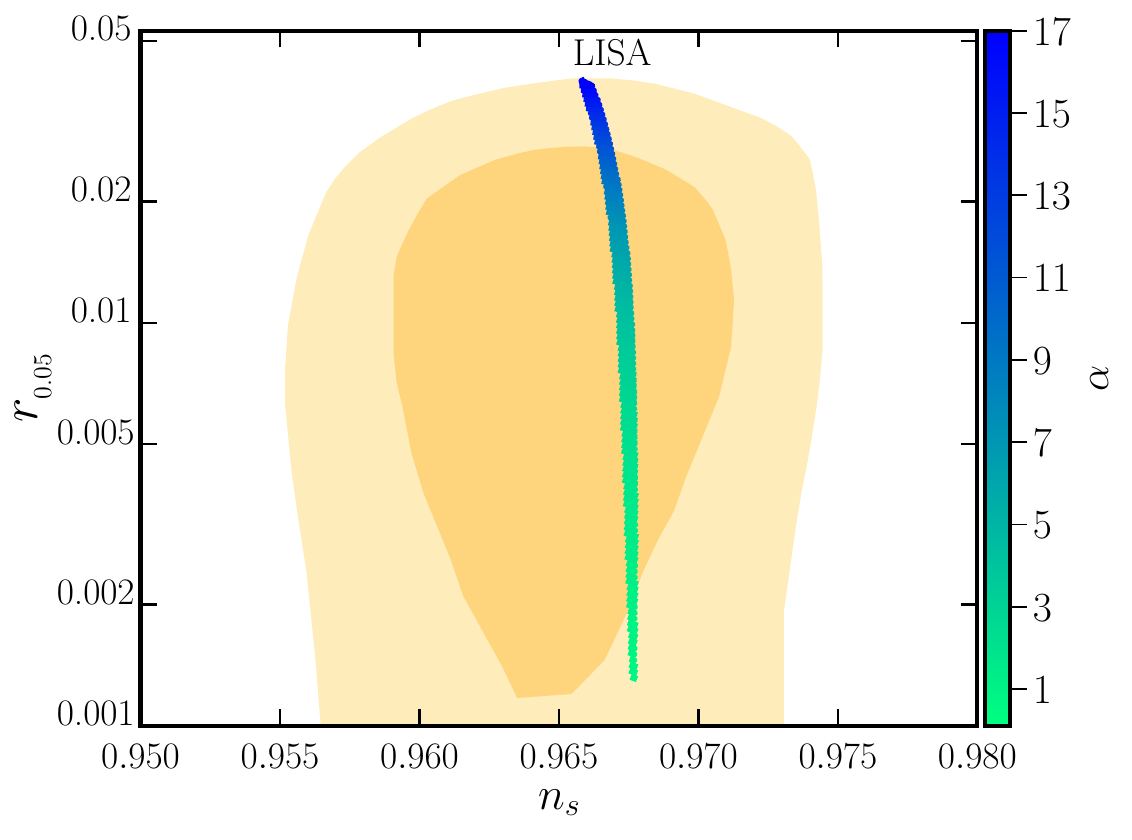}
    \includegraphics[scale=.39]{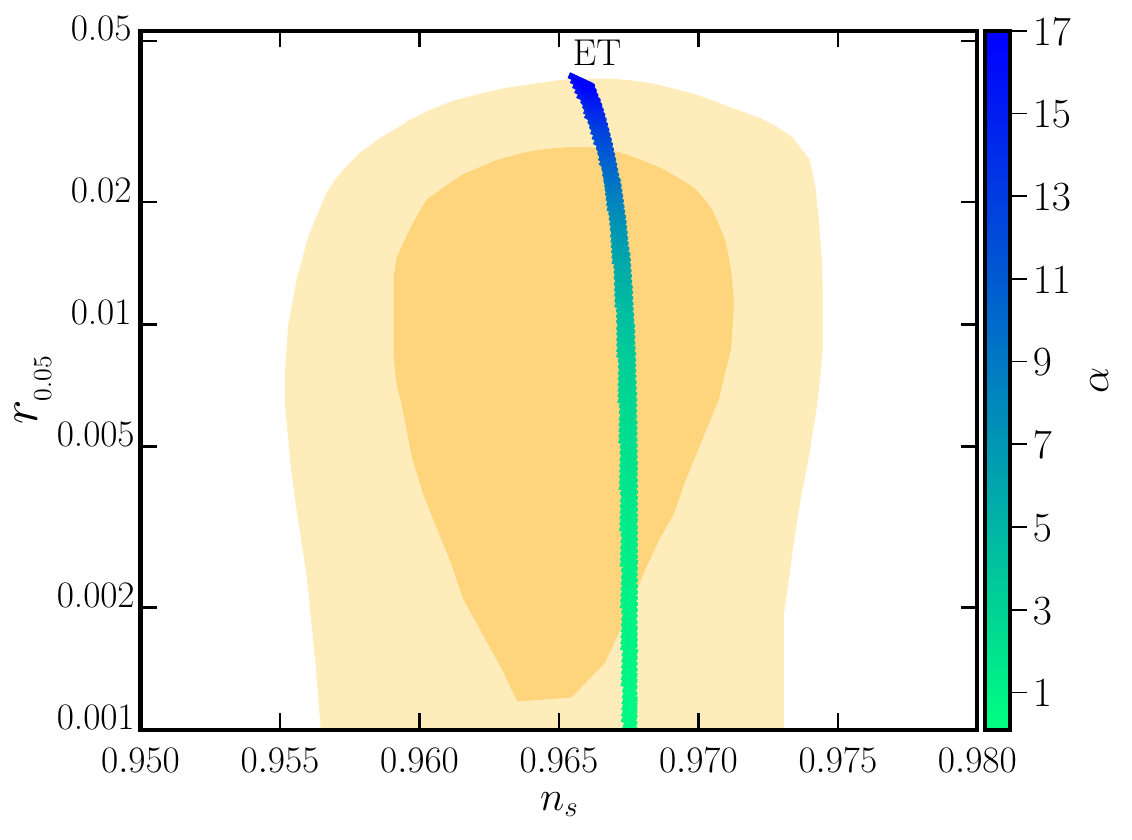}
    \includegraphics[scale=.39]{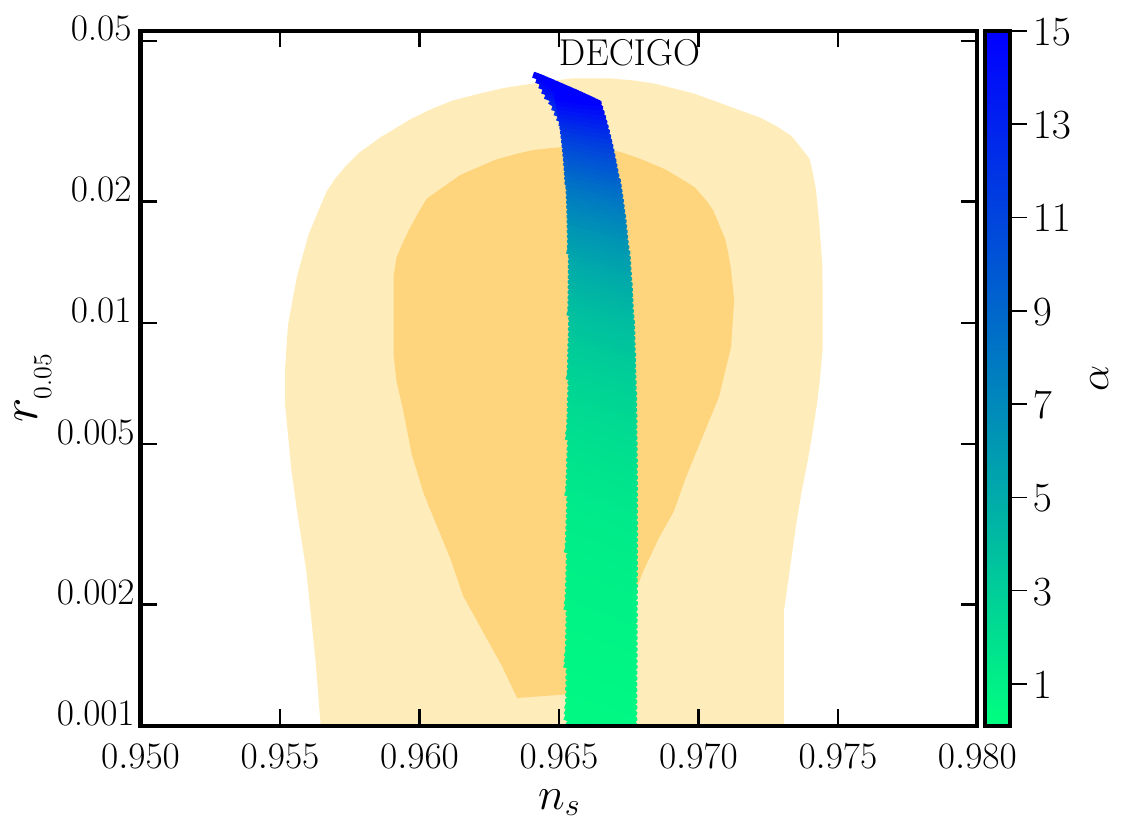}
    \includegraphics[scale=.39]{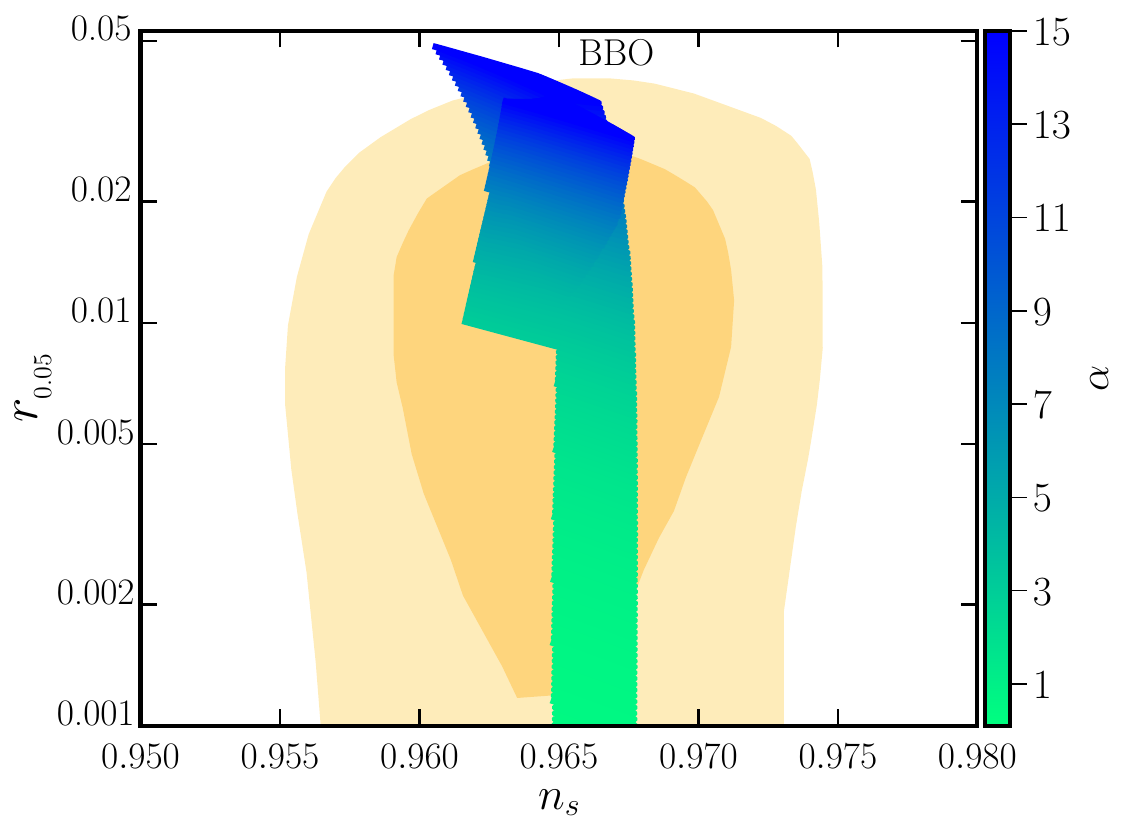}
    \caption{ The detectable region in $n_s-r$ plane for $\alpha$-attractor model of inflation corresponding to the different GW detectors, LISA, E.T., DECIGO, and BBO, projected on the recent {\rm PLANCK + BICEP/$Keck$} data \cite{Tristram:2021tvh}. Deep and light yellow shaded area indicate $1\sigma$ region at 68$\%$ CL and $2\sigma$ region at 95$\%$ C.L., respectively. Note that to find out this parameter space, we vary $n$  from $2-10$ and the reheating temperature from maximum, set by instantaneous reheating to the minimum possible temperature fixed by either $\Delta N_{\rm eff}$ bound or from BBN depending on which one is stronger. }
    \label{fig: ns r}
\end{figure}
We have considered the $\alpha$-attractor model of inflation.  After specifying the inflation model, we have three independent parameters: $n$, $\alpha$, and $\tre$.  The reheating EoS is determined by $n$ (i.e., Eq.~\eqref{eq: were n}), and the inflationary energy scale is set by $\alpha$ and $n$ (i.e., Eqs.~\eqref{Eq:vend} and \eqref{eq: rhoend alpha}). Therefore, the PGWs spectrum depends on $n$, $\tre$, and $\alpha$. 
At first, we show the detectable parameter space in the $\tre-n$ plane for two different values of $\alpha=(1,\,10)$, illustrated in Fig.~\ref{fig:modelplot}.  The parameter space from the detection of LISA, ET, DECIGO, and BBO are plotted in the top left, top right, bottom left, and bottom right panels, respectively.  The value of $n$ varies from $2-20$, corresponding to the variation of $\wf$ as $0-9/11$.  The red shaded area indicates the excluded region from BBN temperature that corresponds to $\tre<4$ MeV. As expected, the parameter space resembles the behavior as given in the model-independent analysis (see, for instance, Fig.~\ref{fig:freeplot}). Such as, the probing parameter space for LISA is smaller compared to others, and with increasing energy scale, which is here related to increasing the value of $\alpha$, the probing parameters space got widened, and among the four detectors we consider, BBO can probe a wider range of parameter space. 
\begin{table}[h!]
  \begin{center}
    \begin{tabular}{||c|c|c|c|c|c||}
      \hline
         Detectors &$\alpha$ & LISA & ET & DECIGO & BBO \\
         \hline\hline
         \multirow{2}{*}{$\wf$}&$1$&$[0.51,0.6]$ & $[0.51,0.64]$&$[0.36,0.8]$&$[0.34,0.8]$\\
         \cline{2-6}
         &$10$&$[0.45,0.6]$ & $[0.47,0.64]$&$[0.33,0.8]$&$[0,0.8]$\\
         \hline
         {$n_{\rm s}$}&&$[0.965,0.967]$ & $[0.965,0.968]$&$[0.963,0.968]$&$[0.961,0.968]$\\
         \hline
    \end{tabular}
    \caption{The probing range of $\wf$ for the $\alpha$-attractor model with different future GWs detection is listed here for $\alpha=1$ and $\alpha=10$. 
 We have also listed the range of $\ns$ for each detector that lies in the $2\sigma$ contour of $n_s-r$.}\label{tabl:ns-r}
  \end{center}
\end{table}

As already discussed, Once we fixed the energy scale of the inflation specifying a reheating temperature and the EoS $\wf$ fixes the duration of reheating and the number of e-folds calculated from the end of inflation to the point where the pivot scale exits the Hubble radius. At the pivot scale, $k_*=0.05$ $\rm Mpc^{-1}$ the $n_s-r$ prediction from Planck+BICEP/$Keck$ data \cite{Tristram:2021tvh} projected by yellow contours of Fig.~\ref{fig: ns r}. Considering the  $\alpha$-attractor model of inflation, we have shown the probing region in the $n_s-r$ plane for different GW detectors. Note that to determine the consistent region in the $n_s-r$ plane, we vary the reheating temperature from the maximum, fixed by the instantaneous reheating, to the minimum possible temperature. For a given value of $\alpha$ and $n$, we can calculate the value of $n_s$ and $r$ at the pivot scale using slow roll approximation, which is given in Eq.~\eqref{eq: ns} and Eq.~\eqref{eq: ttos}. We varied the value of $\alpha$ from {$0.1$ to $17$} for LISA and ET and from {$0.1$ to $15$} for DECIGO and BBO, plotted in the color from green to blue. From Fig.~\ref{fig: ns r}, it is clear that for $\alpha> 10$, the $n_s-r$ value goes out of the $1\sigma$ region of the Plank+BICEP/$Keck$ data. To be consistent with the Planck data, we restrict ourselves from choosing the value of $\alpha>10$ in later analysis. The range of $n_s$ that can be probed through the mentioned GW detectors is tabulated in Tab.~\ref{tabl:ns-r}. As expected, BBO provides a wider region in the $n_s-r$ plane compared to others. 

In the next section, we shall consider the various reheating processes and find out the probing region of the parameters associated with the corresponding reheating process, such as the strength of  couplings in non-gravitational reheating, curvature coupling $\xi$ in gravitational reheating, and PBH parameters ($M_{\rm in},\,\beta$) in case of PBH reheating with two sample values of $\alpha=(1,\,10)$. In non-gravitational and gravitational reheating, we only consider the PGWs; however, for PBH reheating with PBH domination, we add induced GWs due to the density fluctuations in PBH together with the PGWs. 
  
\subsection{Detectability of different reheating parameters}

We shall now consider various processes through which reheating takes place. First, we shall discuss the case where the inflaton is coupled to the SM particles through various non-gravitational couplings in the perturbative reheating setup. Then, we discuss the case where the energy transfer takes place through purely gravitational interaction. Finally, we discuss the case where the thermal bath is produced from the evaporation of PBHs, termed PBH reheating.

\subsubsection*{\it Reheating with non-gravitational coupling}
\begin{figure}[t!]
    \centering
    \includegraphics[scale=.428]{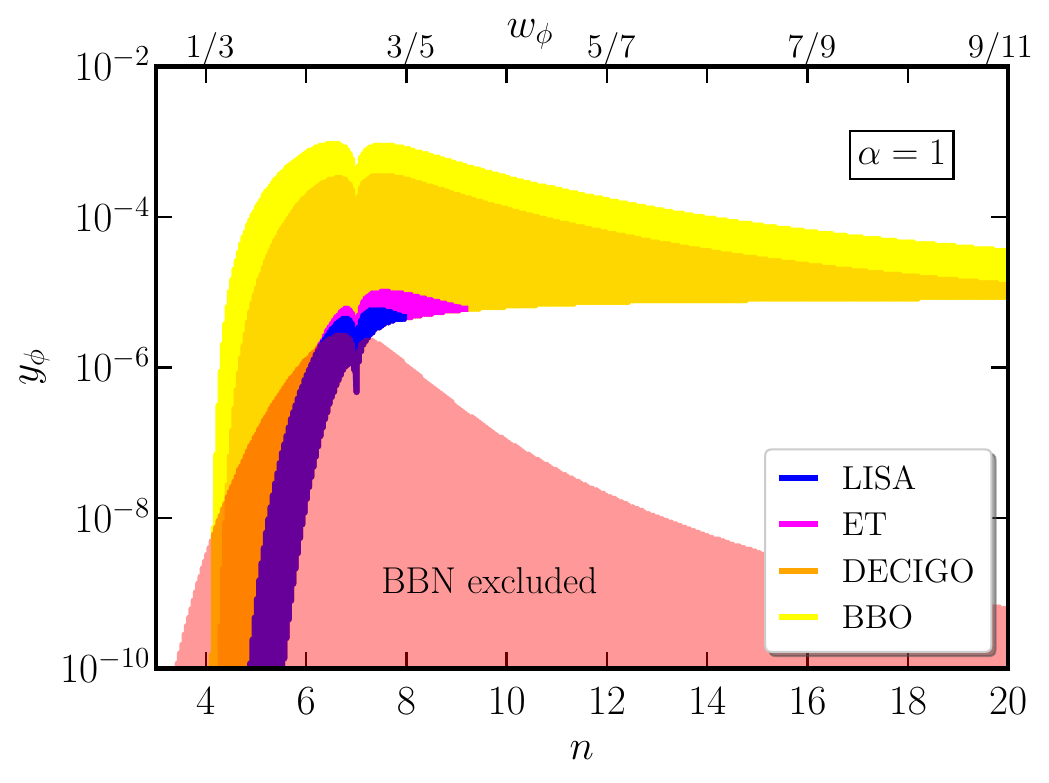}
    \includegraphics[scale=.428]{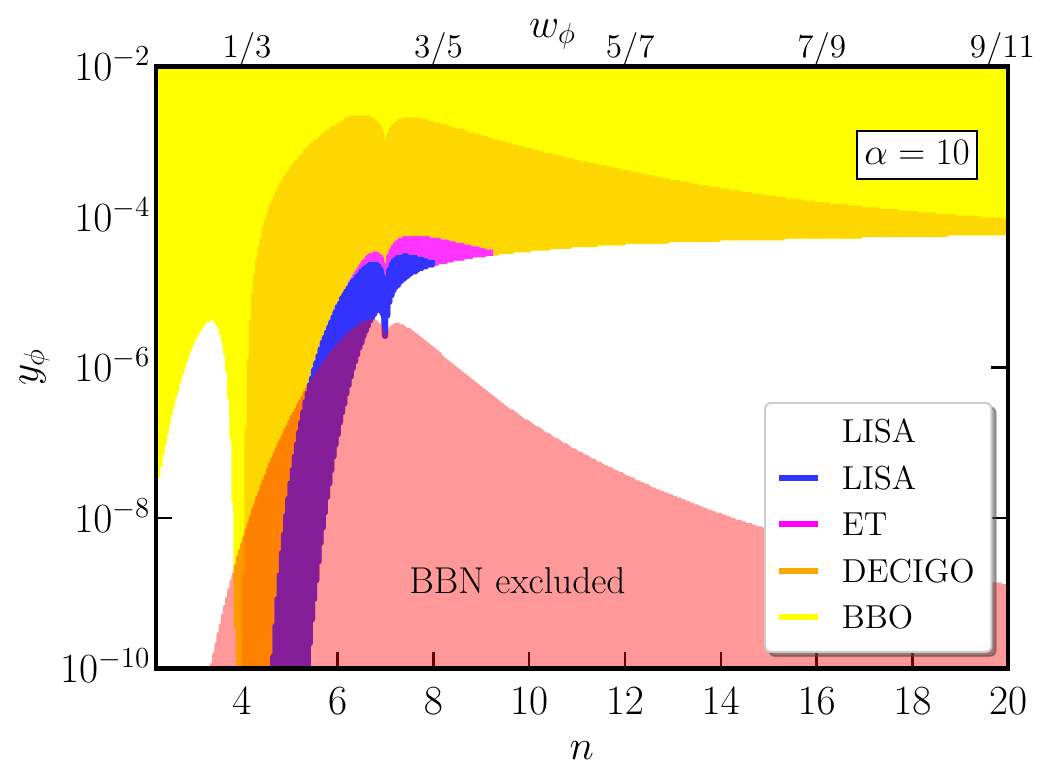}
    \includegraphics[scale=.428]{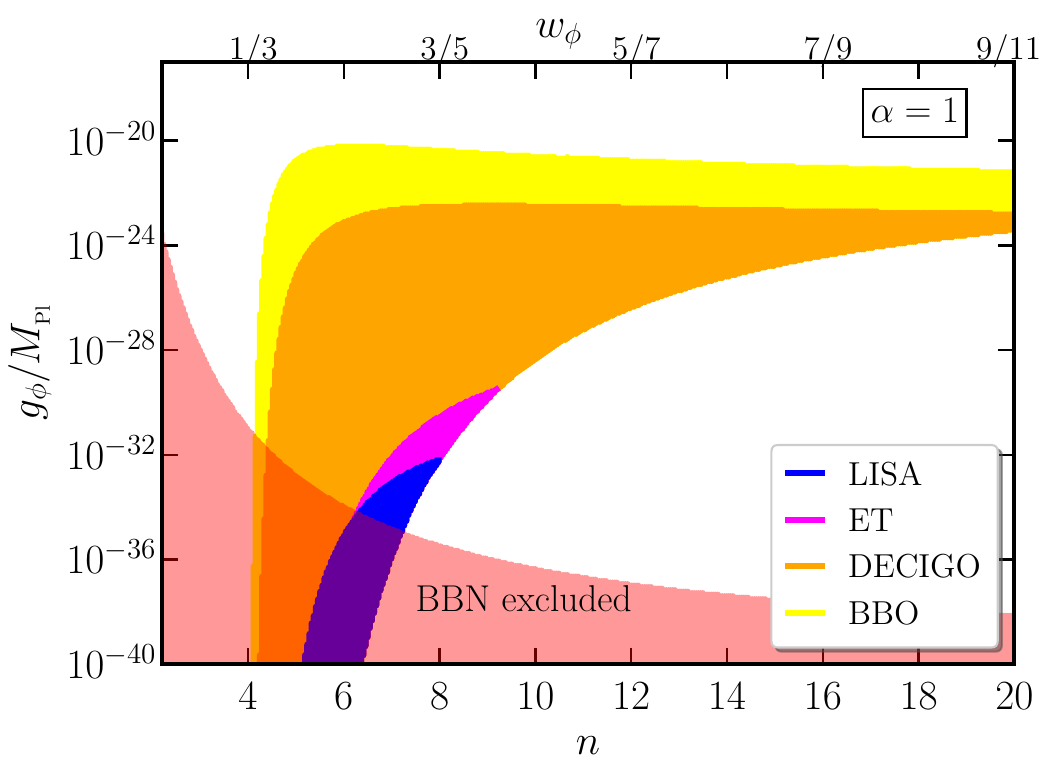}
    \includegraphics[scale=.428]{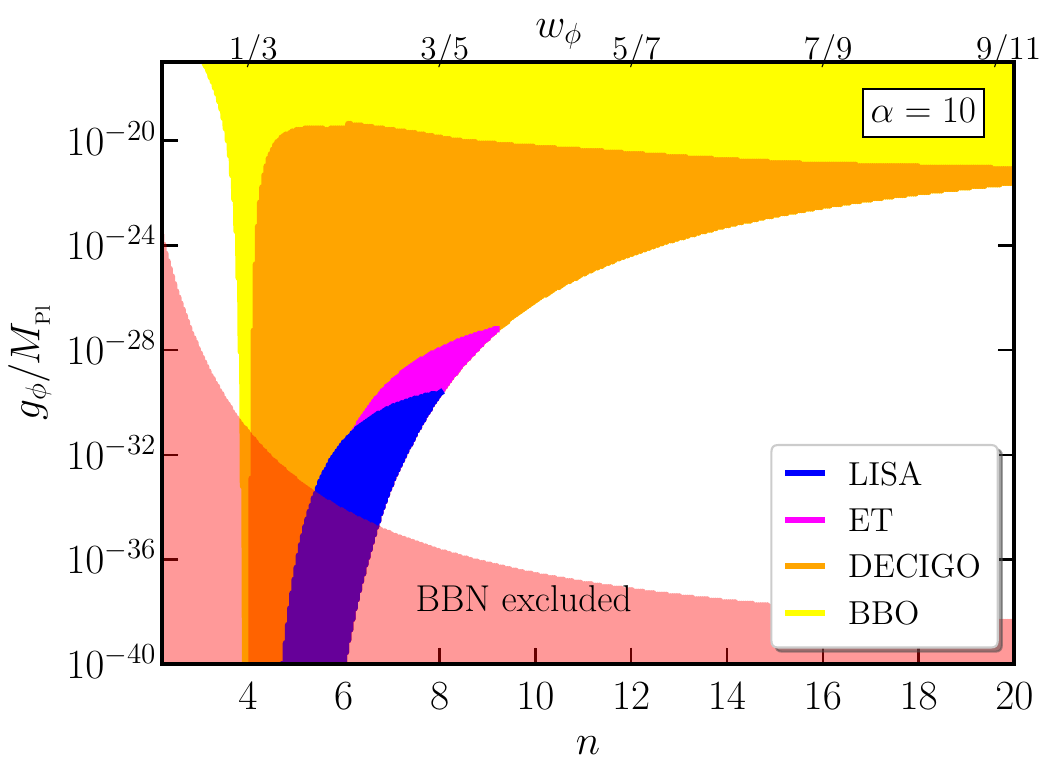}
    \includegraphics[scale=.428]{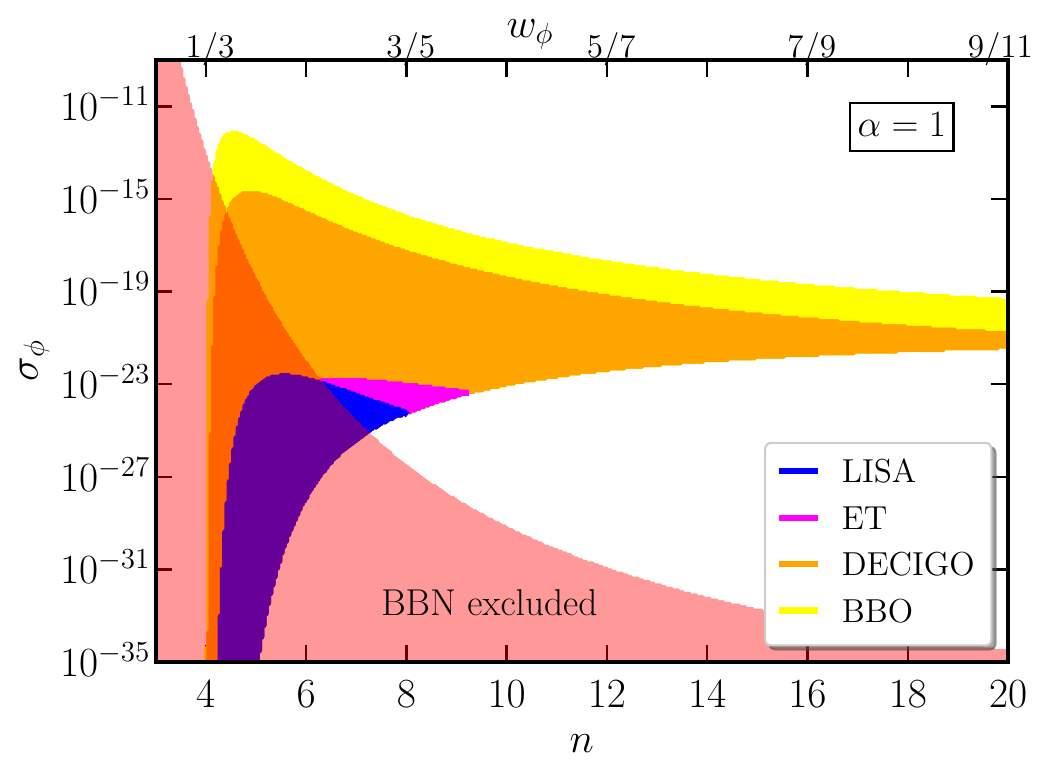}
    \includegraphics[scale=.428]{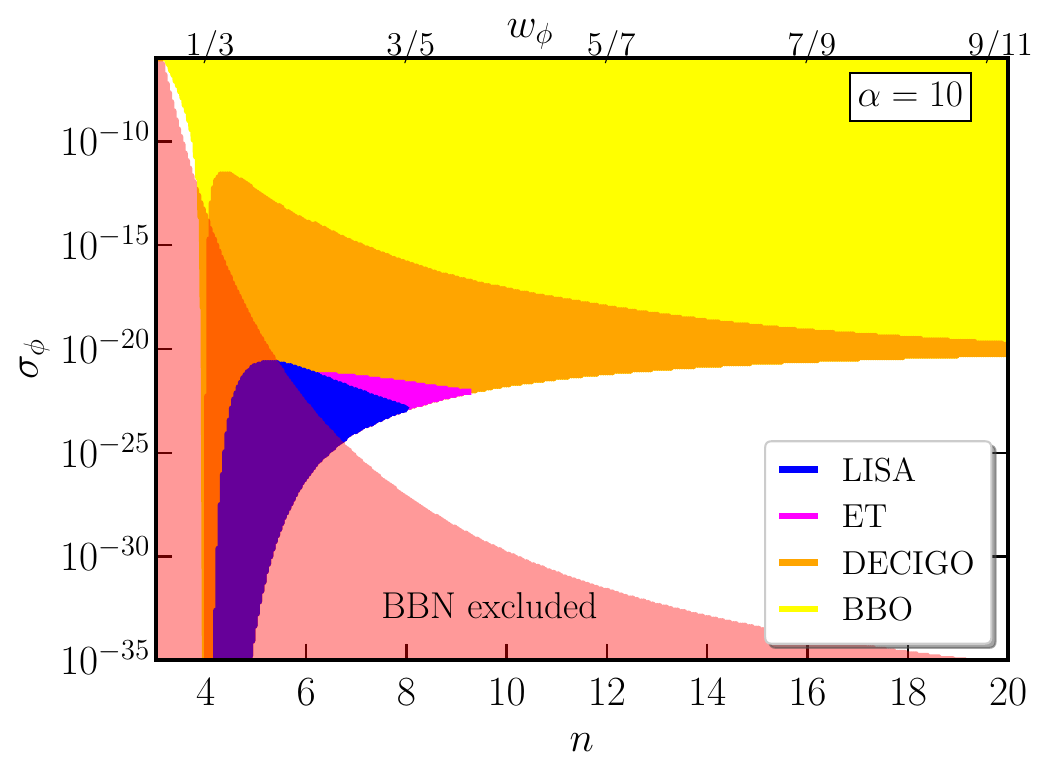}
    \caption{  The possible probing range of different couplings between the inflaton and SM particles are plotted in terms of $n$ corresponding to four GW detectors: LISA, ET, DECIGO, and BBO. The colors blue, magenta, orange, and yellow correspond to the detectors LISA, ET, DECIGO, and BBO, respectively. The couplings $y_\phi$, $g_\phi$ and $\sigma_\phi$ are plotted in the top, middle, and bottom panel respectively. We plotted for $\alpha=1$ and $\alpha=10$ on the left and right panels. The red-shaded regions correspond to reheating temperature $\tre<4$ MeV.  }
    \label{fig:ng-coupling}
\end{figure}

As previously mentioned, we posit three distinct phenomenological decay or annihilation processes when considering non-gravitational interactions. These processes are associated with three types of couplings denoted as $y_\phi$, $g_\phi$, and $\sigma_\phi$, corresponding respectively to decay processes such as $\phi\to \bar f f$, $\phi\to b b$, and $\phi\phi\to  b b$. It is evident from Eqs. \eqref{eq:tre-ng1} and \eqref{eq:tre-ng2} that the magnitude of these couplings can be linked to the reheating temperature and the EoS during reheating, or equivalently, to the values of $n$ in the $\alpha$-attractor model. Furthermore, as previously noted, the value of $\alpha$ and $n$ determines the energy scale of inflation, where a higher value of $\alpha$ signifies a higher energy scale. 
\begin{table}[h!]
  \begin{center}
    \begin{tabular}{||c|c|c|c|c||}
      \hline
        Detectors & $\alpha$ & $y_\phi$ & ${g_\phi}/{\Mpl}$ & $\sigma_\phi$ \\
         \hline \hline
         \multirow{2}{*}{LISA}&$1$&$[10^{-6},6\times 10^{-6}]$ & $[10^{-35},7\times10^{-33}]$&$[6\times 10^{-26},10^{-23}]$\\
         \cline{2-5}
         &$10$ & $[2\times 10^{-6},3\times 10^{-5}]$& $[10^{-35},2\times 10^{-30}]$&$[3\times 10^{-25},10^{-21}]$\\
         \hline
         \multirow{2}{*}{ET}&$1$&$[10^{-6}, 10^{-5}]$ & $[10^{-35},3\times10^{-30}]$&$[6\times 10^{-26},10^{-23}]$\\
         \cline{2-5}
         &$10$ & $[2\times 10^{-6},5\times 10^{-5}]$& $[10^{-35},7\times 10^{-28}]$&$[3\times 10^{-25},4\times10^{-22}]$\\
         \hline
         \multirow{2}{*}{DECIGO}&$1$&$[2\times10^{-8},3\times 10^{-4}]$ & $[10^{-35},10^{-23}]$&$[6\times 10^{-26},10^{-15}]$\\
         \cline{2-5}
         &$10$ & $[7\times 10^{-9},2\times 10^{-3}]$& $[10^{-35},4\times 10^{-20}]$&$[3\times 10^{-25},4\times10^{-12}]$\\
         \hline
         \multirow{2}{*}{BBO}&$1$&$[7\times10^{-9},9\times 10^{-4}]$ & $[10^{-35},7\times10^{-22}]$&$[6\times 10^{-26},10^{-12}]$\\
         \cline{2-5}
         &$10$ & $[2\times 10^{-9},2]$& $[10^{-35},10^{-6}]$&$[3\times 10^{-25},10^{-6}]$\\
         \hline
    \end{tabular}
    \caption{The range of the non-gravitational coupling strength that can be probed by LISA, ET, DECIGO, and
BBO are listed here.}\label{tabl:nongrav}
  \end{center}
\end{table}
In Fig.~\ref{fig:ng-coupling}, we plot the detectable regions of the three couplings attainable through the future forecast of GW detectors such as LISA, ET, DECIGO, and BBO. As expected, even with a higher value of $\alpha$, indicating a high energy scale of the inflation, LISA and ET can detect very small regions in the coupling strength—$ n$ plane, consistent with the detector's forecast. In Tab.~\ref{tabl:nongrav}, we have listed the range of coupling strengths that can be probed with those mentioned detectors. One important point is that except BBO, all other detectors favor reheating with a stiff EoS $\wf>1/3$. As predicted earlier, BBO forecasts a wide range of coupling parameters and reheating equations of state that can be consistent with the observation.
\subsubsection*{{\it Gravitational reheating}}
\begin{figure}
    \centering
    \includegraphics[scale=.37]{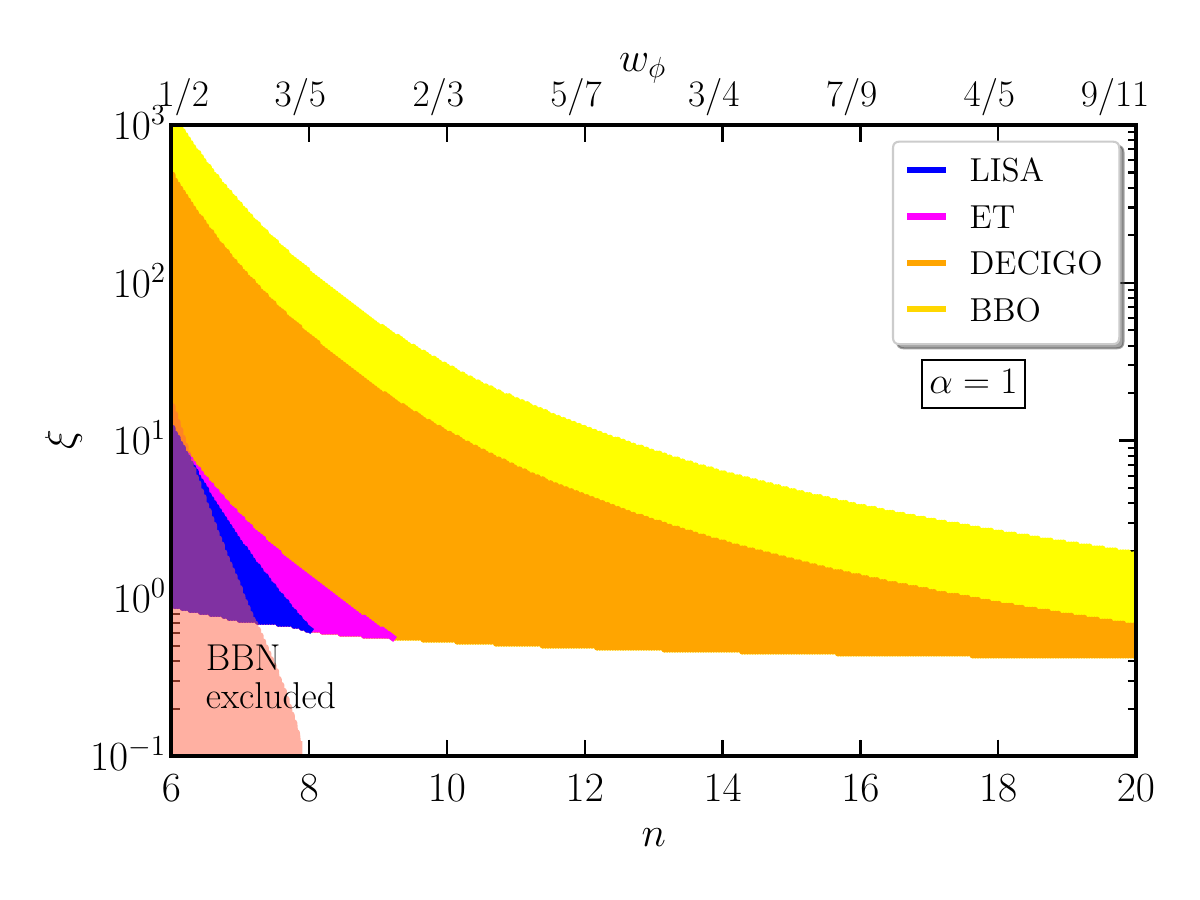}
    \includegraphics[scale=.37]{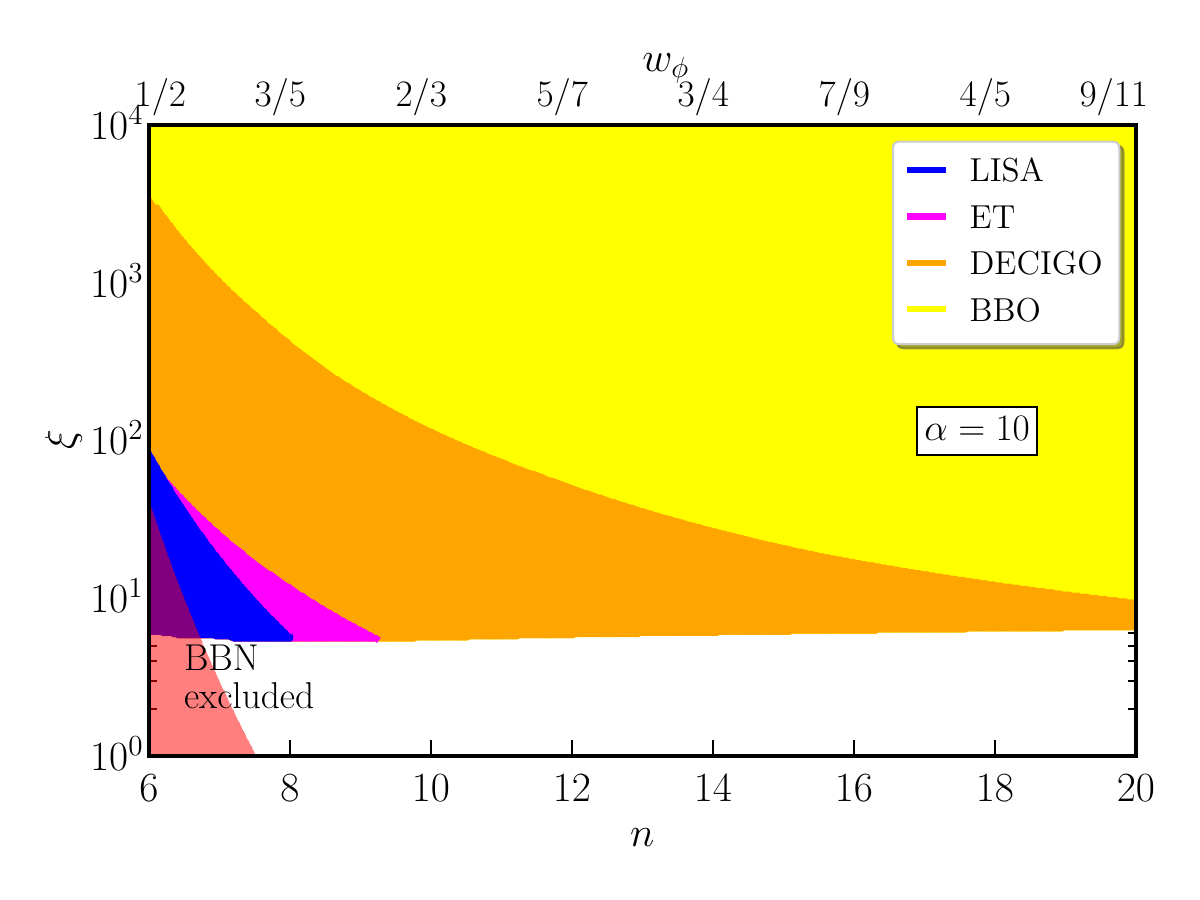}
    \caption{The probable range of couplings between the Higgs and curvature are plotted in terms of $n$ corresponding to different GW observations. The colors blue, magenta, orange, and yellow correspond to the detectors LISA, ET, DECIGO, and BBO. We plotted for $\alpha=1$ and $\alpha=10$ on the left and right panels. The red shaded regions correspond to $\tre<4$ MeV.}
    \label{fig: gravitational reheating}
\end{figure}
\begin{figure}[t!]
    \centering
    \includegraphics[scale=.37]{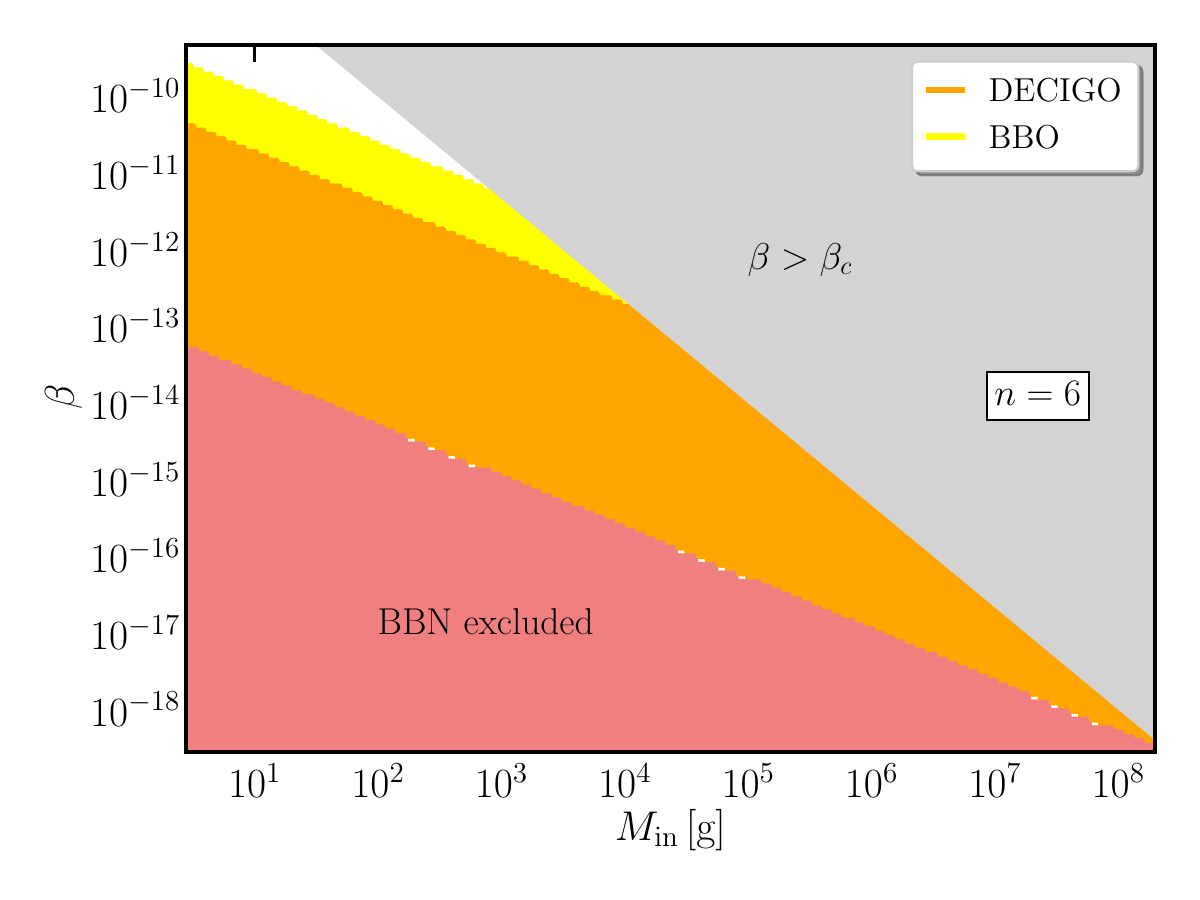}
    \includegraphics[scale=.37]{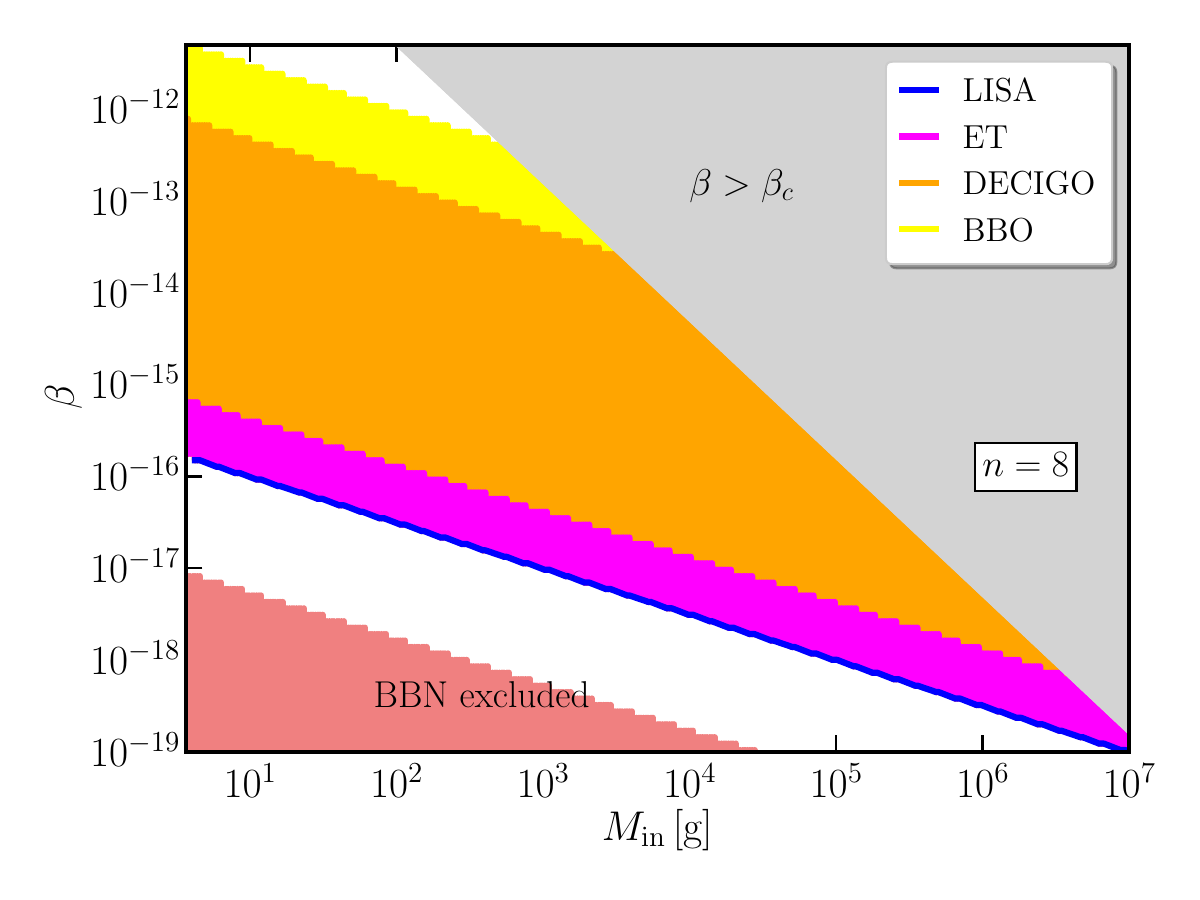}
    \includegraphics[scale=.375]{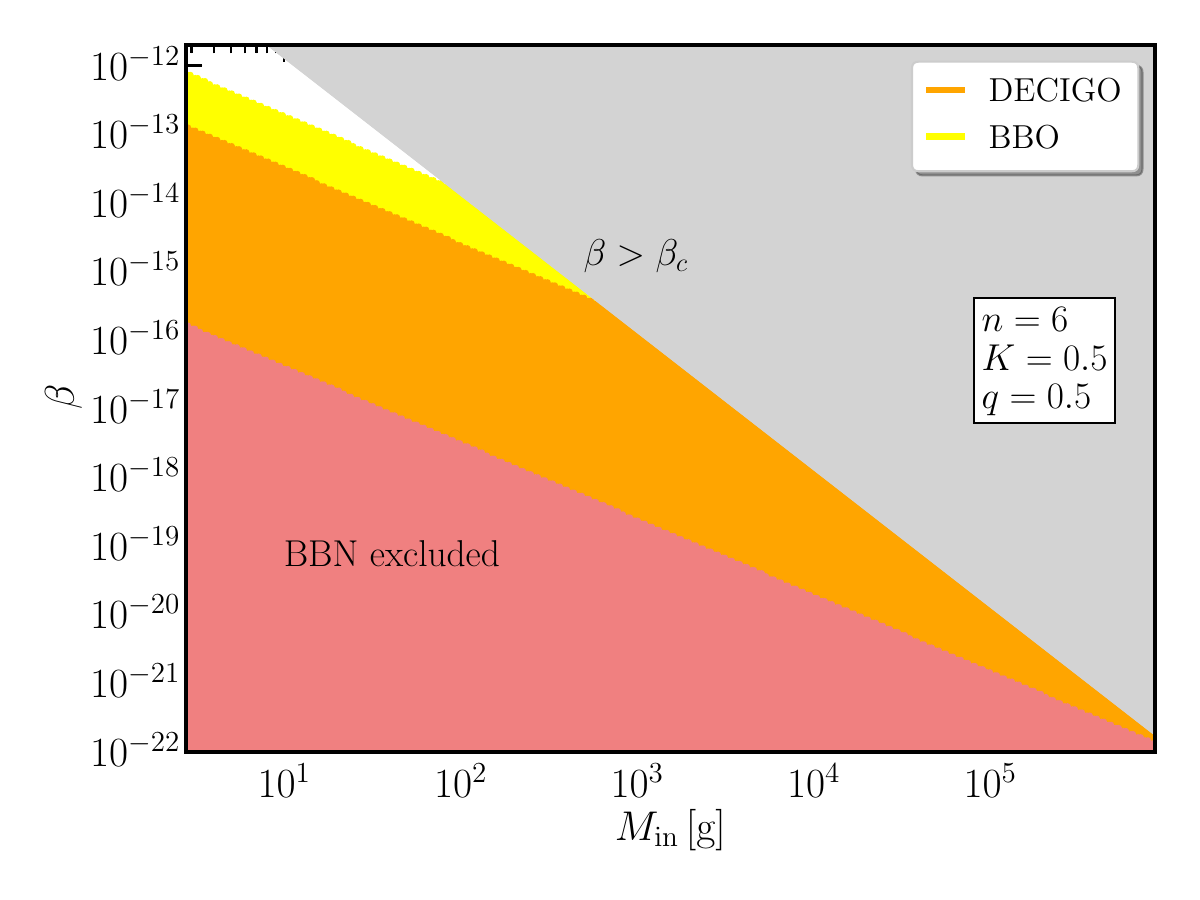}
    \includegraphics[scale=.375]{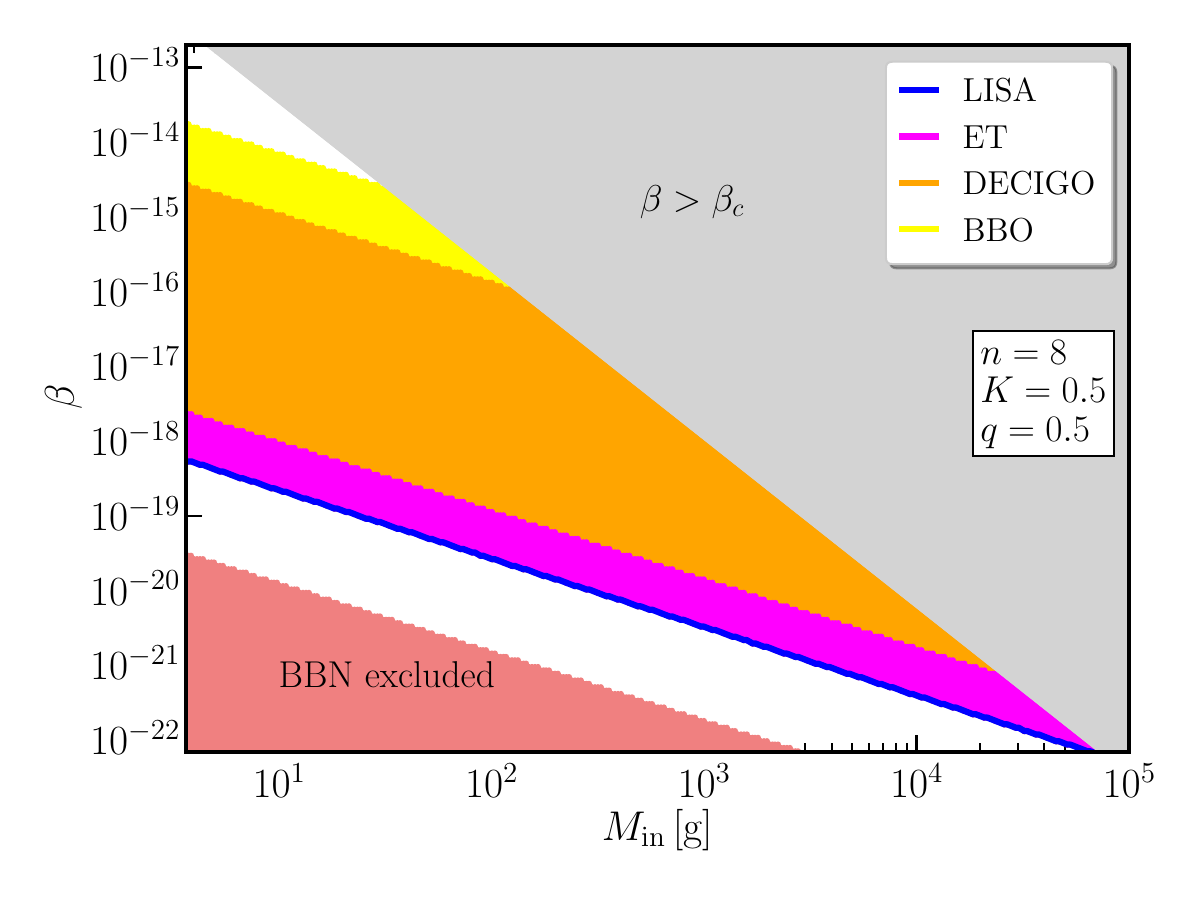}
    \caption{The region that can be probed by different future GW detectors in the \( \beta-M_{\rm in} \) plane is shown for two different sets of values of \( n \) in the case of PBH reheating without domination, i.e., \( \beta < \beta_{\rm c} \). We have fixed \( \alpha = 1 \). The blue, magenta, orange, and yellow regions correspond to the detectors LISA, ET, DECIGO, and BBO, respectively. The top and bottom panels show the results without and with the effects of memory burden, respectively. The cases for \( n = 6 \) and \( n = 8 \) are presented in the left and right panels. The red-shaded regions indicate reheating temperatures \( \tre < 4 \) MeV (a constraint from BBN), while the gray-shaded region corresponds to \( \beta > \beta_{\rm c} \). To compare the results for memory-burdened PBHs with the standard scenario, we have consistently chosen the scenario where \( K = 0.5 \) and \( q = 0.5 \).}
    \label{fig: pbh reheating}
\end{figure}
In the scenario of gravitational reheating, we posit that the transfer of energy from the inflaton to SM particles occurs solely through gravitational interaction. This entails assuming a non-minimal coupling of the Higgs field to curvature, as described by the Lagrangian in Eq.~\eqref{eq:lagrangian_gravitational}. It is worth noting that for coupling strengths $\xi \geq{\cal O}(1)$, the influence of this non-minimal coupling supersedes that of the minimal production mechanism, which is due to the exchange of graviton.
\begin{table}[h!]
  \begin{center}
    \begin{tabular}{||c|c|c|c|c|c||}
      \hline
      Coupling &
         $\alpha$ & LISA & ET & DECIGO & BBO \\
         \hline\hline
         \multirow{2}{*}{$\xi$}&$1$ &$0.6-6$ & $0.52-8$ &$0.48-490$& $0.48-1200$\\
         \cline{2-6}
         &$10$ & $0.16-8$ & $0.16-8$ & $0.16-1500$& $0.16-3\times 10^5$\\
         \hline
    \end{tabular}
    \caption{The range of the curvature coupling strength that can be probed by LISA, ET, DECIGO, and
BBO are listed here. }\label{tabl:grav}
  \end{center}
\end{table}
The strength of the curvature coupling is intricately linked to the reheating temperature and the EoS during reheating, or equivalently, to the values of $n$, as expressed by Eq.~\eqref{eq:tre-grav}. Consistent with expectations, LISA and ET can probe very small regions of the parameter space, even with the higher values of $\alpha$. DECIGO, on the other hand, exhibits a broader exploratory range, while the region probed by BBO is maximal. In Tab.~\ref{tabl:grav}, we provide the range of the curvature coupling strengths that will be consistent with the prediction by the previously mentioned GW detectors. Note that for this gravitational reheating scenario, to get the reheating temperature above the BBN energy scale (roughly $1$ MeV), we have to choose $n\geq 6$. Similarly to the case of non-gravitational coupling, we observe that for BBO with $\alpha=10$, all the allowed coupling ranges are consistent, where the instantaneous reheating and BBN energy scale set the allowed coupling ranges.


\subsubsection*{{\it PBH reheating}}

\begin{figure}[t!]
    \centering
    \includegraphics[scale=.37]{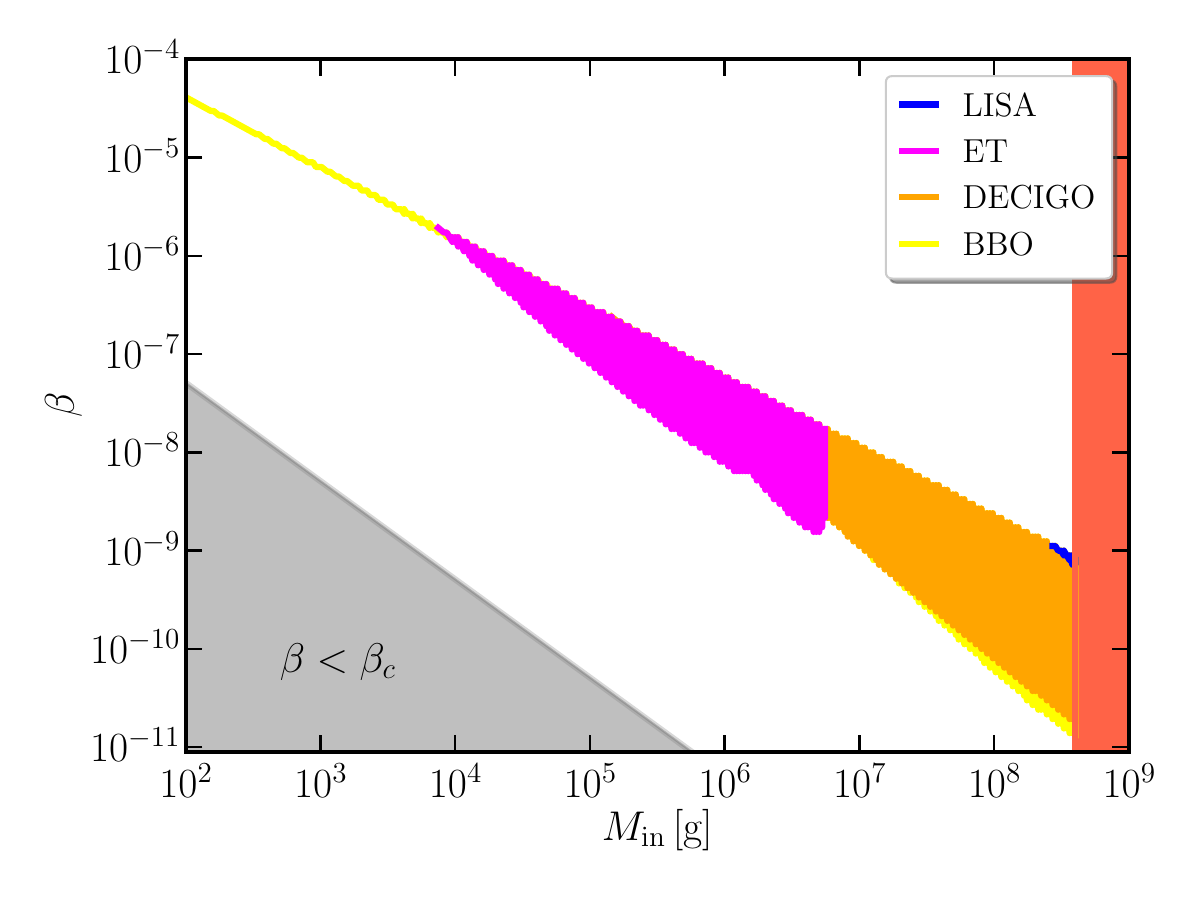}
    \includegraphics[scale=.37]{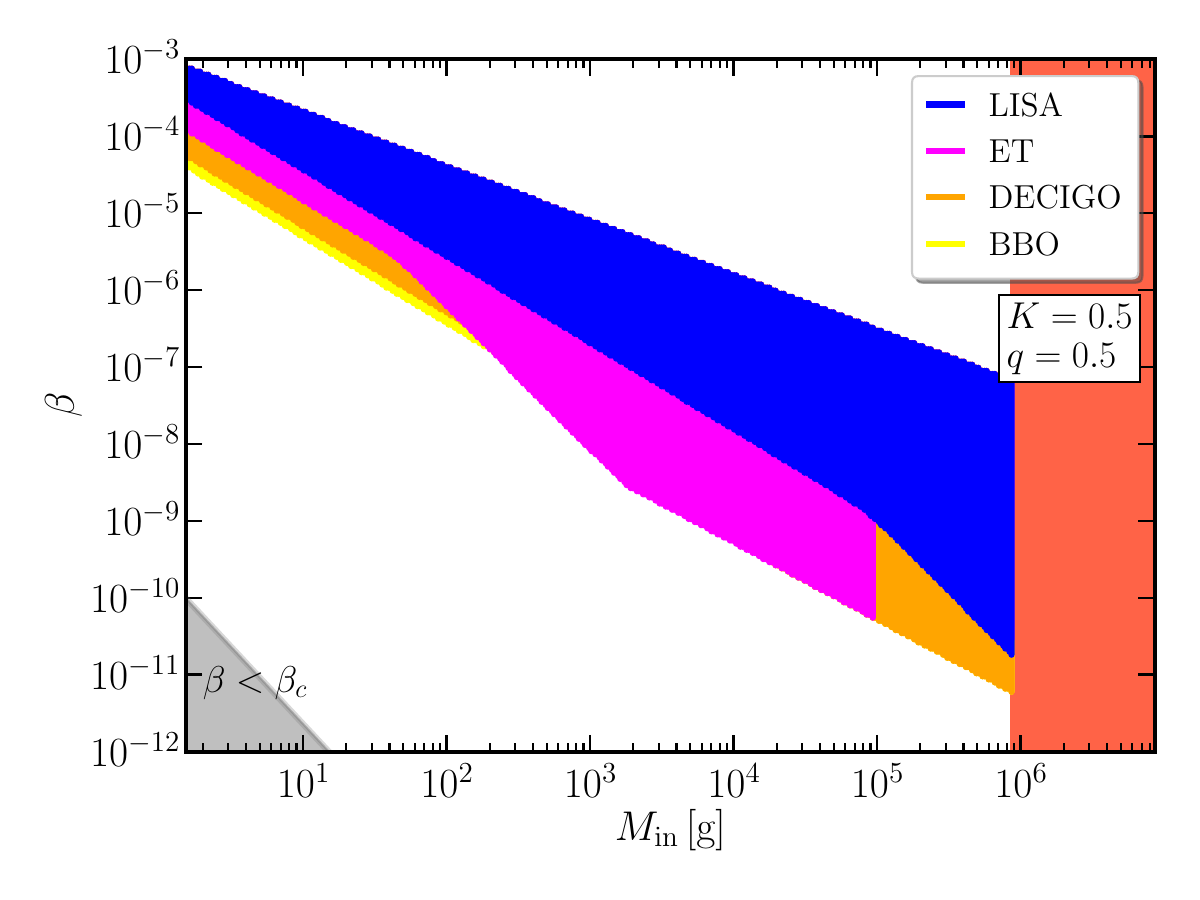}
    \caption{The probing region for different GW detectors in $\beta-\min$ plane are plotted without taking the effects of memory burden (on the left) and with taking the effects of memory burden (on the right). The colors magenta, orange, and yellow correspond to the detectors ET, DECIGO, and BBO, respectively. The grey-shaded region corresponds to the $\beta<\beta_c$, and the red-shaded region indicates the mass region of PBHs that evaporates during or after BBN. Here, we have taken $\alpha=1$ and $n=4$.}
    \label{fig:pbhgwplot}
\end{figure}

 As another possibility, we have discussed that if PBHs are formed at the early reheating stage, these ultralight evaporating PBHs can produce SM particles, which can reheat the universe (see, for instance, Sec.~\ref{subsec: Pbh-reheating}). Recall that we have two distinct scenarios based on the value of~$\beta$. The case with $\beta>\betac$ will eventually lead to an epoch of PBH domination, and for $\beta<\betac$, PBHs will not dominate over the background energy density before it evaporates.
 
{ For $\beta<\betac$} if the background is governed by a steeper EoS than radiation ($\wf>1/3$), the inflaton energy density redshifts faster than radiation, and as a consequence, after a certain time, the radiation energy density starts to dominate over the inflaton, and reheating occurs. Note that here, we must choose a very weak coupling between inflation and SM, such that the production rate is always subdominant compared to the production from PBH. Thus for the case $\beta<\betac$, we shall restrict ourselves in the region $n>4$, and so we take two values of $n$ such as $n=6$ (which correspond to $\wf=1/2$) and $8$ (which correspond to $\wf=3/5$). In Fig.~\ref{fig: pbh reheating}, we have plotted the parameter $\beta$ as a function of the formation mass of PBHs, $M_{\rm in}$ for the range that can be probed with the detectors we considered. Blue, magenta, orange, and yellow correspond to the detectors LISA, ET, DECIGO, and BBO, respectively. 
\begin{table}[h!]
    \centering
    \begin{tabular}{||c|c|c|c|c|c||}
    \hline
    $\beta$ &$\alpha,n$&LISA & ET & DECIGO & BBO\\
    \hline\hline
        \multirow{2}{*}{$<\beta_{\rm c}$} &$1,6$&--&--&$[10^{-18.4},10^{-10.4}]$& $[10^{-18.4},10^{-9.7}]$ \\
         \cline{2-6}
         &$1,8$&$[10^{-19},10^{-16}]$& $[10^{-19},10^{-15.2}]$ & $[10^{-19},10^{-12.2}]$ & $[10^{-19},10^{-11.3}]$ \\
         \hline
        {$>\beta_{\rm c}$} &$1,4$& $[10^{-9.2},10^{-9}]$& $[10^{-8.6},10^{-5.9}]$ &$[ 10^{-10.5},10^{ -7} ]$& $[ 10^{-10.92},10^{-4.2} ]$ \\
         \hline
\multicolumn{6}{||c||}{$k=0.5,~q=0.5$}\\
         \hline
                 \multirow{2}{*}{$<\beta_{\rm c}$} &$1,6$&--&--&$[10^{-21.9},10^{-11.8}]$& $[10^{-21.9},10^{-12.1}]$ \\
         \cline{2-6}
         &$1,8$&$[10^{-22.2},10^{-18.1}]$& $[10^{-22.2},10^{-17.5}]$ & $[10^{-22.2},10^{-14.5}]$ & $[10^{-22.2},10^{-13.7}]$ \\
         \hline
        {$>\beta_{\rm c}$} &$1,4$& $[10^{-10.7},10^{-3.1}]$& $[10^{-10.3},10^{-3.1}]$ &$[ 10^{-11.3},10^{ -3.1} ]$& $[ 10^{-11.3},10^{-3.1} ]$ \\
         \hline
    \end{tabular}
    \caption{The compatible ranges of $\beta$ from the four different GW detector forecasts with the cases $\beta>\betac$ and $\beta<\betac$ are tabulated here.
    }
    \label{tab:my_label}
\end{table}
We see that for $\alpha=1$ and $n=6$, LISA and ET can not probe the PBH reheating scenario for $\beta<\betac$. Even for higher values of  $n=8$, these two detectors can probe very narrow regions of the parameter space. The ranges of $\beta$ that can be probed through the forecast of GW detectors are tabulated in Tab.~\ref{tab:my_label}.

{ For} $\beta>\betac$, there will be an additional contribution to the GWs from the inhomogeneous distribution of PBHs. This induced GW spectrum peaks at a certain frequency $f_{\rm UV}$, as illustrated in Fig.~\ref{fig:igwpbh}. In Fig.~\ref{fig:pbhgwplot}, we have taken the contribution from both the induced GWs and PGWs together and traced the region in $\beta$-$\min$ space that can be probed by the GW detection form LISA, ET, DECIGO, and BBO. 
The PBH parameters that can be consistent with the prediction from the different mentioned GW detectors are shown in Tab.~\ref{tab:my_label}. 
\subsubsection*{\it Effects of memory burden on $\beta-\min$ parameter space:}

Recall that, due to the memory burden, the lifetime of a PBH is extended. The first observable consequence of this burden is that the maximum allowed mass of PBHs that would have evaporated before BBN is lower than the mass threshold for complete evaporation from Hawking radiation alone. Additionally, the critical value of \( \beta \), above which the energy density of PBHs dominates the background energy density, decreases when memory burden is taken into account.

In the bottom plots of Fig.~\ref{fig: pbh reheating}, we show the parameter space of \( \beta - \min \) probed by the observatories for the case \( \beta < \beta_{\rm c} \), incorporating the effects of memory burden. On the other hand, the parameter space of \( \beta - \min \) for \( \beta > \beta_{\rm c} \), after accounting for the memory burden, is illustrated in Fig.~\ref{fig:pbhgwplot}. This parameter space is derived from the peaked spectral energy density resulting from the inhomogeneous distribution of PBHs. As the memory burden parameter increases, the position of the peak shifts slightly to lower frequencies, while the strength of the peak increases significantly, as shown in Fig.~\ref{fig:igwpbhmb}. We find that a higher value of the memory burden parameter leads to a wider range of possible \( \beta \) values. However, as discussed earlier, the maximum allowed value of \( \min \) decreases.

In Tab.~\ref{tab:my_label}, we also present the range of \( \beta \) that can be probed after accounting for the effects of memory burden. For this analysis, we have chosen \( K = 0.5 \) and \( q = 0.5 \). We observe that for \( \beta < \beta_{\rm c} \), the upper limit of \( \beta \) that can be probed has decreased by several orders, ranging from \( 10 \) to \( 10^3 \), while the lower limit has reduced by about \( 10^3 \) orders. In contrast, for \( \beta > \beta_{\rm c} \), the upper limit has increased by several orders, allowing the detectors to probe lower mass PBHs.

\subsubsection*{\it Effects of varying SNR on different couplings and PBH parameters:}
\begin{figure}
    \centering
    \includegraphics[height=3.6cm,width=15.5cm]{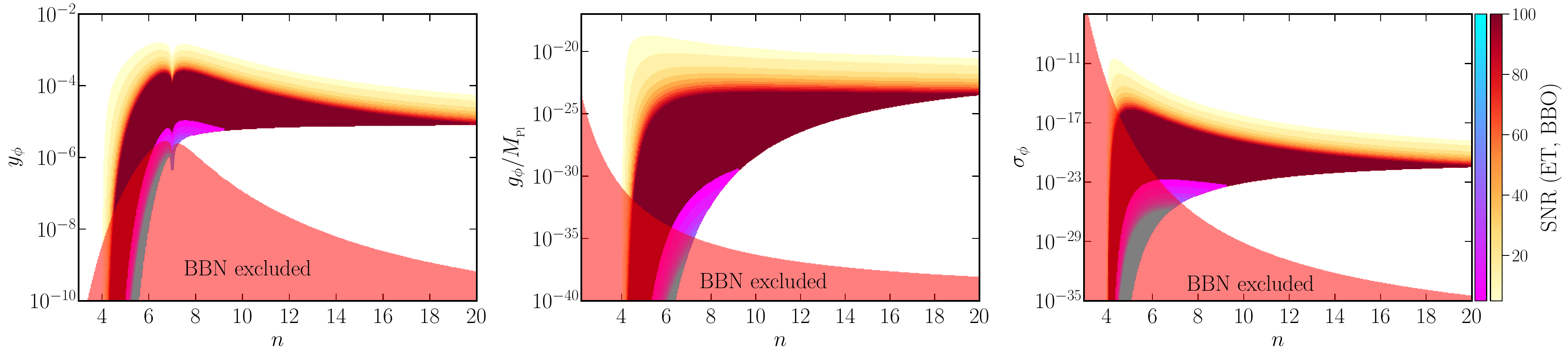}
    \includegraphics[height=3.6cm,width=4.8cm]{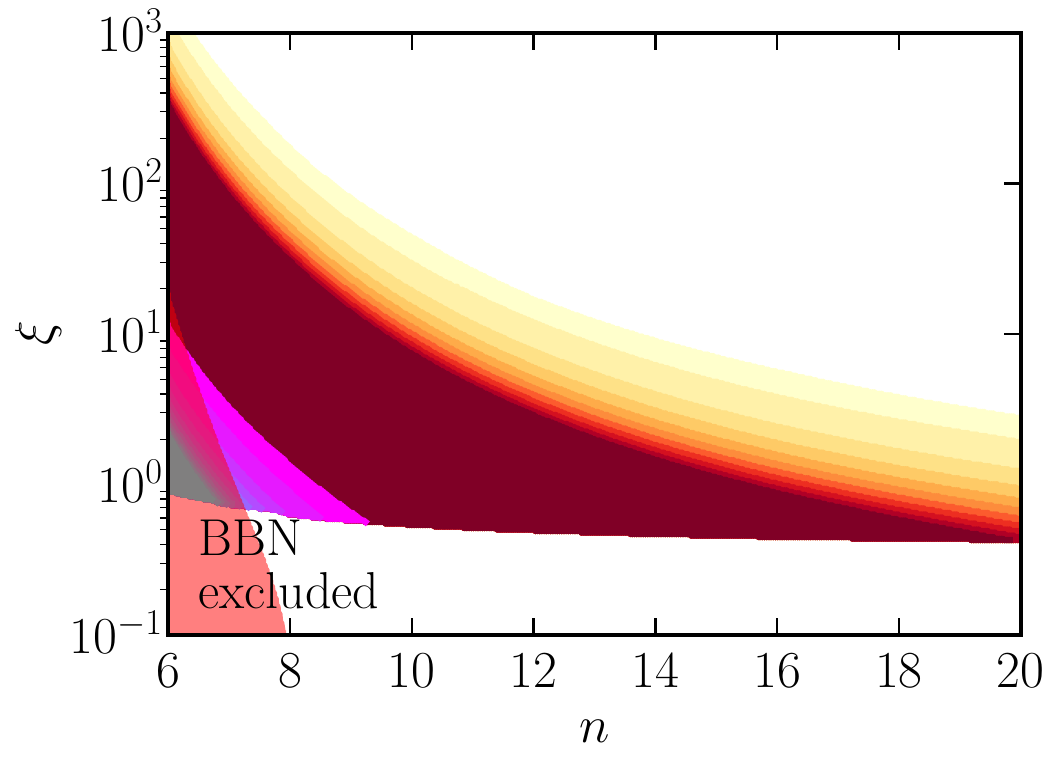}
    \includegraphics[height=3.7cm,width=4.85cm]{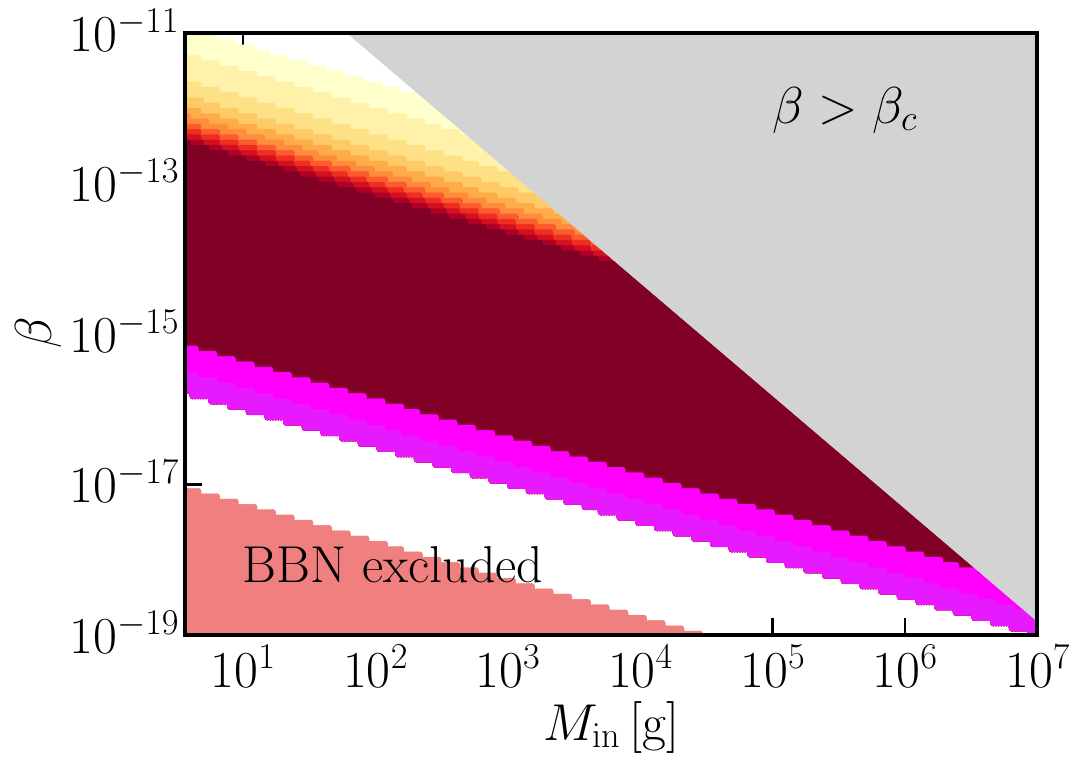}
    \includegraphics[height=3.7cm,width=5.55cm]{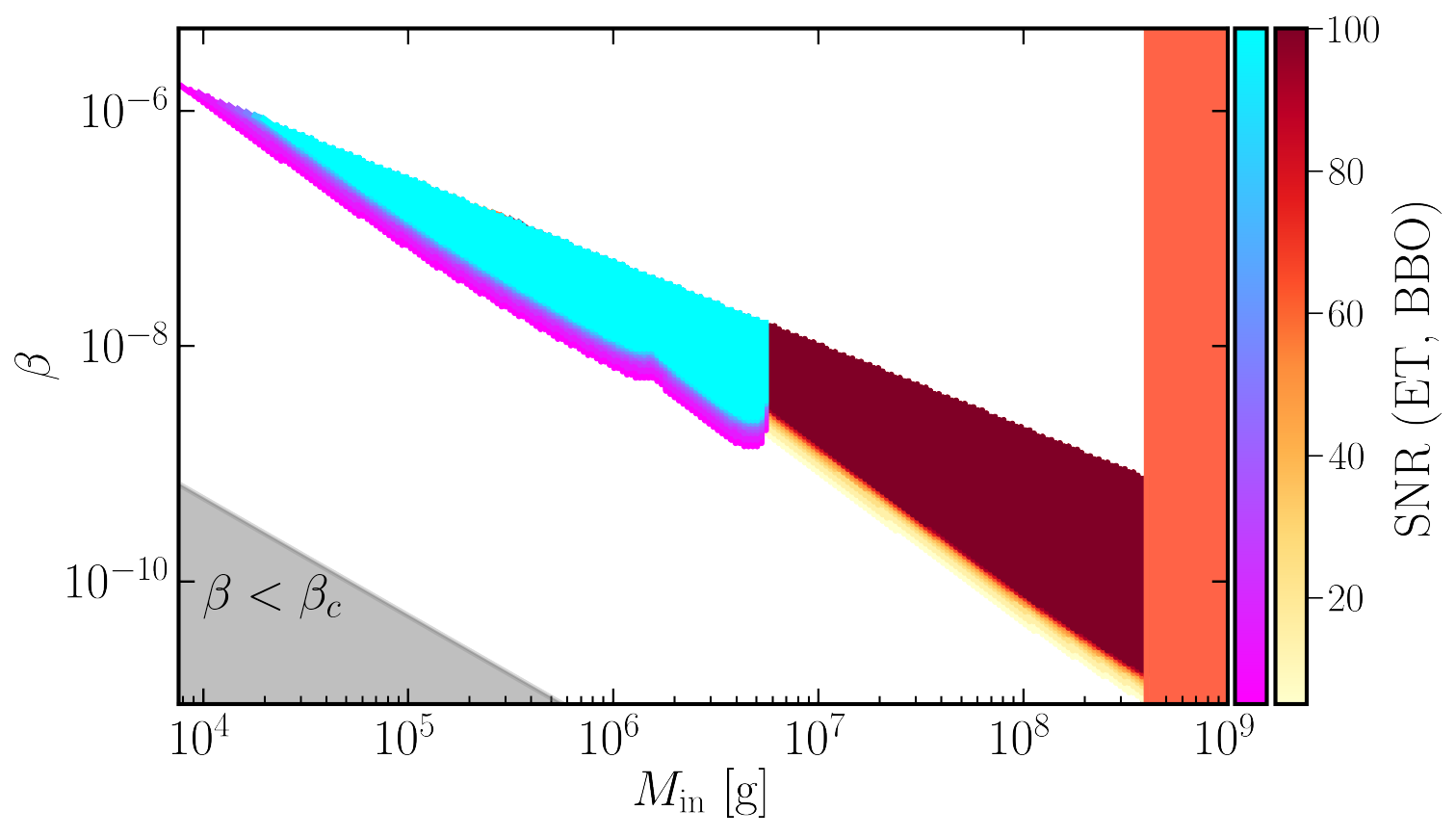}
    \caption{ The effects of changing SNR  on the reheating and PBH parameter space can be probed through the detection of proposed GWs detectors ET and BBO. The color magenta to cyan corresponds to the increase in SNR for ET, whereas yellow to red indicates the same for BBO. We have fixed $\alpha=1$. The top panel corresponds to the non-gravitational coupling, the bottom left panel corresponds to gravitational curvature coupling, and the bottom middle and bottom right panel correspond to PBH reheating. }
    \label{fig: SNR coupling}
\end{figure}
Recall that while to make use of Eq.~\eqref{eq: snr} to calculate the detectability of a GW signal, we need to set the threshold value of $\rho$ above which the signal is considered to be detected.  
In Fig.~\ref{fig: SNR coupling}, we have plotted the effects of changing SNR value on the parameters related different reheating couplings and PBH parameters such as $\min,\,\beta$ those are consistent with the detection of GWs by ET and BBO. We have varied the SNR, i.e., $\rho/\rho_{\rm th}$ from $1$ to $100$, which is denoted by the color bar on the right of Fig.~\ref{fig: SNR coupling}. We set $\alpha=1$, to do this analysis. Note that increasing the value of SNR will eventually reduce the parameter space that can be probed in each of the cases. For ET, an SNR $>50$ will omit the possibility of detecting the signal in any of the cases discussed above. 
In the next section, we shall discuss the range of DM mass compatible with the GWs observation where the DM production takes place through the gravitational interaction, which we have mentioned in Sec.~\ref{sec: DM production}. 

\subsection{Detectability of the DM mass }


\begin{figure}
    \centering                                
    \includegraphics[height=3.9cm,width=5cm]{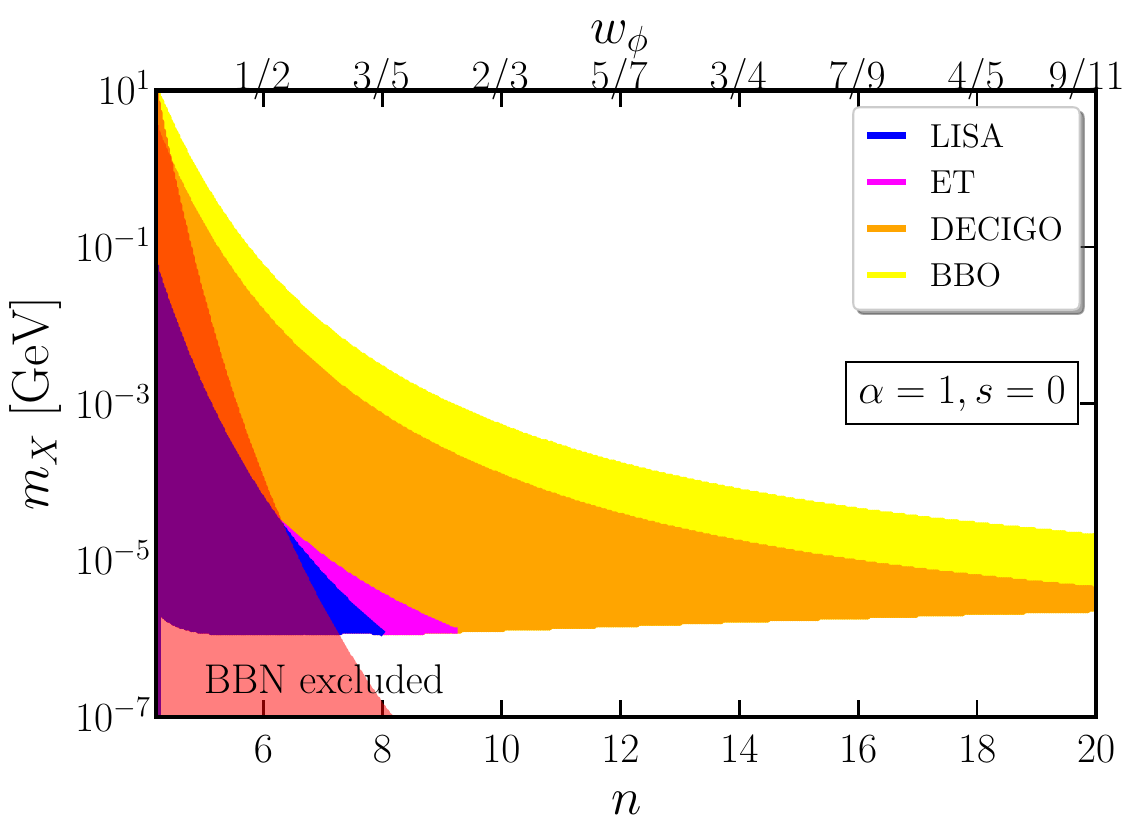}
    \includegraphics[height=3.9cm,width=5cm]{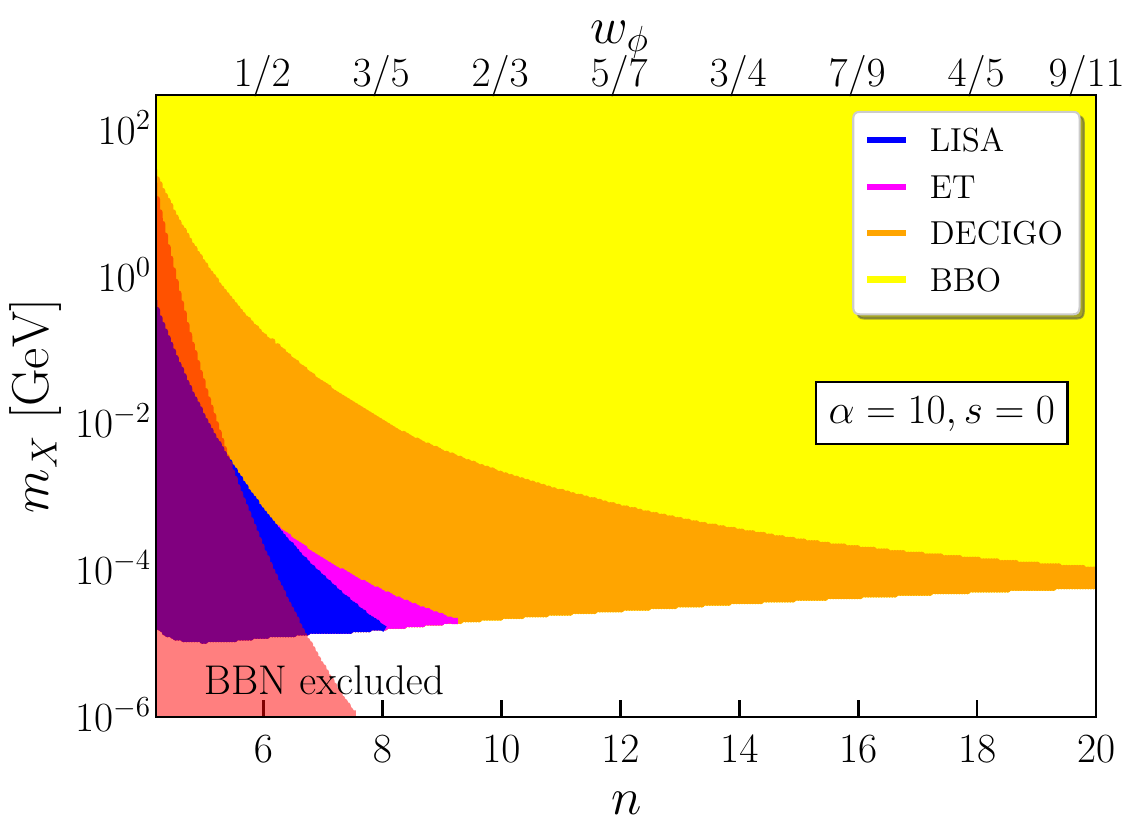}
    \includegraphics[height=3.6cm,width=5cm]{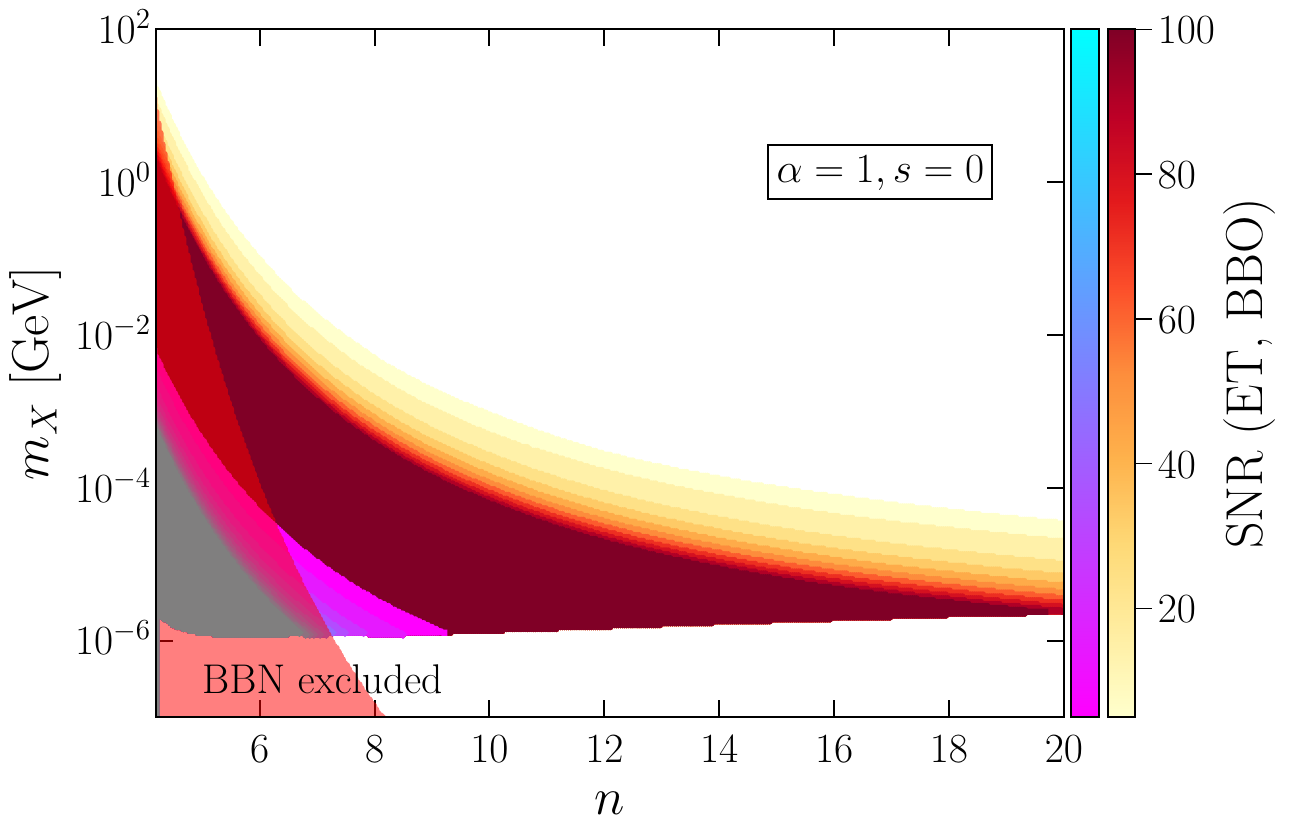}
    \includegraphics[height=3.9cm,width=5cm]{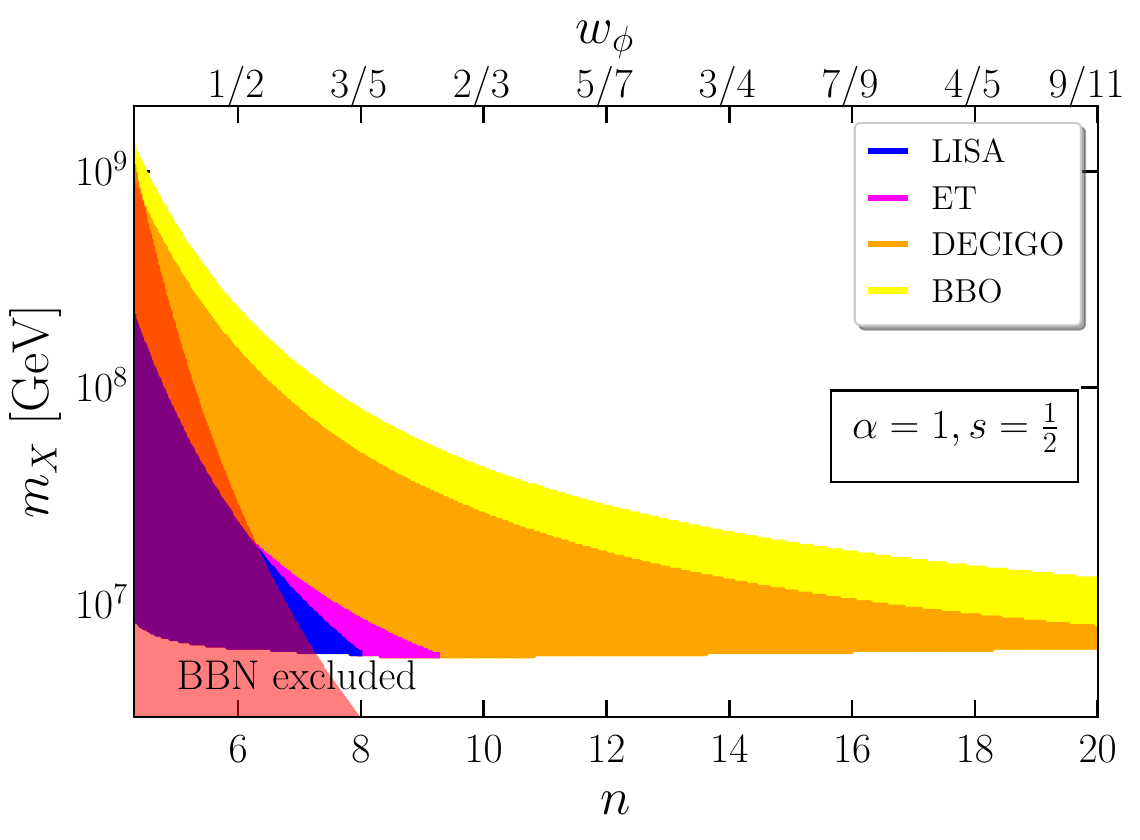}
    \includegraphics[height=3.9cm,width=5cm]{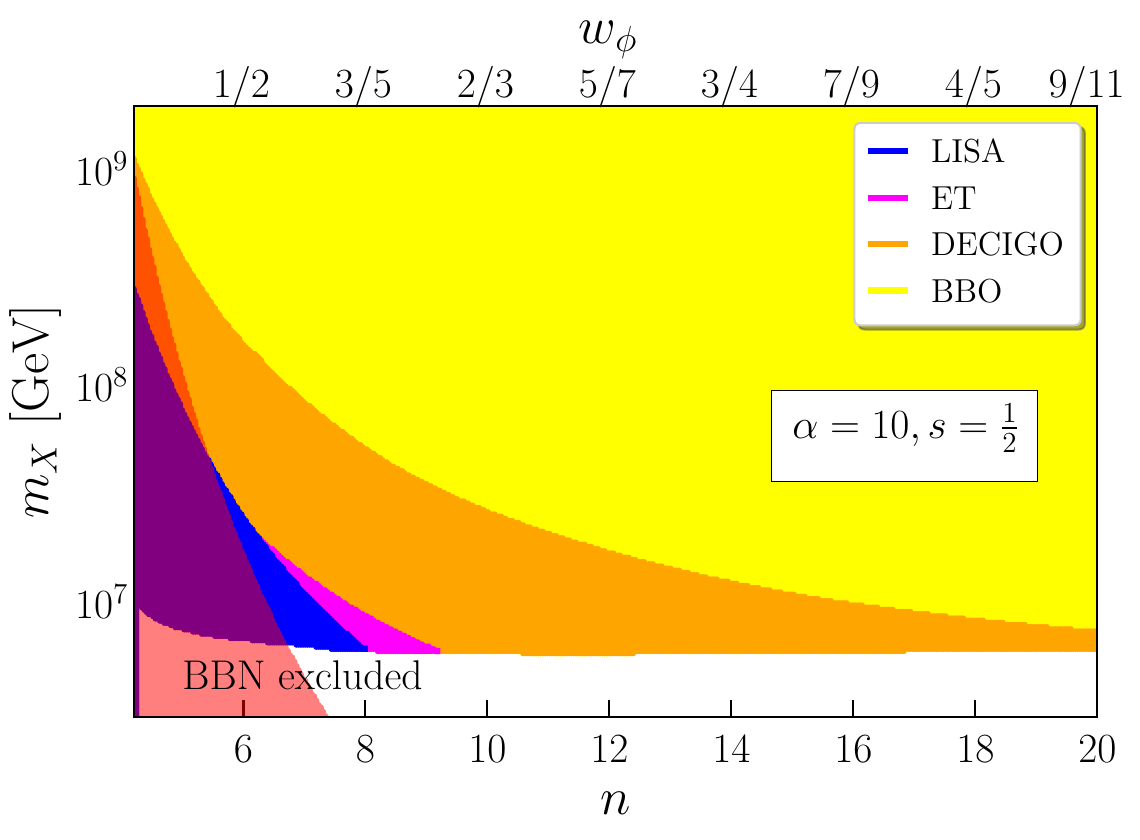}
    \includegraphics[height=3.6cm,width=5.2cm]{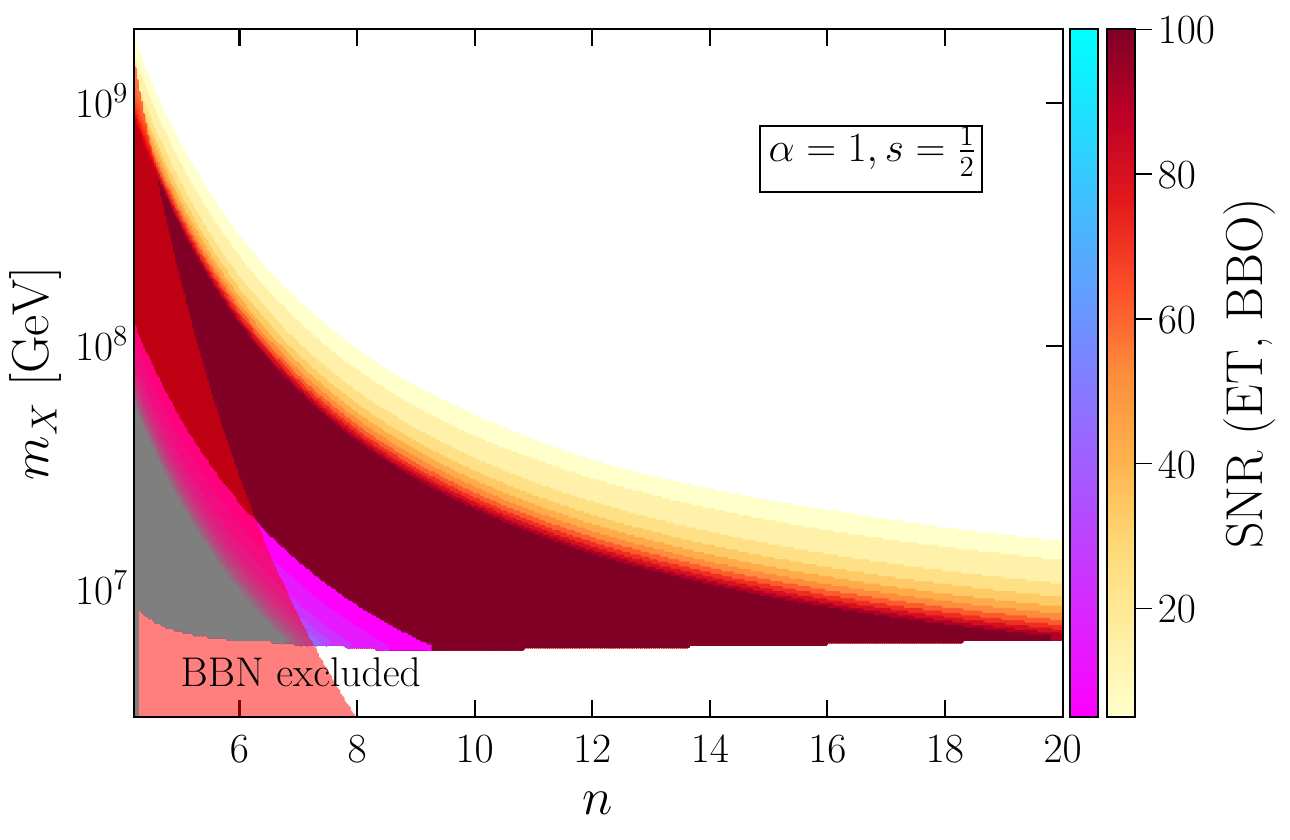}
    \caption{ The range of gravitationally produced DM mass $m_X$ that can be explored from the detection of PGWs considering four different detectors is plotted as a function of $n$. The colors blue, magenta, orange, and yellow correspond to the detectors LISA, ET, DECIGO, and BBO. On the top and bottom panels, we have plotted for scalar DM and fermionic DM. The left plots correspond to $\alpha=1$, the middle plots correspond to $\alpha=10$, and the right panel shows the effect of SNR for both scalar and fermionic DM with $\alpha=1$.}
    \label{fig: SNR DM}
\end{figure}

\vskip 5pt \noindent
Let us estimate the DM mass by detecting PGWs where DM is generated from the scattering of inflation through the exchange of graviton.
Following Eqs.~\eqref{Eq: DM-scalar}, \eqref{Eq: DM-fermion} and \eqref{Eq:omegah2m} in Sec.~\ref{sec: DM production} we can calculate the DM mass, for a given $n$, $\alpha$ and $\tre$, where the energy density $\rho_{\rm end}$ can be written in terms of $n$ and $\alpha$. Considering only the gravitational production of the DM, our analysis focuses on the DM mass that satisfies the present DM relic density.
\begin{table}[h!]
  \begin{center}
    \begin{tabular}{||c|c|c|c|c|c||}
      \hline
        DM type & Strength & LISA & ET & DECIGO & BBO \\
         \hline \hline
         \multirow{2}{*}{Scalar (GeV)}&
         $\alpha=1$ & $[10^{-6},10^{-5}]$& $[10^{-6},10^{-5}]$&$[10^{-6},1]$ & $[10^{-6},10]$\\
         \cline{2-6}
         &$\alpha=10$ & $[10^{-4.8},10^{-2.5}]$& $[10^{-4.8},10^{-3}]$& $[10^{-4.8},25]$ & $[10^{-4.8},200]$\\
         \hline
         \multirow{2}{*}{Fermion (GeV)}&
         $\alpha=1$ & $[10^{6.7},10^{7.2}]$ & $[10^{6.7},10^{7.2}]$&  $[10^{6.7},10^{8.8}]$ & $[10^{6.7},10^{9.2}]$\\
         \cline{2-6}
         &$\alpha=10$ & $[10^{6.7},10^{7.7}]$ & $[10^{6.7},10^{7.4}]$&  $[10^{6.7},10^{9.1}]$ & $[10^{6.7},10^{9.2}]$\\
         \hline
    \end{tabular}
    \caption{The range of the DM mass (in GeV) that is produced via gravitational interaction and which can be probed by LISA, ET, DECIGO, and
BBO are listed here.}\label{tabl:dmmass}
  \end{center}
\end{table}
In Fig.~\ref{fig: SNR DM}, we have plotted the mass ranges of DM, consistent with the detection of GWs for four GW detectors LISA, ET, DECIGO, and BBO. We projected our result in the $m_X-n$ plane.
Note that for scalar DM, considering the detection from BBO and $\alpha=1$, the consistent mass range lies within $(10^{-6}-10~{\rm GeV})$ that is much lower than the mass range of fermionic DM $(10^{7}-10^9~{\rm GeV})$. As in the other cases, we see that LISA and ET can probe a minimal range of the parameter space, whereas the region probed by BBO is the maximum. Also, for changing the value of $\alpha$ from $1$ to $10$, only BBO would show a significant change in the parameter space, whereas all the other detectors show mild dependency on $\alpha$. For the detailed range of DM mass that can be probed through the mentioned GW detectors with the variation of $\alpha$  tabulated in Tab.~\ref{tabl:dmmass}.
On the right panels in Fig.~\ref{fig: SNR DM}, we have also shown the effect of changing SNR for both scalar and fermionic DM with $\alpha=1$ considering the detection from ET and BBO. We notice that the increases in the value of SNR reduce the mass ranges for both scalar and fermionic DM, and more importantly, for SNR $>50$, ET can not probe any DM mass from the PGW detection.


\section{Conclusions \label{sec: conclusions}}
GWs offer a unique opportunity to examine the dynamics of our cosmos in its very early phases, which appears to be challenging to observe through other known ways. However, detecting GWs can be difficult due to the extremely weak nature of gravitational force. The recent detection of stochastic GWs by PTA collaborations has heightened anticipation for the potential discovery of GWs~\cite{NANOGrav:2023gor, NANOGrav:2023hde,Antoniadis:2023lym,EPTA:2023fyk,Zic:2023gta, Reardon:2023gzh} at nano Hz frequency originating from diverse sources. However, the prospect of detecting PGWs remains largely unexplored. Directly detecting PGWs within the frequency range of PTA poses significant challenges and requires a blue-tilted tensor power spectrum. Generating this kind of blue-tilted spectrum is not possible for most of the standard models of inflation as they predict scale invariant tensor-power spectrum. This kind of blue-tilted spectrum can be generated due to a non-vacuum initial state~\cite{Wang:2014kqa} and violation of the null energy condition (NEC) during inflation~\cite{Tahara:2020fmn,Vagnozzi:2023lwo, Cai:2022lec,Cai:2020qpu,Cai:2023uhc}. However, in this study, our focus was on a standard scenario where the GW spectrum generated during inflation maintains a scale-invariant tensor power spectrum, and the tilt in the GW spectrum comes from the post-inflationary dynamics due to inflaton domination is named the reheating phase. In order to test different reheating strategies, such as non-gravitational, gravitational reheating, and reheating by PBH evaporation, we aim to investigate inflationary primordial GW spectral shapes at interferometer-based future GW missions such as LISA, ET, DECIGO, and BBO. Throughout our analyses, we enforced a minimum reheating temperature of $4$ MeV and employed the SNR approach outlined in Eq.~\eqref{eq: snr} to assess the detectability of the signal. Additionally, we incorporated constraints from the $\Delta N_{\rm eff}$ on the intensity of the GW spectrum to eliminate certain parameter regions resulting from the spectrum's detection. 
Initially, we investigated a general reheating scenario, wherein we mapped out the parameter space of $\tre$ and $\wf$ based on the detection of PGWs through different GW detectors. This was followed by examining the same, considering a specific inflationary model, the so-called $\alpha$-attractor model with different values of model parameter $\alpha$. Subsequently, we delved into various reheating mechanisms, encompassing both non-gravitational and gravitational couplings, aiming to outline the parameter space of each coupling compatible with GW detection. Additionally, we considered a reheating scenario involving the production of the SM particles through the evaporation of PBHs, wherein we mapped out the parameter space of $\beta$ and $\min$.
Lastly, we deal with DM production via gravitational coupling and showcase the range of DM masses that can be detectable through future GW missions, where the generated DM contributes to the total DM relic density. Let us summarize our main findings in a few points :
\begin{itemize}
    \item 
In each scenario plotted, we have utilized distinct color schemes, assigning blue, magenta, orange, and yellow to denote detections via LISA, ET, DECIGO, and BBO, respectively. The regions shaded in red across all figures signify excluded areas from the BBN bound on the reheating temperature where $\tre <4$ MeV. Strikingly, LISA consistently delineates the strongest prediction among the other mentioned detectors on the reheating parameter space, such as $\tre$ and $\wf$. As a result, there is a very narrow band of non-gravitational, gravitational coupling, and the range of $\beta$ is consistent with the future forecast of LISA. Contrary to that, BBO provides a loose prediction and makes a wider range of parameters consistent, which is also evident in Figs.~\ref{fig:modelplot}-
\ref{fig:pbhgwplot}. The detailed range of parameters consistent with the forecast of different GW detectors is shown in Tabs.~\ref{tabl:ns-r}- 
\ref{tab:my_label}. 
\item 
Notably, within a reheating framework involving non-gravitational and gravitational interaction, we find that LISA, ET, and DECIGO necessitate $n>4$ or $\wf>1/3$ for effective detection. At the same time, BBO can probe instances with $n<4$, particularly under higher $\alpha$ or sufficiently high inflationary energy scale. In the PBH reheating scenario with both cases, PBH domination $\beta>\betac$ and without PBH domination $\beta<\betac$, getting a detection from the forecast of LISA is very difficult. 
For $\beta<\betac$ with $\alpha=1$ and $n=8$, we get a very narrow parameter space in $\beta-\min$ plane. 
\item
If we take into account the effects due to memory burden, the maximum allowed value for $\min$ decreases. We got that for $\beta<\betac$, the parameter space of $\beta-\min$ is shifted to a lower value in both $\beta$ and $\min$. 
For \( \beta > \beta_{\rm c} \), the contribution arises from the peak of the spectral energy density of GWs generated by the inhomogeneous distribution of PBHs. When accounting for the memory burden, we observe that although the maximum allowed value of \( \min \) decreases, the detectors can now probe a broader range of \( \beta \) values and are capable of detecting very small mass PBHs.

\item
Furthermore, we extend our analysis to the parameter space of DM concerning gravitational production. Most importantly, GW missions like LISA and ET predict a minimal range of fermionic and scalar DM mass. As an example for fermionic DM, forecast from both LISA and ET  indicates mass range $(7\times10^6,\,2\times 10^7)$ GeV for $\alpha=1$, whereas increasing $\alpha$ value to $\alpha=10$ the upper limit sifted to $10^{8}$ GeV. The range of DM mass can be probed through the mentioned four GW detectors tabulated in Tab.~\ref{tabl:dmmass}. We have also delved into the effects of SNR across all scenarios. Higher SNR values correlate with reduced parameter space, indicating enhanced precision in detection. Notably, for SNR values above 50, ET and LISA exhibit limitations in probing the reheating phase.
\end{itemize}
Our analysis provides a comprehensive overview of the potential parameter ranges related to inflation, reheating, PBHs, and DM that future GW missions such as LISA, ET, DECIGO, and BBO could explore. In our analysis, we consider the intensity of the GW spectrum. It is important to note that the actual detection of GWs will involve many additional characteristics, including spectral tilt, polarization, and potential additional features. Furthermore, detection through multiple detectors will afford us a more nuanced understanding of the spectrum across various scales. These additional characteristics will yield valuable insights into the dynamics of the phase of reheating.

\black

\section*{Acknowledgements}
The authors wish to thank L. Sriramkumar for useful comments and suggestions. 
  SM wishes to thank the Indian Institute of Technology (IIT) Madras, Chennai, India, for support through the Exploratory Research Project RF22230527PHRFER008479.
  MRH wishes to acknowledge support from the Science and Engineering Research 
Board (SERB),  Department of Science and Technology (DST), Government of 
India (GoI), through the National Postdoctoral Fellowship~PDF/2022/002988. We thank the anonymous referee for their valuable comments, which have helped us improve our manuscript.

\bibliographystyle{JHEP}
\bibliography{references}
\end{document}